	\documentclass[12pt,twoside,openany,titlepage,letter]{ksuthesis-play}
	\usepackage{fancyhdr}
	\usepackage[Bjarne]{fncychap}
	\usepackage{titlesec}
	\titleformat{\section}{\large\bfseries}{\thesection}{1em}{\hrule}

	\renewcommand{\headrulewidth}{0.5pt}
	
\usepackage{amsmath} % AMS Math Package
\usepackage{amsthm} % Theorem Formatting
\usepackage{amssymb}	% Math symbols such as \mathbb
\newcommand{\ket}[1]{\left| #1 \right>} % for Dirac bras
\newcommand{\bra}[1]{\left< #1 \right|} % for Dirac kets
\let\baraccent=\= % rename builtin command \= to \baraccent
\renewcommand{\=}[1]{\stackrel{#1}{=}} % for putting numbers above =

\theoremstyle{definition}

\theoremstyle{remark}

	\usepackage{doublespace} 

	\usepackage{amsmath}

	\usepackage{amssymb}

	\usepackage{ifthen}
	\usepackage[dvips]{graphicx}
	\graphicspath{{.}{/}}	% look in this directory and the subdirectory called Figures

	\usepackage{epsfig}
	\usepackage{psfrag}

	\usepackage{textcase}

	\usepackage{xspace}

	\usepackage{subfigure}

	\usepackage{calc}

	\usepackage{tabularx}
	
	\usepackage{enumerate}

	\usepackage{dcolumn}

	\usepackage{makeidx}
	\makeindex
	
	\usepackage{afterpage}

	\usepackage{import}
	
	\usepackage{lastpage}
	
	\usepackage{sidecap}
	\sidecaptionvpos{figure}{c}
	
	\usepackage{wrapfig}
	
\begin{document}							% Begin the document
	\normalsize								% Return to Normal font size
	\pagestyle{plain}
%	\begin{spacing}{1.0}							% Single Space the Text
	\begin{spacing}{2.0}							% Double Space the Text
	\frontmatter
\begin{titlepage}
\thispagestyle{empty}
\begin{center}
%{\Large {\bf{PHANTOM \\ Polarized $^3$He($e,e'n$) Asymmetries in Three Orthogonal Measurements}}}
POLARIZED $^3$He($e,e'n$) ASYMMETRIES IN THREE ORTHOGONAL MEASUREMENTS
%PHANTOM \\ Polarized $^3$He($e,e'n$) Asymmetries in Three Orthogonal Measurements
\end{center}
 \vspace {6cm}
\begin{center}
A dissertation submitted\\ to Kent State University in partial\\ fulfillment of the requirements for the\\ degree of Doctor of Philosophy
%A dissertation submitted to\\ Kent State University\\ in partial fulfillment of the\\ requirements for the degree of\\ Doctor of Philosophy
\end{center}
 \vspace {3cm}
\begin{center}
by\\Elena Amanda Long\\August, 2012
\end{center}
\end{titlepage}					% Include the title page
	\begin{titlepage}
%\begin{right}
%{\Large{ABSTRACT}}
%\end{right}
\hfill\Large{ABSTRACT}
\vspace{40 pt}

\normalsize\begin{singlespace}
%Elena Amanda Long, Ph.D., August 2012 ~~~~~~~~~~~~~~~~~~~~~~~~~~~~~~~~~~~~~~~~~~~~ Physics\\
\noindent Elena Amanda Long, Ph.D., August 2012 \hfill Physics\\ \\
{\bf{POLARIZED $^3$He($e,e'n$) ASYMMETRIES IN THREE ORTHOGONAL MEASUREMENTS}}  (\pageref{LastPage} pp.)
\\ \\
Directors of Dissertation: Bryon D. Anderson and Douglas W. Higinbotham\\[1.2cm]
\end{singlespace}

Asymmetry measurements were conducted in Jefferson Lab's experimental Hall A through electron scattering from a polarized $^3$He target in the quasi-elastic $^3\mathrm{He}(e,e'n)$ reaction. Measurements were made with the target polarized in the longitudinal direction with respect to the incoming electrons ($A_L$), in a transverse direction that was orthogonal to the beam-line and parallel to the q-vector ($A_T$), and in a vertical direction that was orthogonal to both the beam-line and the q-vector ($A_y^0$). The experiment measured $A_y^0$ at four-momentum transfer squared (Q$^2$) of 0.127 (GeV/$c$)$^2$, 0.456 (GeV/$c$)$^2$, and 0.953 (GeV/$c$)$^2$. The $A_T$ and $A_L$ asymmetries were both measured at Q$^2$ of 0.505 (GeV/$c$)$^2$ and 0.953 (GeV/$c$)$^2$. This is the first time that three orthogonal asymmetries have been measured simultaneously. Results from this experiment are compared with the plane wave impulse approximation (PWIA) and Faddeev calculations. These results provide important tests of models that use $^3$He as an effective neutron target and show that the PWIA holds above Q$^2$ of 0.953 (GeV/$c)^2$.

\end{titlepage}					% Include the abstract
	\frontmatter
\newpage
\pagestyle{empty}
\setcounter{page}{5}
\begin{center}
Dissertation written by\\Elena Amanda Long\\
\begin{singlespace}B.S., Juniata College, 2006\\M.A., Kent State University, 2008\\Ph.D., Kent State University, 2012\end{singlespace}
\end{center}
\vspace{2.5mm}
\begin{center}
Approved by
\end{center}

\begin{singlespace}
\begin{tabbing}
	\makebox[2in]{\hrulefill}, \=Dr. Bryon D. Anderson, \\
	\>Co-chair, Doctoral Dissertation Committee \\ \\[8pt]
	\makebox[2in]{\hrulefill}, \>Dr. Douglas W. Higinbotham, \\
	\>Co-chair, Doctoral Dissertation Committee \\ \\[8pt]
	\makebox[2in]{\hrulefill}, \>Dr. D. Mark Manley, Professor, Physics \\ \\[8pt]
	\makebox[2in]{\hrulefill}, \>Dr. Peter C. Tandy, Professor, Physics \\ \\[8pt]
	\makebox[2in]{\hrulefill}, \>Dr. Diane Stroup, Professor, Chemistry \\ \\[8pt]
\end{tabbing}
\end{singlespace}
\begin{center}
	Accepted by
\end{center}
\begin{singlespace}
\begin{tabbing}
	\makebox[2in]{\hrulefill}, \=Dr. James T. Gleeson, Chair, \\
	\> Department of Physics\\ \\[8pt]
	\makebox[2in]{\hrulefill}, \>Dr. Timothy Chandler, Dean, \\
	\>College of Arts and Sciences\\ \\[8pt]
\end{tabbing}
\end{singlespace}

% Dr. Bryon D. Anderson, ~~~~~~~~~~~~~~~Co-chair, Doctoral Dissertation Committee \\[10pt]
% Dr. Douglas W. Higinbotham, ~~~~~~~Co-chair, Doctoral Dissertation Committee \\[10pt]
% Dr. D. Mark Manley,  ~~~~~~~~~~~~~~~~~~~Member, Doctoral Dissertation Committee \\[10pt]
% Dr. Peter C. Tandy,   ~~~~~~~~~~~~~~~~~~~~Member, Doctoral Dissertation Committee \\[10pt]
% Dr. Diane Stroup,  ~~~~~~~~~~~~~~~~~~~~~~~Member, Doctoral Dissertation Committee \\[10pt]
%\begin{center}
%Accepted by
%\end{center}
% Dr. James T. Gleeson,   ~~~~~~~~~~~~~~~Chair,  Department of Physics\\[10pt]
% Dr. Timothy Moerland,   ~~~~~~~~~~~~~~~~~~~Dean, College of Arts and Sciences\\[10pt]

% \_\hrulefill\ , Chair,  Department of Physics~~~~~~~~~~~~~~~~~ \\[10pt]
% \_\hrulefill\ , Dean, College of Arts and Sciences~~~~~~~~~~~~~~~~
%\begin{center}
%Approved by
%\end{center}
% \_\hrulefill\ , Chair, Doctoral Dissertation Committee ~~~~ \\[10pt]
% \_\hrulefill\ , Co-chair, Doctoral Dissertation Committee \\[10pt]
% \_\hrulefill\ ,  Members, Doctoral Dissertation Committee \\[10pt]
% \_\hrulefill\ ~~~~~~~~~~~~~~~~~~~~~~~~~~~~~~~~~~~~~~~~~~~~~~~~~~~~~~~~~~~\\[10pt]
% \_\hrulefill\ ~~~~~~~~~~~~~~~~~~~~~~~~~~~~~~~~~~~~~~~~~~~~~~~~~~~~~~~~~~~\\[10pt]				% Include the approval page

	\newpage
%	\setlength{\topmargin}{-0.1875 in}
%	\setlength{\oddsidemargin}{0.4375 in} 
%	\setlength{\evensidemargin}{0.0625 in}
%	\setlength{\textheight}{8.5 in}
%
%	\ChTitleVar{\vspace{-0.75 in}\raggedleft \Large\rm}

	\tableofcontents							% Make a table of contents
	\listoffigures								% Make a list of figures
	\addcontentsline{toc}{chapter}{List of Figures}
	\listoftables								% Make a list of tables
	\addcontentsline{toc}{chapter}{List of Tables}
	\newpage
%	\addcontentsline{toc}{chapter}{Acknowledgements}
	%\noindent{\hfill\Large\bf{Acknowledgement}}\\
%\begin{center}\noindent{{Acknowledgements}} \end{center}
%\vspace{0.5625 in}
\begin{acknowledgements}
%\vspace{2 in}
%\hfill\Large{ACKNOWLEDGEMENTS}\normalsize
%\vspace{40 pt}
%\vspace{0.4375 in}

The journey that has brought me to this accomplishment would not have been possible without the help and support of countless people. Any listing of these people will undoubtably leave some incredible people out, but I do wish to acknowledge those who stand out in this path.

My parents, Barry and Susan, were paramount in setting me up with a good start in life. They fostered my curiosity from a young age, even if it occasionally made them shake their heads. They taught me lessons on life and learning, and will always be my role models. I would not have reached this point in life without their love and support. 

Every path that leads down the long road of academia does so because of incredible teachers and mentors along the way. During high school, Herman Fligge opened my eyes to the beauty in mathematics. In college, Jamie White, James Borgardt, Norm Siems, Mark Pearson, Barry Bruce, and Mary Atchley turned my curiosity about physics into a passion. Without their teachings, mentorship, and support, graduate school would not have been an option for me. This passion has begun to mature through the advice of Bryon Anderson and Doug Higinbotham. They have opened the doors of exploration for me.

Navigating a journey is difficult without a guide. I would have been lost in the process many times over if it weren't for help of Kim Birkner, Loretta Hauser, Chris Kurtz, and Cindy Miller at Kent State and Stephanie Harris at Jefferson Lab. 

Friendship and support from colleagues has also been essential in providing a welcoming climate in the field. Many nights were spent working on problems and just beginning to think about the universe with Michael Best, Alison Earnhart, Dan Sidor, Jeremy Ellden, and every other student in the Juniata physics department. In graduate school, Tanya Ostapenko, Kelly Reidy, and Jason Ellis gave me their friendship and support. My life has truly been richer from getting to know each and every one of them. Moving to and working at a national laboratory brought with it new challenges and new friends. Vince Sulkowsky and Aidan Kelleher kindly shared both their experience and friendship. Our experiment would also not have been possible without the countless hours spent in the counting house, working on equipment, and analyzing data by my fellow graduate students Jin Ge, Miha Mihovilovic, and Yawei Zhang. 

Colleagues in physics, but distant to the work presented, have proved crucial towards making physics open and accepting. In particular, I would like to thank Nicole Ackerman, Timothy Atherton, Wouter Deconinck, Michael Falk, Savannah Garmon, and Edward Henry for their continued hard work. I can honestly say that I've never worked with a finer group of people so dedicated on making physics available to everyone.

In order to work efficiently, time must be spent on centering and exploration of self. During my college years, this was spent with a group of friends, the Island of Misfit Toys. In my early graduate years,  Joyce Murton, Stacy Pugh, Jamie Roberts, and Amy House gave me a space to be myself. I am grateful for their friendship. I have Reg Richburg, De Sube, Michelle-Marie and Cameron McKay, and Deb Wilson to thank for turning Virginia into a home.

Of all those who have been with me on this path, none has known me better than my husband, Julian. He has been infinitely patient and caring throughout every challenge we've faced. I could not ask for a better friend or more loving companion to continue this journey with me. I am thankful for every moment.

As I mentioned earlier, there are innumerable people who have supported me which has led to accomplishing this dissertation. I wish to acknowledge that this work also belongs to them. All achievements are reached because of the community that made them possible. This work is yours as much as it is mine.

\vspace{10pt}
\hfill\today
\end{acknowledgements}			% Include the acknowledgements
	\mainmatter								% Reset the Numbering to Arabic 1
	%\makehalftitle\thispagestyle{empty}			% Make the half page title

%	\ChTitleVar{\vspace{0 in}\raggedleft \Large\rm}
%	\setlength{\oddsidemargin}{0.5625 in} 
%	\setlength{\evensidemargin}{-0.03125 in}
%	\setlength{\topmargin}{0.375 in}
%	\setlength{\textheight}{8.0 in}

\makeatletter
\renewcommand*{\@makechapterhead}[1]{%
  \vspace*{-0.50625 in}%
  {\parindent \z@ \raggedright \normalfont
    \ifnum \c@secnumdepth >\m@ne
      \if@mainmatter%%%%% Fix for frontmatter, mainmatter, and backmatter 040920
        \DOCH
      \fi
    \fi
    \interlinepenalty\@M
    \if@mainmatter%%%%% Fix for frontmatter, mainmatter, and backmatter 060424
      \DOTI{#1}%
    \else%
      \DOTIS{#1}%
    \fi
  }
  }
% For the case \chapter*:
\renewcommand*{\@makeschapterhead}[1]{%
  \vspace*{5\p@}%
  {\parindent \z@ \raggedright
    \normalfont
    \interlinepenalty\@M
    \DOTIS{#1}
    \vskip 20\p@
  }}
\makeatother

	% Include Chapters
	% *********************************************************************
%	\large
	\pagestyle{fancy}
%	\input{test_page}				% Test Page full of Lorem Ipsum
	% vvvvvvvvvvvvvvvvvvvvvvvvvvvvvvvvvvvvvvvvvvvvvvvvvvvvvvvvvvvvvvvvvvvvvvvvvvvvvvvv
% Chapter 1 (chapter_1.tex)
%
% Chapter 1 of Elena Long's Ph.D. Dissertation
%
% To be completed: March, 2012
%
% ^^^^^^^^^^^^^^^^^^^^^^^^^^^^^^^^^^^^^^^^^^^^^^^^^^^^^^^^^^^^^^^^^^^^^^^^^^^^^^^^

\chapter{Introduction}	% Chapter Title
\label{introduction}		% Chapter Label
\normalsize			% Return to Normal font size

% Introduction
% vvvvvvvvvvvvvvvvvvvvvvvvvvvvvvvvvvvvvvvvvvvvvvvvvvvvvvvvvvvvvvvvvvvvvvvvvvvvvvvv
Atoms, which constitute most of normal matter, are made of electrons ($e$) and nuclei. Particles that make up nuclei are called nucleons ($N$), which can be of two types: protons ($p$) and neutrons ($n$). The nucleons themselves consist of smaller particles known as quarks that are held together by the exchange of particles of the strong nuclear force, called gluons. Nucleons are primarily comprised of two different flavors of quarks: up and down. In the simplified constituent quark model the proton is comprised of two up quarks and one down quark and the neutron is composed of one up and two down quarks.

Interactions between these particles are used to study the internal structure of nucleons. For example, consider that $^3$He nuclei, which are comprised of two protons and one neutron, are impinged upon by a beam of electrons. If the incoming electron interacts with the $^3$He nucleus with low energy and transferred momentum (usually by the exchange of a single virtual photon, $\gamma^*$) such that the nucleus remains intact and in its ground state (lowest energy state) after the interaction, this is called elastic scattering. In another case, the electron may interact with the $^3$He nucleus with higher energy and transferred momentum and a single nucleon is knocked free from the nucleus, but the nucleon remains intact and in its ground state. This is called quasi-elastic scattering. At even higher energy and momentum transferred, the electron can interact directly with a single quark, which can break the nucleon apart. This is called deep inelastic scattering. 

A shorthand notation is often used to describe the interaction channel that is measured. For example, assume that an electron beam is incident upon a $^3$He nucleus and knocks out a neutron that is detected. The notation for this would be $^3$He($e,e'n$), where $^3$He represents the target, $e$ represents the incoming electron, $e'$ represents the scattered electron, and $n$ represents the scattered neutron. 

One of the observables that is well suited to extracting structure information of nucleons is spin asymmetry. Each of the particles mentioned carry a quantum property known as spin \cite{springerlink:10.1007/BF01397326}, which is mathematically similar to classical rotational angular momentum but with quantized properties. Nucleons are spin 1/2 particles and can be in one of two states called spin up and spin down. The direction of the spin can be controlled and measured through the use of magnetic fields. An asymmetry measurement is useful in determining if one of the spin states dominates the other one. A simplified example would be
\begin{equation}
A = \frac{N_{\uparrow} - N_{\downarrow}}{N_{\uparrow} + N_{\downarrow}},
\end{equation}
where $N_{\uparrow}$ is the number of detected particles with spin up, $N_{\downarrow}$ is the number of detected particles with spin down, and $A$ is the asymmetry. A similar asymmetry can be made with helicity, which is simply the projection of the spin ($\vec{s}$) onto the direction of momentum ($\hat{p}$) and is written as
\begin{equation}
h = \vec{s} \cdot \hat{p}.
\end{equation}

Electron scattering is a well understood process that is useful for probing the internal structure of nucleons \cite{Walecka:2001gs}.  The Thomas Jefferson National Accelerator Facility (Jefferson Lab) is a prime location to conduct these experiments due to its ability to produce a highly polarized continuous-wave electron beam. Experimental Hall A at Jefferson Lab is particularly suited to perform asymmetry measurements due to its polarized $^3$He target and high resolution spectrometers. 

This dissertation is organized into six chapters. Chapter \ref{introduction} discusses the theoretical motivation for the measurements taken and places them within a historical context. Chapter \ref{theory} provides an overview of the theoretical calculations that are being tested. Chapter \ref{experimentsetup} describes the equipment used throughout the experiment. This includes both information about the electron beam and the equipment within Jefferson Lab's experimental Hall A. The methods used for particle identification are described in Chapter \ref{particleid}. Correction factors adjusting the asymmetry measurement, such as dilutions, as well as the error analysis methods used are discussed in Chapter \ref{dilution-uncertainties}. Results from the measurements are presented in Chapter \ref{results}. Supplemental material that describes the cuts used on the neutron detector is in Appendix \ref{hand-vetoes} and a list of collaborators is in Appendix \ref{Collaboration_list}.

% ^^^^^^^^^^^^^^^^^^^^^^^^^^^^^^^^^^^^^^^^^^^^^^^^^^^^^^^^^^^^^^^^^^^^^^^^^^^^^^^^

\large
\section {Motivation}
\label{ch1-motivation}
\normalsize
% Theoretical Motivation
% vvvvvvvvvvvvvvvvvvvvvvvvvvvvvvvvvvvvvvvvvvvvvvvvvvvvvvvvvvvvvvvvvvvvvvvvvvvvvvvv

\begin{figure}
	\centering
	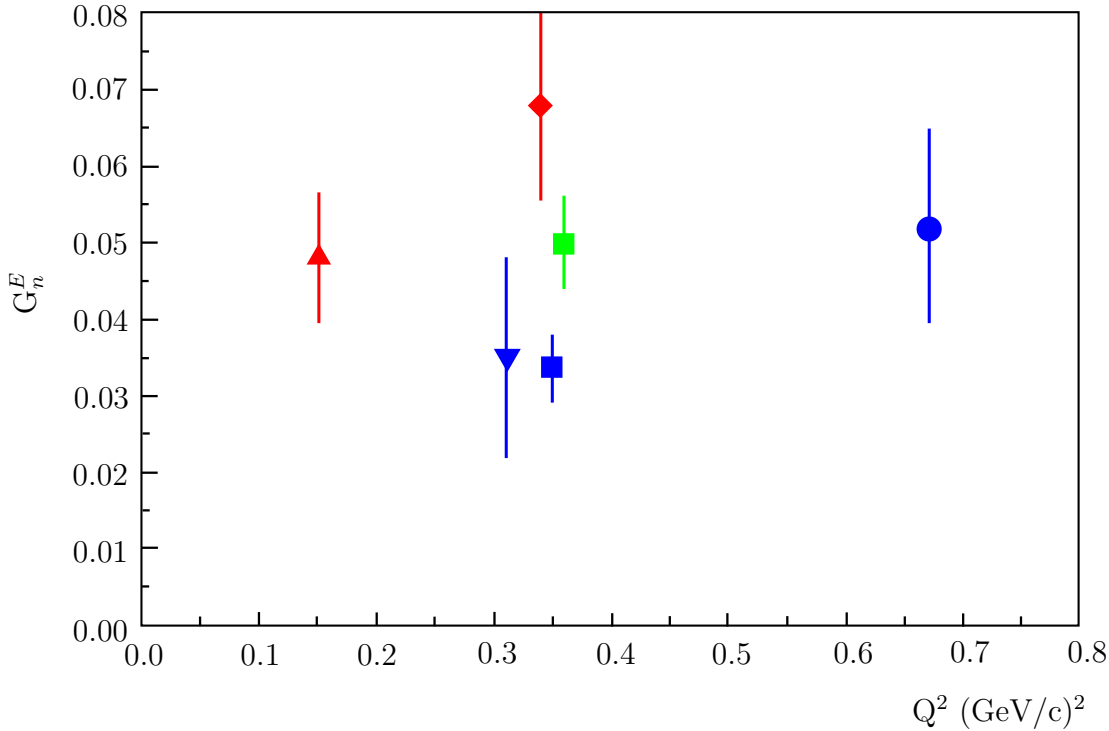
	\caption [Discrepancy Between $^3$He and $^2$H G$^n_E$ extractions Circa 1999] {This figure emphasizes the discrepancy between $^3$He and $^2$H G$^n_E$ extractions circa 1999. The red $\blacktriangle$ \cite{Herberg:1999ud} and $\blacklozenge$ \cite{Ostrick:1999xa} correspond to extractions of $G^n_E$ from deuterium where the blue $\blacktriangledown$ \cite{Meyerhoff:1994ev}, $\blacksquare$ \cite{Becker:1999tw}, and $\bullet$ \cite{Rohe:1999sh} correspond to extractions from $^3$He using PWIA models with small contributions from FSI and MEC. The green $\blacksquare$ \cite{Golak:2000nt} corresponds to $^3$He data using models that include larger contributions from FSI.}
	\label{3he-d-discrepancy}
\end{figure}

Information of the charge and magnetization carried by nucleons is described by the electromagnetic nucleon form factors \cite{Perdrisat:2006hj}. In the non-relativistic case, the form factors are simply the Fourier transforms of the rest frame spatial distributions of the charge and magnetization \cite{Miller:2008jc}. The nucleon form factors are not direct observables and thus must be extracted from observables through the use of theoretical models. Assumptions made in producing the models can have a large effect on the extraction of the neutron form factors. For example, there was a discrepancy between extractions of the electric form factor of the neutron, $G_E^n$, obtained from deuterium scattering and those from $^3$He as seen in Figure \ref{3he-d-discrepancy} \cite{Golak:2000nt}.

The original models were based on the plane-wave impulse approximation (PWIA) to extract the form factors from asymmetry measurements. At low momentum transfer, the simple PWIA is known not to describe experimental results accurately due to the effects of meson exchange currents (MEC) and final-state interactions (FSI). Meson exchange currents are used to describe nucleon-nucleon interaction potentials as the exchange of virtual massive particles, namely mesons \cite{Yukawa:1935xg, Kemmer04051938}. MEC contributions to electron scattering are valid within certain energies which the data taken in this experiment fall within \cite{PTP.6.581}. Thus, apart from the quasi-free scattering amplitude, there will be contributions from direct coupling to the electromagnetic currents of exchanged mesons. Final-state interactions are also important since the final state is a system of three interacting nucleons rather than simple plane waves \cite{Cahill:1974zz}. To leading order, FSI can be considered as rescattering of the struck nucleon (the neutron here) by the residual nucleus.  

\begin{figure}
	\centering
%	\import{./}{nikhef-asym.eps_tex}
	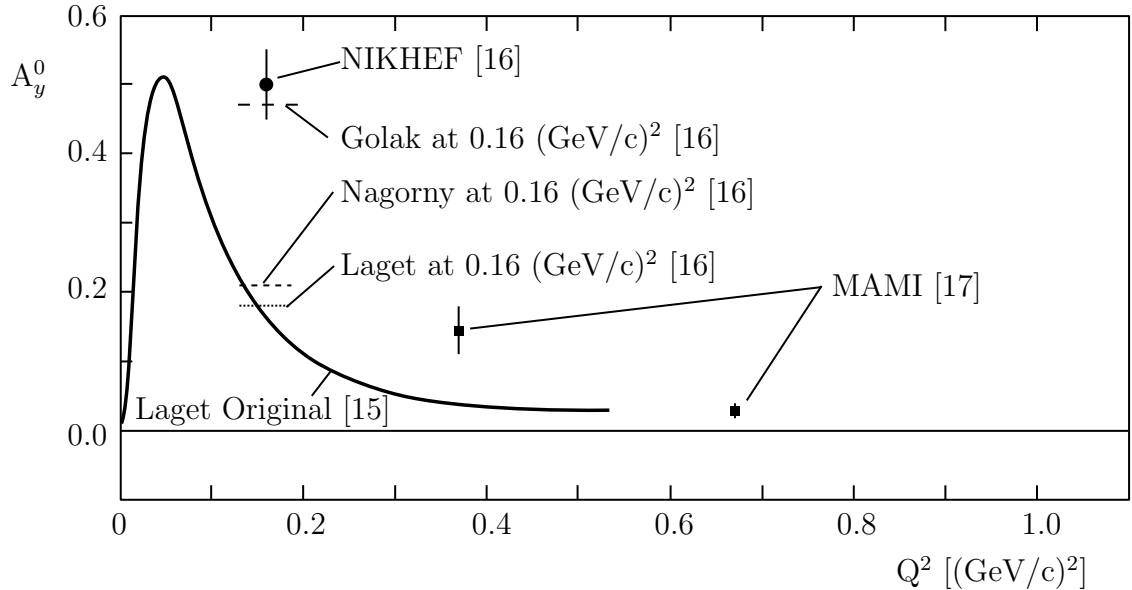
	\caption [Previous A$_y^0$ Asymmetry Measurements] {Vertical single-spin asymmetry measurements, $A_y^0$, from NIKHEF at Q$^2$=0.16 (GeV/$c)^2$, MAMI at Q$^2$=0.37 and 0.67 (GeV/c)$^2$, and various theoretical models are plotted. The Bochum group used Fadeev calculations to calculate the FSI whereas the others are modified PWIA. The PWIA predicts this asymmetry to be exactly zero.}
	\label{nikhef-asym}
	\end{figure}

In the PWIA, a single spin asymmetry transverse to the scattering plane has been calculated to be exactly zero. Early predictions expected contributions from FSI and MEC to be small above a squared momentum transfer (Q$^2$) of 0.2 (GeV/$c$)$^2$, as can be seen by Laget's original calculation \cite{Laget:1991pb} in Figure \ref{nikhef-asym}. In the same figure, there is a data point from an experiment that was done at the Nationaal Instituut voor Kernfysica en Hoge-Energiefysica (NIKHEF), which showed this asymmetry to be larger than expected. The Bochum theoretical group, which correctly predicted the observed asymmetry, used full Fadeev calculations that correctly incorporated the significant effects of FSI \cite{Poolman:1999uf}. Extractions of the electric form factor of the neutron need to take these corrections into account, which led to a re-analysis of the data in Figure \ref{3he-d-discrepancy} and largely removed the discrepancy between $^2$H and $^3$He data as can be seen by the green $\blacksquare$. Another measurement was later made by MAMI \cite{Bermuth:2003qh} and is also shown on Figure \ref{3he-d-discrepancy}.

A target single-spin asymmetry has not previously been measured at high Q$^2$, leaving contributions from FSI and MEC in this region largely unknown. The current experiment measured this spin asymmetry (A$_y^0$) at Q$^2$=0.127, 0.456, and 0.953 (GeV/$c$)$^2$. 

Extractions of the electric form factor of the neutron can be made from a double-spin asymmetry, where the beam is polarized with helicity along the beam-line and the target spin is polarized along the direction of the quark q-vector. The experiment also measured this transverse asymmetry (A$_T$) at Q$^2$= 0.505 and 0.953 (GeV/$c)^2$. 

$^3$He($e,e'n$) asymmetry measurements in three orthogonal directions have never been previously measured simultaneously. The current experiment is also the first to measure the longitudinal beam helicity asymmetry (A$_L$) at Q$^2$=0.505 and 0.953 (GeV/$c)^2$. These measurements provide significantly improved tests of the various theoretical predictions, which are discussed in detail in Chapter \ref{theory}.

% ^^^^^^^^^^^^^^^^^^^^^^^^^^^^^^^^^^^^^^^^^^^^^^^^^^^^^^^^^^^^^^^^^^^^^^^^^^^^^^^^

\large
\section {Experiment Overview} 
\label{ch1-expoverview}
\normalsize
% Experiment Overview
% vvvvvvvvvvvvvvvvvvvvvvvvvvvvvvvvvvvvvvvvvvvvvvvvvvvvvvvvvvvvvvvvvvvvvvvvvvvvvvvv

\begin{figure}
	\centering
	\includegraphics{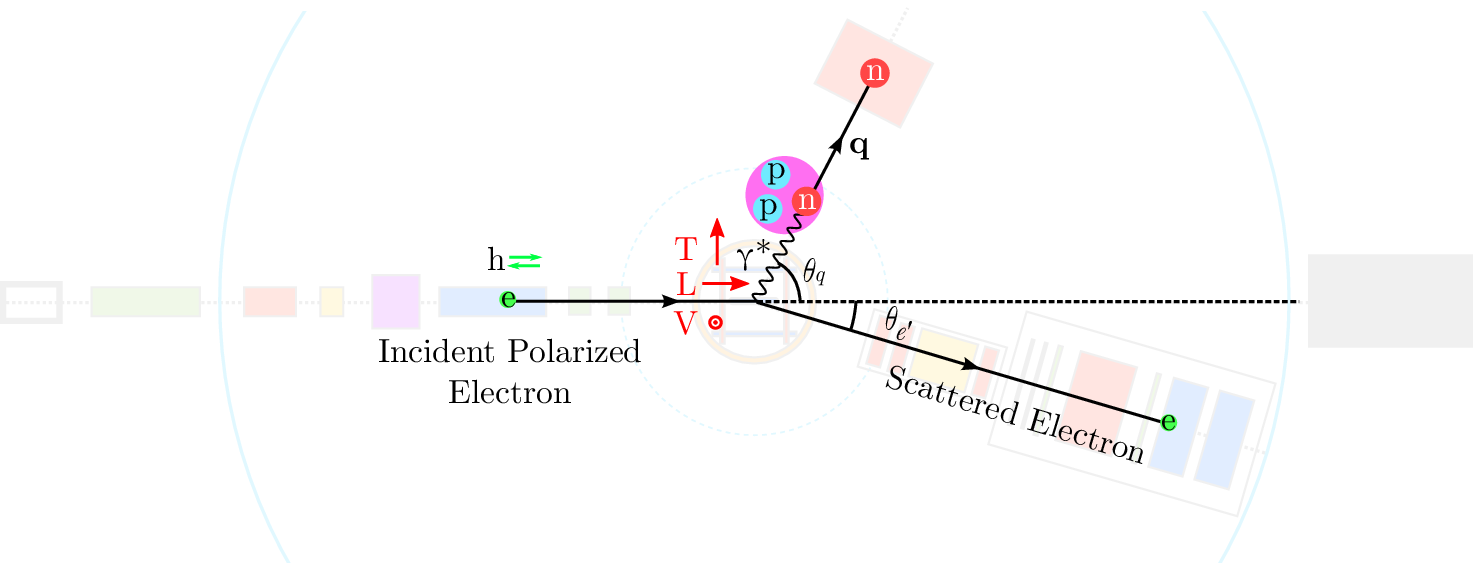}
	\caption [Definition of Polarization Directions] {Definition of Polarization Directions. Vertical target polarization (V) was used to measure A$_y$, transverse target polarization (T) was used to measure A$_z$, and longitudinal target polarization (L) was used to measure A$_x$.}
	\label{eenDefineDirections}
\end{figure}

The present experiments were performed at the Thomas Jefferson National Accelerator Facility in Newport News, Virginia in experimental Hall A. Experiments (E05-015 \cite{e05015}, E05-102 \cite{e05102}, and E08-005 \cite{e08005}) were conducted to learn about the polarized $^3$He states as well as interactions that occur in electron scattering on $^3$He. A $^3$He target was used that could be polarized in three orthogonal directions. The first, defined as longitudinal, was parallel to the incident electron beam. The second, defined as vertical, was orthogonal to both the incident electron beam as well as the neutron trajectory. The third, defined as transverse, was orthogonal to the incident electron beam and parallel to the neutron trajectory. Each of these is indicated in Figure \ref{eenDefineDirections}.

\begin{figure}
	\centering
	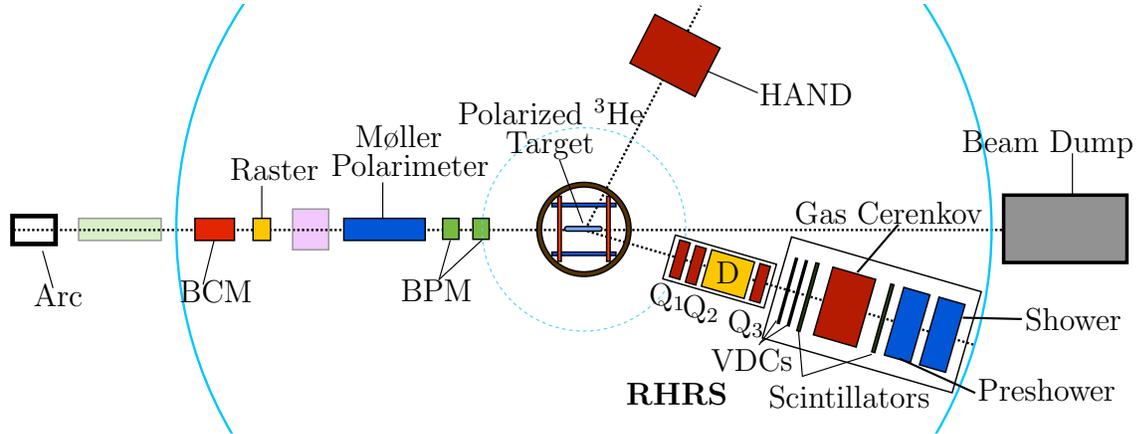
	\caption [Hall A Equipment used for $^3$He($e,e'n$) Measurements] {Hall A equipment used for $^3$He($e,e'n$) measurements}
	\label{eenHallLayout}
\end{figure}

The incoming electron beam had a polarization of approximately 80\%. The beam, at energies of 1.2, 2.4, and 3.6 GeV, was incident on a 40 cm long $^3$He cell that was capable of being polarized up to 60\% in the vertical, longitudinal, or transverse directions. The scattered electrons were detected in a High Resolution Spectrometer (HRS) that consisted of three focusing quadrupole magnets, one bending dipole magnet, and a pair of scintillators, wire tracking chambers, a gas Cerenkov detector, and lead-glass calorimeters used for particle identification, as shown in Figure \ref{eenHallLayout}. The knocked out neutrons were detected by the Hall A Neutron Detector (HAND), which consisted of a matrix of 88 plastic scintillator bars, each 10 cm thick and arranged in four layers with a veto layer in front that consists of 64 2-cm thick scintillator bars. HAND has a total thickness of 40 cm. 

A coincidence measurement was performed between the HRS and HAND that correlated the scattered electrons with the knocked-out neutrons. The target had repeated spin-flips throughout the experiment where the polarization of the $^3$He was rotated by 180$^{\circ}$, giving `up' and `down' states oriented in the vertical, longitudinal, and transverse directions. The asymmetries were measured with the target polarized in each of these three directions. Of particular importance was the vertical and transverse asymmetries. The measurement of the vertical single-spin asymmetry provided new constraints on models of G$^n_E$, as discussed in Section \ref{ch1-motivation}, while a measurement of the transverse double-spin asymmetry allows an extraction of G$^n_E$ to be made.

% ^^^^^^^^^^^^^^^^^^^^^^^^^^^^^^^^^^^^^^^^^^^^^^^^^^^^^^^^^^^^^^^^^^^^^^^^^^^^^^^^

					% Include the Chapter 1 text (chapter_1.tex)
	% vvvvvvvvvvvvvvvvvvvvvvvvvvvvvvvvvvvvvvvvvvvvvvvvvvvvvvvvvvvvvvvvvvvvvvvvvvvvvvvv
% Chapter 2 (chapter_2.tex)
%
% Chapter 2 of Elena Long's Ph.D. Dissertation
%
% To be completed: March, 2012
%
% ^^^^^^^^^^^^^^^^^^^^^^^^^^^^^^^^^^^^^^^^^^^^^^^^^^^^^^^^^^^^^^^^^^^^^^^^^^^^^^^^

\chapter{Theory}	% Chapter Title
\label{theory}		% Chapter Label
\normalsize			% Return to Normal font size
% Theory
% vvvvvvvvvvvvvvvvvvvvvvvvvvvvvvvvvvvvvvvvvvvvvvvvvvvvvvvvvvvvvvvvvvvvvvvvvvvvvvvv

\large
\section {$^3$He Ground State}
\label{3he-ground-state}
\normalsize
% Theoretical Motivation
% vvvvvvvvvvvvvvvvvvvvvvvvvvvvvvvvvvvvvvvvvvvvvvvvvvvvvvvvvvvvvvvvvvvvvvvvvvvvvvvv
The experimental study of the internal structure of the proton is relatively straight-forward due to readily available free proton targets. This is not the case for studying the internal structure of the neutron, since free neutron targets are not available. As such, low-A targets, where the nucleons are weakly bound, are often used to approximate a free neutron target. This is most often done using $^2$H or $^3$He. Deuterons are advantageous in that they provide the closest approximation to a free neutron target and are extremely useful for cross-section measurements. $^3$He is uniquely suited for measurements that involve the spin of the neutron since the dominant state of the $^3$He wave-function is the ground-state configuration, where the two protons have anti-parallel spins with respect to each other. This causes the spin of the entire nucleus to be approximately the spin of the neutron. There are complications, as illustrated in Figure \ref{3he-ground-state}, however the S-state makes up ${\raise.17ex\hbox{$\scriptstyle\sim$}}90\%$ of the $^3$He target wave function \cite{PhysRevC.48.38}. Additionally, the magnetic moment of the neutron is almost identical to that of $^3$He \cite{Nakamura:2010zzi, RevModPhys.77.1}.

\begin{SCfigure}
	\centering
	\includegraphics[width=3.75 in]{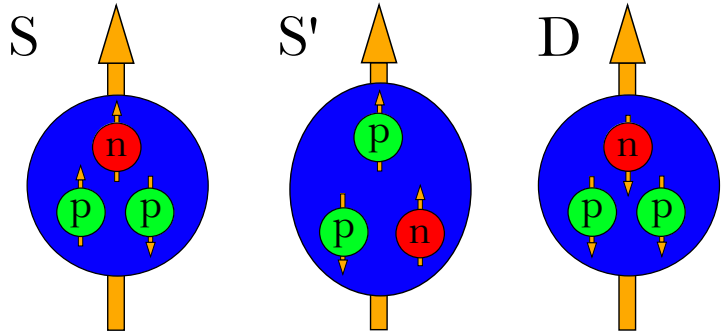}
	\caption [$^3$He States] {This cartoon is a representation of the three most common states of the $^3$He target. The S-state, where the proton spins are aligned anti-parallel to each other, makes up approximately 90$\%$ of the $^3$He wave function, which makes this nucleus an ideal candidate for studying neutron spin physics \cite{PhysRevC.48.38}.}
	\label{3he-ground-state}
\end{SCfigure}

The simplest method of describing the $^3$He($e,e'n$) reaction is with the plane-wave impulse approximation (PWIA), which is discussed in detail in Section \ref{pwia}. Due to using multi-nucleon targets, extra effects from final-state interactions (FSI) and meson-exchange currents (MEC) must be taken into account. This is especially true at lower momentum-transfer where the contributions of each is amplified. These reactions are discussed in detail in Section \ref{fsi-mec}. Full three-body calculations, known as Faddeev calculations, are very well suited to describing the $^3$He states at low momentum-transfer (Q$^2\lesssim 0.5$ (GeV/$c)^2$). These calculations are discussed in detail in Section \ref{faddeev}. As momentum-transfer is increased, relativistic effects must be taken into account. Full Faddeev calculations are not available in this kinematic region.

% ^^^^^^^^^^^^^^^^^^^^^^^^^^^^^^^^^^^^^^^^^^^^^^^^^^^^^^^^^^^^^^^^^^^^^^^^^^^^^^^^

\large
\section {Formalism}
\label{formalism}
\begin{SCfigure}
	\centering
	\includegraphics[width=4 in]{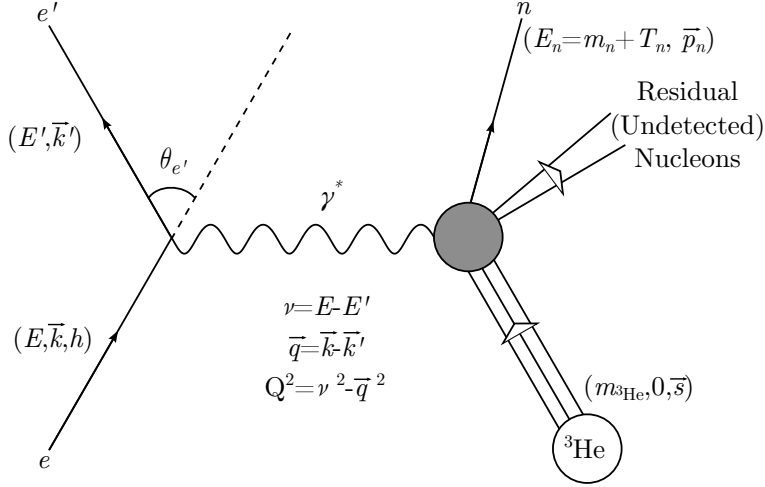}
	\caption [Scattering Definitions] {This diagram represents the $^3$He($e,e'n$) reaction where an incident electron, $e$, knocks a neutron, $n$, out of a $^3$He nucleus by exchange of a virtual photon, $\gamma^*$, through energy (momentum) exchange $\nu$ ($\vec{q}$) . The scattered electron is deflected at angle $\theta_{e'}$. The incident (scattered) electron carries energy $E$ ($E'$), momentum $\vec{k}$ ($\vec{k'}$), and helicity $h$. The $^3$He nucleus is initially at rest with spin $\vec{s}$. The knocked-out neutron, with energy $E_n$ and momentum $\vec{p}$, is detected while the residual nucleons are not.}
	\label{basic-reaction}
\end{SCfigure}

\normalsize
% Theoretical Motivation
% vvvvvvvvvvvvvvvvvvvvvvvvvvvvvvvvvvvvvvvvvvvvvvvvvvvvvvvvvvvvvvvvvvvvvvvvvvvvvvvv
In order to discuss the ideas presented in this dissertation, a number of definitions must be made. Figure \ref{basic-reaction} demonstrates the reaction channel where an incident electron, $e$, with energy $E$, momentum $\vec{k}$, and helicity $h$ interacts with a $^3$He nucleon at rest through a virtual photon, $\gamma^*$. The scattered electron, $e'$, is deflected at an angle $\theta_{e'}$, has energy $E'$, and momentum $\vec{k'}$.

The electron loses some energy through the interaction of the exchanged photon, which has energy $\nu=E-E'$ and momentum transfer vector $\vec{q}=\vec{k}-\vec{k'}$. For each of the asymmetries presented in this dissertation, the energy transfer, $\nu$, is a useful quantity for showing how the asymmetry changes. The square of the four-vector momentum transfer is defined as Q$^2=\nu^2-\vec{q}~^2$ and is a useful quantity for showing differences of $A_y^0$ values. Another useful quantity is the Bjorken scaling variable $x_{Bj}$, which is defined as
\begin{equation}
	x_{Bj}=\frac{\mbox{Q}^2}{2m_N \nu},
\end{equation}
where $m_N$ is the mass of a nucleon. Quasi-elastic scattering occurs in the energy range where $\nu \approx \mbox{Q}^2/2m_N$ or, equivalently, where $x_{Bj}\approx1$.

\begin{figure}
	\centering
	\includegraphics[width=15 cm]{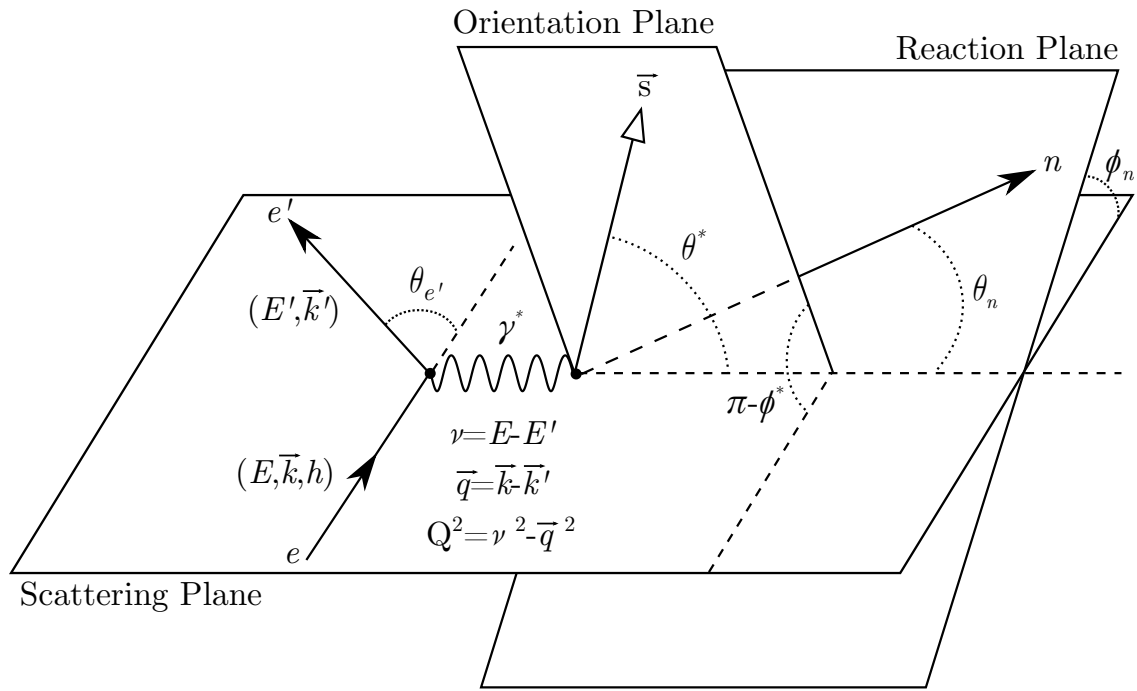}
	\caption [Angle Definitions] {This diagram represents the lab frame and defines angles of the reaction described in Figure \ref{basic-reaction}. The incident electron, $e$, exchanges a virtual photon, $\gamma^*$, with a nucleon, $n$. The scattering plane is defined as the plane created from the trajectory of $e$ and the the trajectory of the scattered electron, $e'$. The spin orientation plane is defined as the plane that contains the target spin direction, $\vec{s}$, and a line parallel to the $\vec{q}$ vector. The reaction plane is defined as the plane that contains the trajectory of $n$ and a line parallel to the $\vec{q}$ vector. Angles $\theta^*$ and $\phi^*$ describe the spin orientation plane with respect to the scattering plane. Angles $\theta_n$ and $\phi_n$ describe the orientation of the reaction plane to the scattering plane. This diagram is adapted from Reference \cite{Poolman:1999uf}.} 
	\label{angle-definitions}
\end{figure}

In addition, the polar ($\theta^*$) and azimuthal ($\phi^*$) angles of the target spin direction with respect to the q-vector are imperative to translate experimental asymmetries to theoretical calculations. Figure \ref{angle-definitions} represents these angles. 
The asymmetries measured in this experiment are of the form
\begin{equation}
	A(\theta^*,\phi^*)=\frac{1}{P}\cdot\frac{Y_{\uparrow}-Y_{\downarrow}}{Y_{\uparrow}+Y_{\downarrow}},
\end{equation}
where $P$ is the polarization of the target ($P_t$) for single-spin asymmetries or the polarization of the target times the polarization of the beam ($P_t \cdot P_b$) for double-spin asymmetries, and $Y_{\uparrow (\downarrow)}$ are the yields of spin-up (spin-down) events. 

Double-spin asymmetries are commonly used in the extraction of the neutron form factors. In particular, the asymmetries $A_{\parallel}=A(0^{\circ},0^{\circ})$ and $A_{\perp}=A(90^{\circ},0^{\circ})$ can be used to extract the electric form factor of the neutron ($G_E^n$). In the PWIA, this takes the form of
\begin{equation}
	G_E^n = \frac{b}{a}\cdot G_M^n \frac{(P_b P_t V)_{\parallel}}{(P_b P_t V)_{\perp}}\frac{A_\perp}{A_\parallel},
\end{equation}
where $G_M^n$ is the magnetic form factor of the neutron, $a$ and $b$ are kinematic factors, and $V_{\parallel(\perp)}$ are dilution factors \cite{Bermuth:2003qh}.

It is important to note that due to experimental constraints, the asymmetries measured and discussed in this dissertation, $A_T$ and $A_L$ deviate from $A_{\parallel}$ and $A_{\perp}$ (respectively) by a small rotation. However, the vertical target-spin asymmetry measured in this dissertation, $A_y^0$, is identical to the theoretical $A(90^{\circ},90^{\circ})$.

% ^^^^^^^^^^^^^^^^^^^^^^^^^^^^^^^^^^^^^^^^^^^^^^^^^^^^^^^^^^^^^^^^^^^^^^^^^^^^^^^^

\large
\section {Plane-Wave Impulse Approximation}
\label{pwia}
\normalsize
% Theoretical Motivation
% vvvvvvvvvvvvvvvvvvvvvvvvvvvvvvvvvvvvvvvvvvvvvvvvvvvvvvvvvvvvvvvvvvvvvvvvvvvvvvvv

\begin{SCfigure}
	\centering
	\includegraphics[width=4 in]{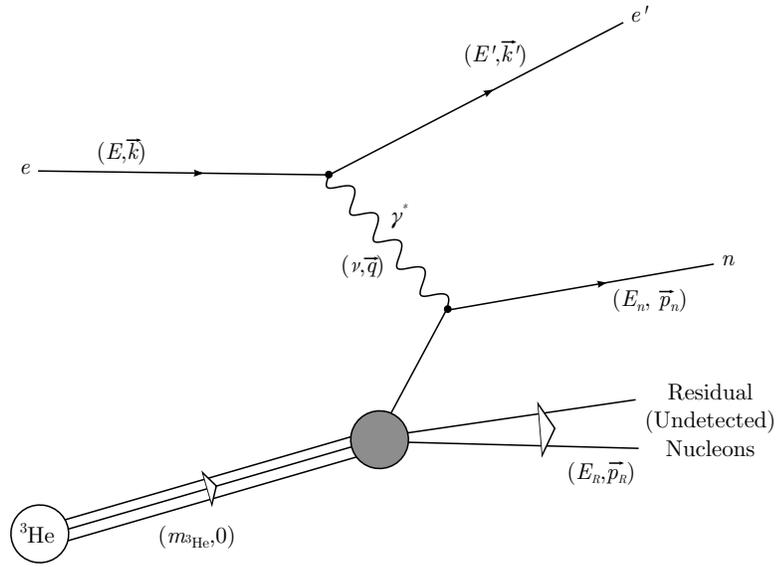}
	\caption [PWIA Diagram] {This diagram describes the plane-wave impulse approximation where an incident electron, $e$, scatters off of a single nucleon, $n$, in a $^3$He nucleus by exchanging a virtual photon, $\gamma^*$. In this approximation, no other interactions are taken into account. This diagram is adapted from Reference \cite{frois1991modern}.}
	\label{pwia-diagram}
\end{SCfigure}

The plane-wave impulse approximation is a model for describing electron scattering. In the PWIA for a knock-out reaction, it is assumed that a nucleon is cleanly knocked-out of a nucleus due to scattering from an incident electron without rescattering with the residual nucleus. This mechanism is presented in diagrammatic form in Figure \ref{pwia-diagram}. For scattering off of $^3$He, the residual nucleus is either a deuteron in the case of two-body break-up, or two unbound nucleons in the case of three-body break-up. 

In order to understand the PWIA, we start with the differential cross section of the electron-nucleon reaction, which can be written as the contraction of two tensors: the leptonic tensor, $\eta_{\mu\nu}$, and the hadronic tensor, $W^{\mu\nu}$, such that
\begin{equation}
	\frac{d\sigma}{dE'd\Omega_e d\Omega_N dE_N} = \frac{2 \alpha^2}{Q^4} \frac{p_NM_nM_b}{(2\pi)^3E_R}\eta_{\mu\nu}W^{\mu\nu}\delta(E_N+E_R-M_{^3\mathrm{He}}-\nu).
	\label{eq:lepton-hardron-cs}
\end{equation}
The hadronic tensor describes all of the nuclear structure and dynamics, which stems from the product of the nuclear electromagnetic transition currents $J^{\mu}(Q)^*_{fi}J^{\nu}(Q)_{fi}$. The leptonic tensor has been described in extensive detail in References \cite{Donnelly:1985ry} and \cite{Raskin:1988kc}. In the extreme relativistic limit, where $\gamma=\frac{1}{\sqrt{1-\beta^2}} \gg1$, the helicity of the electron only appears in the antisymmetric part of the tensor. The resulting expression can be separated into a symmetric and antisymmetric part by interchanging the indices $\mu$ and $\nu$.  Both tensors are then contracted by 
\begin{equation}
	2\eta_{\mu\nu}W^{\mu\nu}=v_0(R_{fi}+hR'_{fi}),
\end{equation}
where
\begin{equation}
	v_0\equiv 4EE'\cos^2{\frac{\theta_e}{2}}.
\end{equation}

The leptonic tensor can be projected onto the coordinate system described in Figure \ref{angle-definitions} such that $\hat{z}\parallel \vec{q}$, $\hat{y}\parallel (\vec{k}\times\vec{k'})$, and $\hat{z}\parallel (\hat{y}\times\hat{z})$. This projection yields the kinematic factors $v_k$ and $v_{k'}$ where $k=L,T,TL,TT$ and $k'=T',TL'$ such that the energy transfer $\nu$, the four-momentum square Q$^2$, and the electron-scattering angle $\theta_{e'}$ are within these factors. From this, the six-fold differential cross section can be described as
\begin{align}
	\frac{d\sigma^h}{dE'd\Omega_e d\Omega_N dE_N} &= \frac{p_NM_nM_b}{(2\pi)^3M_{^3\mathrm{He}}}\sigma_{\mathrm{Mott}}(R_{fi}+hR'_{fi}) \nonumber \\
	& \equiv \Sigma_{fi}+h\Delta_{fi},
	\label{eq:cross-section-sigma-delta}
\end{align}
which is the sum of a helicity-independent part ($\Sigma_{fi}$) and a helicity-dependent part ($\Delta_{fi}$). The polarized and unpolarized cross-sections can be parametrized by two helicity-dependent (primed) and four helicity-independent (unprimed) response functions defined as
\begin{equation}
	R_{fi} = v_LR^L_{fi} + v_TR^T_{fi} + v_{TL}R^{TL}_{fi} + v_{TT}R^{TT}_{fi},
\end{equation}
\begin{equation}
	R'_{fi} = v_{T'}R^{T'}_{fi} + v_{TL'}R^{TL'}_{fi}.
\end{equation}

The response functions $R^{T'}_{fi}$ and $R^{TL'}_{fi}$ can be separated by changing the kinematic factors $v$. In the case where only the initial state of the electrons and target are polarized, and where the final state does not have polarization determined, it is possible to describe the components of the cross section in Equation \ref{eq:cross-section-sigma-delta} in terms of nine structure functions such that
\begin{align}
	\Sigma_{fi} \sim ~&v_LW^L_{fi}(\Delta\phi) + v_TW^T_{fi}(\Delta\phi) \nonumber \\
	&+ v_{TL}\left[ \cos{\phi_N}W^{TL}_{fi}(\Delta\phi)+ \sin{\phi_N}\widetilde{W}^{TL}_{fi}(\Delta\phi)\right] \\
	&+ v_{TT}\left[ \cos{2\phi_N}W^{TT}_{fi}(\Delta\phi) + \sin{2\phi_N}\widetilde{W}^{TT}_{fi}(\Delta\phi)\right] \nonumber,
\end{align}
\begin{equation}
	\Delta_{fi} \sim v_{T'}\widetilde{W}^{T'}_{fi}(\Delta\phi) + v_{TL'}\left[\sin{\phi_N}W^{TL'}_{fi}(\Delta\phi) + \sin{2\phi_N}\widetilde{W}^{TL'}_{fi}(\Delta\phi)\right],
\end{equation}
where the structure functions are dependent on the kinematic variables $q$, $\nu$, $\theta_N$, $p_N$, $E_N$, and the target spin orientations $\theta^*$ and $\Delta\phi\equiv \phi^*-\phi_N$. In the PWIA, the terms $\widetilde{W}^L_{fi}, ~\widetilde{W}^{TT}_{fi},$ and $W^{TL'}_{fi}$ are equal to zero \cite{Caballero:1992tt}. If the target is unpolarized, then all terms with a ``${\raise.17ex\hbox{$\scriptstyle\sim$}}$" are also equal to zero. Measuring these response functions provides a test for the PWIA as well as any perturbations to the approximation that could be caused by FSI or MEC.

In the PWIA, the electromagnetic current of the nucleus is the sum of currents of $A$ free nucleons. These nucleons are bound inside the nucleus, which causes them to be off-shell and results in the current conservation being broken. Due to this, the PWIA is an ambiguous formalism and arbitrary choices are made for an off-shell extrapolation of the PWIA on-shell vertex \cite{Naus:1989em}. The half-off-shell $\gamma NN$ vertex generally involves the four form factors \cite{Naus:1990em}, which can be extrapolated to the Pauli and Dirac form factors, or the two Sachs form factors, of the nucleon \cite{DeForest:1983vc}. Details of these form factors are discussed in Appendix \ref{ap:form-factors}.

Various extrapolations \cite{Caballero:1992tt, DeForest:1983vc} have been presented in order to find an expression for the spin-dependent off-shell electron-nucleon cross section, $\sigma^{eN}_{\hat{\sigma}}$, which results from the fact that the electromagnetic current is a one-body operator in the PWIA \cite{Frullani:1984nn}. From these descriptions, $\sigma^{eN}_{\hat{\sigma}}$ is reduced to the single-nucleon cross section where the kinematics are on-shell. In the PWIA, this cross section connects the leptonic tensor to part of the hadronic tensor (from Equation \ref{eq:lepton-hardron-cs}) that depends on the $\gamma NN$ vertex presented in Figure \ref{pwia-diagram} and the beginning of this section. The general cross section can now be described in terms of the product of $\sigma^{eN}_{\hat{\sigma}}$ and the spin-dependent spectral function $S_{\hat{\sigma}}^N (\vec{p}, E_s, \Omega^*)$ \cite{Caballero:1992tt, Schulze:1992mb} by
\begin{equation}
	\frac{d\sigma^h}{dE'd\Omega_e d\Omega_N dE_N} = \frac{p_N M_N M_{\mathrm{rec}}}{E_{\mathrm{rec}}}\sum_{\hat{\sigma}}\sigma^{eN}_{\hat{\sigma}}S^{N}_{\hat{\sigma}}(\vec{p},E_s,\Omega^*),
	\label{eq:spectral-cross-section}
\end{equation}
where $S_{\hat{\sigma}}^N (\vec{p}, E_s, \Omega^*)$ is the probability density of finding a nucleon $N$ with separation energy $E_s$, three-momentum $\vec{p}$, and spin projection, $\hat{\sigma}=+(-)$, parallel (antiparallel) to the spin of the $^3$He nucleus. The general form of the spectral function \cite{Schulze:1992mb} is 
\begin{align}
	S_{\hat{\sigma}}^N (\vec{p}, E_s, \Omega^*) = &\frac{1}{2} \bigg\{ f_0^N(p,E_s) + f_1^N(p,E_s)\sigma_N\cdot \sigma_{^3\mathrm{He}} \nonumber \\
	&  + f_2^N(p,E_s)\left[ (\sigma_N\cdot \hat{p}) (\sigma_{^3\mathrm{He}}\cdot\hat{p}) - \frac{1}{3}\sigma_N\cdot\sigma_{^3\mathrm{He}} \right] \bigg\}, 
\end{align}
where $f_0^N(p,E_s)$ is a spin-averaged function and $f_1^N(p,E_s)$ and $f_2^N(p,E_s)$ are two spin-dependent functions. Each of these is described in detail in Reference \cite{Schulze:1992mb} in terms of the momentum-space partial waves of the $^3$He ground-state wave function. The spectral function is directly related to the tri-nucleon bound state and can be described \cite{Schulze:1992mb} by
\begin{multline}
	S_{\hat{\sigma}}^N (\vec{p}, E_s, \Omega^*) = \\ 
	\frac{1}{(2\pi)^3}\sum_A\mathcal{P}(A) \sum_B \left\{ \bra{\psi_{^3\mathrm{He}}}a^+_{p\hat{\sigma}'}\ket{\psi_B}\bra{\psi_B}a_{p\hat{\sigma}}\ket{\psi_{^3\mathrm{He}}} \right\} \delta (E_s - E_{^3\mathrm{He}} - E_B),
	\label{spectral-function}
\end{multline}
where $\ket{\psi_{^3\mathrm{He}}}$ is the $^3$He bound-state solution with binding energy $E_{^3\mathrm{He}}$, $\ket{\psi_{B}}$ is the wave function of the remaining nucleons with internal excitation energy $E_B$, and $a^+_{p\hat{\sigma}'}~(a_{p\hat{\sigma}})$ is the creation (annihilation) operator. Summing over $B$ takes all nucleon subsystems of the final state into account and summing over $A$ weighted by $\mathcal{P}(A)$ yields the distribution of the ground-state angular momentum, $J_A$, over the nuclear substates $M_{J_A}$. The result of Equation \ref{spectral-function} can be used to determine the six-fold differential cross section.

In order to relate the cross section of Equation \ref{eq:spectral-cross-section} to measurable observables, this cross section can be written as
\begin{equation}
	\frac{d\sigma(h,S)}{d\Omega_edE_ed\Omega_n dp_n} = \frac{d\sigma^0}{d\Omega_edE_ed\Omega_n dp_n}\times[1+\vec{s}\cdot \vec{A}^0 + h(A_e+\vec{s}\cdot \vec{A}')],
	\label{asym-cs}
\end{equation}
where $h$ is the helicity of the electron, $\vec{s}$ is the spin of the $^3$He target, $\sigma^0$ is the spin-averaged cross section, $\vec{A}^0\equiv (\vec{A}_x^0,\vec{A}_y^0,\vec{A}_z^0)$ are the target analyzing powers, $A_e$ is the electron analyzing power, and $\vec{A'}\equiv(\vec{A}_x',\vec{A}_y',\vec{A}_z')$ are the spin-correlation asymmetries. In this calculation, $A'_y = A^0_x = A^0_z = 0$. In the PWIA, due to a combination of time-reversal invariance and hermiticity of the transition matrix,  $A_y^0$ is exactly zero \cite{Conzett:1998mm}. As such, any measurement of $A_y^0$ that is non-zero is indicative of higher-order effects such as FSI and MEC.

% ^^^^^^^^^^^^^^^^^^^^^^^^^^^^^^^^^^^^^^^^^^^^^^^^^^^^^^^^^^^^^^^^^^^^^^^^^^^^^^^^
\newpage
\large
\section {Final-State Interactions and Meson-Exchange Currents}
\label{fsi-mec}
\normalsize
% Theoretical Motivation
% vvvvvvvvvvvvvvvvvvvvvvvvvvvvvvvvvvvvvvvvvvvvvvvvvvvvvvvvvvvvvvvvvvvvvvvvvvvvvvvv
Since no free neutron target is available, multi-nucleon targets must be used. The reaction mechanism of neutron scattering from these nuclei must take into account effects from the nuclear medium. In particular, one must account for FSI and MEC.

\begin{SCfigure}
	\centering
	\includegraphics[width=3in]{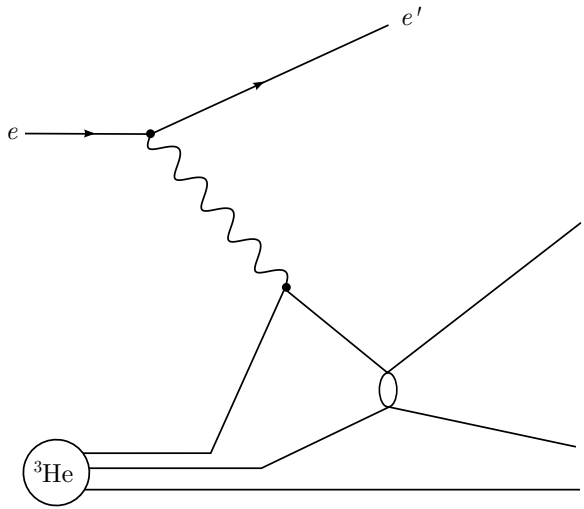} 
	\caption [FSI Example] {The diagram presented here is an example of final-state interactions where the recoiling nucleon interacts with one of the remaining nucleons in the nuclear system after the initial scattering from the incident electron. This diagram is adapted from Reference \cite{Poolman:1999uf}. Further examples, including those used in the original Laget calculation, are described in Figure \ref{laget-diagrams}.}
	\label{fsi-example}
\end{SCfigure}

Final-state interactions occur when the knocked-out nucleon interacts with the remaining nucleons. An example diagram of this type of interaction is presented in Figure \ref{fsi-example}. Naively, as the momentum-transfer is increased, the amount of time in which such interactions can occur is decreased and so it is expected that the FSI decrease at higher Q$^2$. The PWIA does not include such effects, although Laget has perturbed the approximation \cite{Laget:1991pb} to include them as discussed in Section \ref{laget}. They are calculated exactly in full Faddeev calculations that are discussed in detail in Section \ref{faddeev}.

\begin{figure}
	\centering
	\includegraphics[width = 2in]{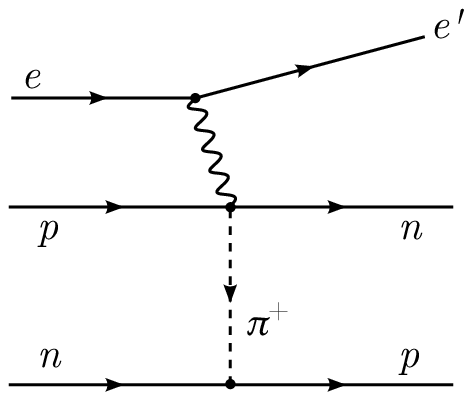}\includegraphics[width = 2in]{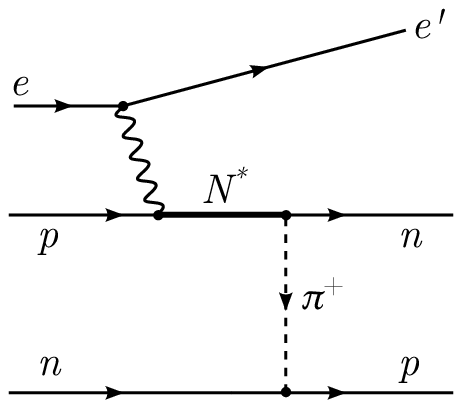}\includegraphics[width = 2in]{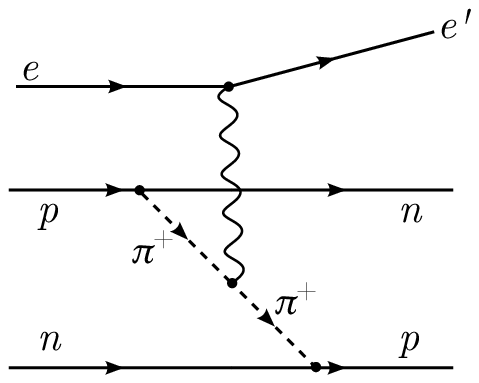} 
	\caption [MEC Examples] {MEC Examples. The diagrams presented here are examples of meson-exchange currents, which describe nucleon-nucleon potentials. An incident electron can interact with the mesons exchanged during this process, which needs to be accounted for when attempting to use a multi-nucleon system as an effective single-nucleon target.}
	\label{mec-example}
\end{figure}

Meson-exchange currents occur described nucleon-nucleon potentials as the exchange of mesons, such as $\pi$- and $\rho$-mesons. Understanding these effects is important when considering $^3$He as an effective neutron target. The contribution of these effects is expected to be much smaller than FSI, especially at lower Q$^2$, although still important in understanding the interactions that occur in $^3$He($e,e'n$) scattering. In the case of Laget's calculation, MEC are taken as a further perturbation of PWIA. In the full Faddeev calculations, MEC are included in the the nucleon-nucleon potential. Each of these cases only accounts for $\pi$- and $\rho$-meson exchange currents, ignoring heavier mesons. Diagrammatic examples of MEC are presented in Figure \ref{mec-example}.

% ^^^^^^^^^^^^^^^^^^^^^^^^^^^^^^^^^^^^^^^^^^^^^^^^^^^^^^^^^^^^^^^^^^^^^^^^^^^^^^^^

\large
\section {Original Laget Calculations}
\label{laget}
\normalsize
% Theoretical Motivation
% vvvvvvvvvvvvvvvvvvvvvvvvvvvvvvvvvvvvvvvvvvvvvvvvvvvvvvvvvvvvvvvvvvvvvvvvvvvvvvvv

In the early 1990s, Laget was working on calculations to estimate the effects of FSI and MEC in the $^3$He($e,e'n$) reaction. This work was based on the PWIA and included effects from FSI and MEC as perturbations. Although his calculations at the time underestimated the effects from FSI and MEC, the qualitative understanding of these calculations still holds.

The general expression of the cross section for ($e,e'n$) reactions is described by Equation \ref{asym-cs}. The components of the spin-transfer polarization of the outgoing nucleon are of the form
\begin{equation}
	\sigma^0 P'_y=\left( \frac{-q^2\epsilon(1-\epsilon)}{2\nu^2} \right)^{1/2} \sin{\phi} \sigma'_{TL}(y),
\end{equation}
\begin{equation}
	\sigma^0P'_{x,z}=-\left(\frac{-q^2\epsilon(1-\epsilon)}{2\nu^2} \right)^{1/2} \cos{\phi}\sigma'_{TL}(x,z) + (1-\epsilon^2)^{1/2}\sigma'_{TT}(x,z),
\end{equation}
where $\nu$, $q^2$, and $\epsilon$ are the energy, squared momentum, and the polarization of the virtual photon respectively, and $\sigma'_{TT(TL)}$ are the transverse-transverse (transverse-longitudinal) interference cross sections. In coplanar geometry, $P'_y=P^0_x=A'_y=A^0_x=A^0_z=0$ due to the sine and cosine terms. 

\begin{SCfigure}
	\centering
	\includegraphics{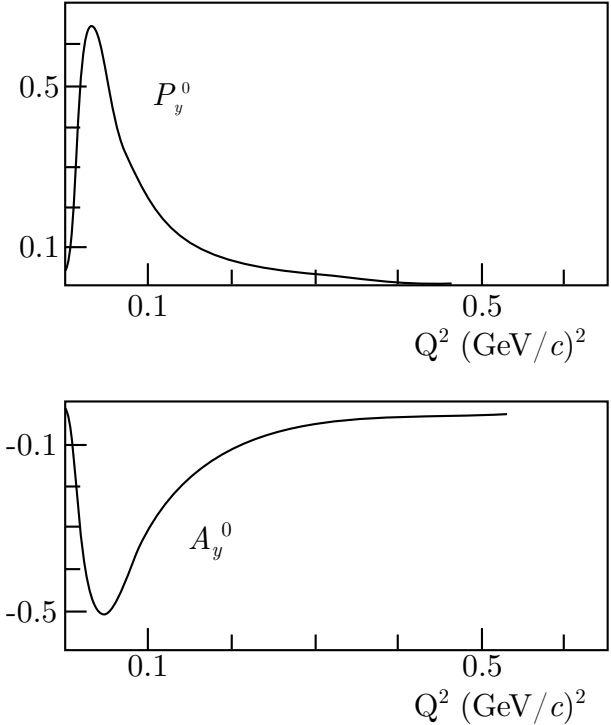}
	\caption [Original Laget Calculations] {This plot shows the results from the original Laget calculations for $P^0_y$ and $A_y^0$. Although Laget has since updated his calculations to include larger effects due to FSI and MEC, they are only calculated for individual Q$^2$ values and not for the range presented here \cite{Laget:1991pb}.}
	\label{laget-original-ay}
\end{SCfigure}

\begin{figure}
	\centering
	\includegraphics[width=15 cm]{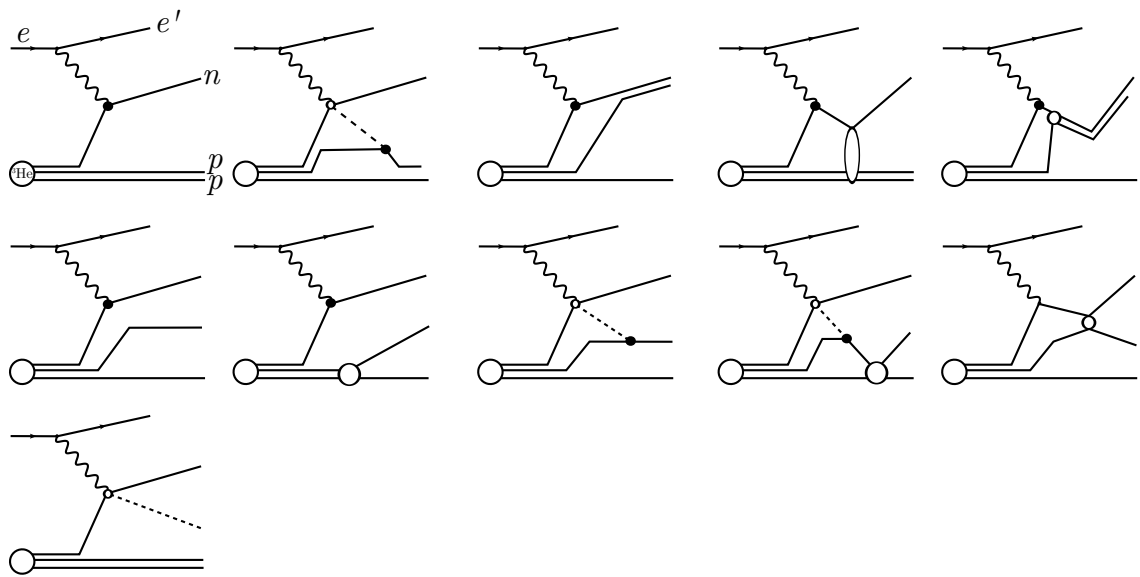} 
	\caption [FSI and MEC Diagrams Chosen by Laget] {The diagrams presented here those that Laget included in his analysis of $A_y^0$. The top row consists of diagrams of two-body break-up, the center row of three-body break-up, and the bottom diagram is for pion electroproduction. These diagrams are adapted from Reference \cite{Poolman:1999uf}.}
	\label{laget-diagrams}
\end{figure}

In the PWIA, due to a combination of time reversal-invariance and hermiticity of the transition matrix, $P^0_y=A^0_y=0$ \cite{Conzett:1998mm}. Laget perturbed the PWIA by including the FSI and MEC diagrams shown in Figure \ref{laget-diagrams}, resulting in $P^0_y\neq0$ and $A^0_y\neq0$. Before $A_y^0$ was experimentally measured, Laget estimated these effects from FSI and MEC to play an important role at low Q$^2$ and to drop off as the momentum-transfer increased as shown in Figure \ref{laget-original-ay} \cite{Laget:1991pb}.

Since then, Laget has updated his calculations to meet with experimental constraints, however, the full range has not been recalculated as it is shown in Figure \ref{laget-original-ay}. Although the magnitude of the effects of FSI and MEC are larger than was originally expected, qualitatively this understanding of $A_y^0$ still holds.

% ^^^^^^^^^^^^^^^^^^^^^^^^^^^^^^^^^^^^^^^^^^^^^^^^^^^^^^^^^^^^^^^^^^^^^^^^^^^^^^^^

\large
\section {Faddeev Calculations}
\label{faddeev}
\normalsize
% Theoretical Motivation
% vvvvvvvvvvvvvvvvvvvvvvvvvvvvvvvvvvvvvvvvvvvvvvvvvvvvvvvvvvvvvvvvvvvvvvvvvvvvvvvv
Faddeev calculations are full calculations of the three-body Schr\"{o}dinger equation in non-relativistic kinematics. Processes such as MEC are absorbed into the nucleon-nucleon potential. They consist of a set of coupled integral equations that have unique solutions for three-body scattering. These calculations have been done by the Bochum group for $A_y^0$, $A_{\parallel}$, and $A_{\perp}$ at low Q$^2$ \cite{Golak:2001ge}.

Faddeev showed that rearranging the perturbation series of the scattering $T$-matrix will lead to unique solutions of the three-bodiednucleon (3N) Schr\"{o}dinger equation \cite{faddeev1993quantum}. This system includes two-body and three-body interactions, all of which have a finite range beyond which the force acting on all three nucleons becomes negligible. 

\begin{wrapfigure}{r}{0.4\textwidth}
	% \capstart
	\centering
	\includegraphics{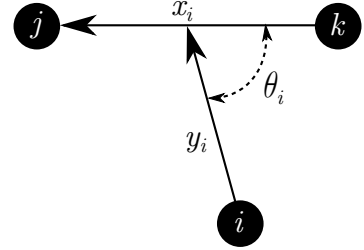}
	\caption [Three-Body Coordinates] {This diagram identifies the three independent Jacobi coordinates for the three-body system where particles $j$ and $k$ interact and particle $i$ is a spectator. The spatial relations, $x_i$ and $y_i$, also correspond to the momenta $\vec{p}_i$ and $\vec{q}_i$, respectively. This diagram is adapted from Reference \cite{Poolman:1999uf}.}
	\label{jacbobi-coordinates}
\end{wrapfigure}

The three particles are labeled $i$, $j$, and $k=1,2,3$ such that $i\neq j\neq k$ as shown in Figure \ref{jacbobi-coordinates}, where $j$ and $k$ are interacting and $i$ is a spectator. In these coordinates, the center of mass of the system is fixed by setting the total momentum, $\vec{P}$, equal to zero. The momentum of the spectator particle, $\vec{q}_i$ is defined with respect to the center of mass of the interacting particles with momentum $\vec{p}_i$. The masses of the particles are defined as $m_i$. In momentum space, the independent variables are
\begin{equation}
	\vec{P}=\sum_{i=0}^3 \vec{k}_i\equiv 0, ~~\vec{p}_1=\frac{\vec{k}_2-\vec{k}_3}{2}, ~~\vec{q}_1=\frac{2\vec{k}_1-(\vec{k}_2+\vec{k}_3)}{3},
\end{equation}
where $\hbar=1$, $m_1=m_2=m_3$, $M=3m$, $\mu_p=m/2$ is the reduced mass of the interacting particles and $\mu_q=2m/3$ is the reduced mass of the entire system. The non-relativistic Schr\"{o}dinger equation for this system is defined as 
\begin{equation}
	(E-H)\ket{\Psi} = (E-H_0 - V)\ket{\Psi}=0,
	\label{eq:nonrel-schrodinger}
\end{equation}
where
\begin{equation}
	H_0 = \sum_{i=1}^3 \frac{\vec{k}_i\cdot  \vec{k}_i}{2m_i} = \frac{P^2}{2M} + \frac{p_i^2}{2\mu_p} + \frac{q_i^2}{2\mu_q}~\mbox{and}
\end{equation}
\begin{equation}
	V = \sum_{i=1}^3 V_i = V_0 + V_1 + V_2 + V_3.
	\label{eq:nuclear-potential}
\end{equation}
The free Hamiltonian is described above by $H_0$, the interaction between the particles is defined as $V$, which is the sum of three independent nucleon-nucleon potentials ($V_i$ where $i=1,2,3$) and one three-body potential, $V_0$. In order to keep the computation relatively simple, $V_0$ is usually neglected. This is the case in Equation \ref{eq:nonrel-schrodinger}. Although not described in detail here, it should be noted that the Coulomb potential is also included in the full Faddeev calculations by the Bochum group \cite{Poolman:1999uf}  

In order to solve the Schr\"{o}dinger equation, Green's functions are introduced that take the form
\begin{equation}
	G(z)\equiv (z-H)^{-1} ~~\mbox{and}~~	G_0(z)\equiv (z-H_0)^{-1}, 
\end{equation}
where $z$ is a variable with dimensions of energy. These functions are related by
\begin{align}
	G(z)&=G_0(z) + G_0(z)VG(z) \nonumber \\
	 &= G_0 + G_0VG_0 + G_0VG_0VG_0+ \cdot \cdot \cdot.
\end{align}
The transition operator, $T(z)$, is related to the potential $V$ by the Lippman-Schwinger equation \cite{PhysRev.79.469} such that
\begin{align}
	T(z)&=V+VG_0(z)T(z) \nonumber \\
	 &= V+VG(z)V.
	 \label{eq:lippman-schwinger}
\end{align}

\begin{figure}
	\centering
	\includegraphics{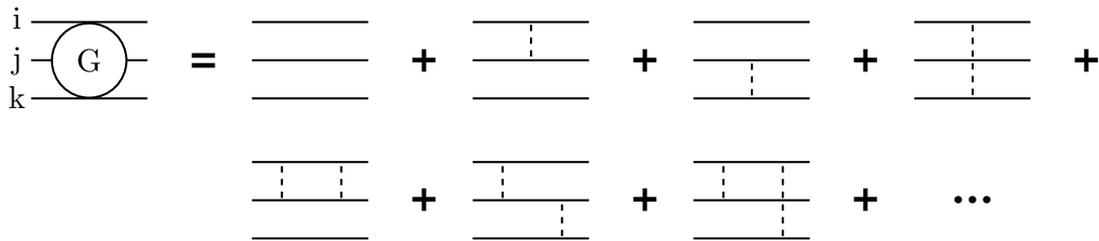}
	\caption [Green's Function Expansion] {This series of diagrams is the expansion of the Green's function operator, $G(z)$, in terms of the free propagator, $G_0$, and two-body interactions. This diagram is from Reference \cite{Poolman:1999uf}.}
	\label{greens-expansion}
\end{figure}

If the potential $V$ from Equation \ref{eq:nuclear-potential} is substituted into Equation \ref{eq:lippman-schwinger}, then the equation can be expanded into an infinite series, where the operator $G_0$ is the propagator of the non-interacting three-body system and the two-body interaction $V_i$ is an intermediary connecting particles $j$ and $k$. The corresponding Green's function to this expansion is defined as 
\begin{align}
	G=& G_0 + G_0\sum_i V_i G_0  + G_0 \sum_i V_i G_0\sum_j V_j G_0 \nonumber \\
	&  + G_0 \sum_i V_i G_0\sum_j V_j G_0\sum_k V_k G_0 + \cdot\cdot\cdot,
\end{align}
which can be expressed diagrammatically as presented in Figure \ref{greens-expansion}.

\begin{figure}
	\centering
	\includegraphics{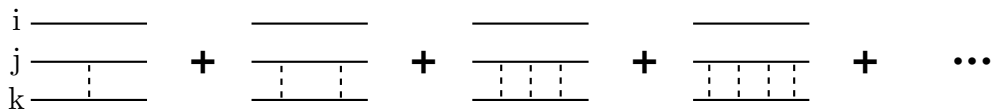}
	\caption [Disconnected Diagrams] {The infinite series displayed in Figure \ref{greens-expansion} contains three subsets that each contain an infinite number of disconnected diagrams where the spectator particle does not interact with the other two. This is an example of one of those series where $i$ is the non-interacting particle. This diagram is from Reference \cite{Poolman:1999uf}.}
	\label{disconnected-diagrams}
\end{figure}

\begin{figure}
	\centering
	\includegraphics{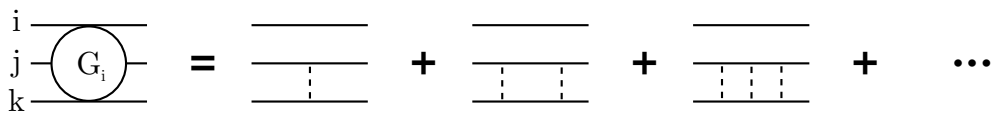}
	\caption [Channel Operator Diagrams] {The series displayed here corresponds to the infinite series of disconnected diagrams for the channel operator $G_i$. This diagram is from Reference \cite{Poolman:1999uf}.}
	\label{channel-greens}
\end{figure}

Within the expansion displayed in Figure \ref{greens-expansion}, there are three infinite series of disconnected diagrams. One of these is displayed in Figure \ref{disconnected-diagrams}. The non-interacting particle has an unchanging momentum that causes a $\delta$-function to remain in the momentum representation. 
The series of diagrams where one particle is disconnected corresponds to the two-body $T$-matrix in 3N-space. The two-body transition operator, $T_i$, can similarly be defined as
\begin{equation}
	T_i\equiv V_i+V_iG_0(z)T_i.
\end{equation}
From this, the channel Green's function can be defined as
\begin{equation}
	G_i\equiv (z-H_0-V_i)^{-1} 
\end{equation}
and is presented in diagrammatic form is Figure \ref{channel-greens}.

\begin{figure}
	\centering
	\includegraphics{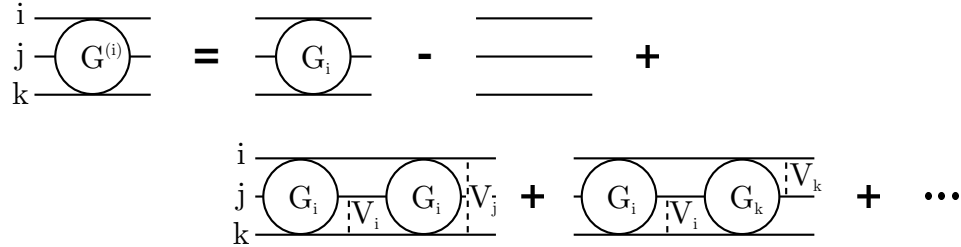}
	\caption [Faddeev Component] {The series displayed is an expansion of one of the Faddeev components, $G^{(i)}(z)$, in terms of the channel Green's function $G_i$. This diagram is from Reference \cite{Poolman:1999uf}.}
	\label{faddeev-components}
\end{figure}

Faddeev described the full operator as the composition of four pieces defined as
\begin{equation}
	G(z)=G_0(z)+G^{(1)}(z)+G^{(2)}(z)+G^{(3)}(z),
	\label{eq:faddeev-components}
\end{equation}
where $G_0(z)$ is the free propagator and $G^{(i)}(z)$ are the three Faddeev components. These components are displayed in diagrammatic form in Figure \ref{faddeev-components}. Through the use of $G_i$, all subsets where only two particles interact are defined in one term.

\begin{figure}
	\centering
	\includegraphics{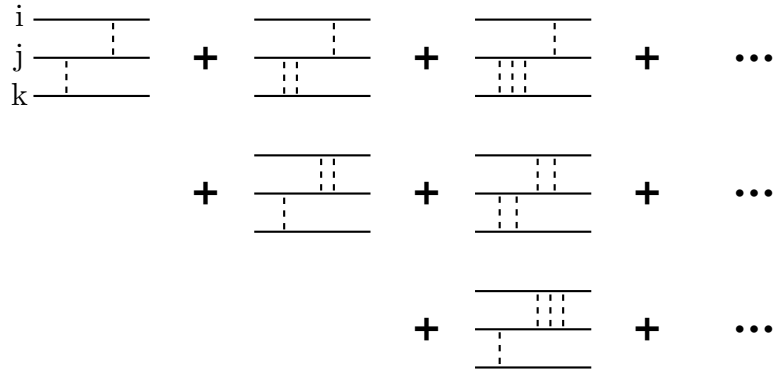}
	\caption [Multiple Interaction Term Expansion] {The final diagram in Figure \ref{faddeev-components} is expanded in detail here. The Faddeev component always starts with an interaction between particles $i$ and $j$ and ends with an interaction between particles $i$ and $j$ or $i$ and $k$. This diagram is from Reference \cite{Poolman:1999uf}.}
	\label{multiple-interactions}
\end{figure}

The final term in Figure \ref{faddeev-components} can be expanded out as represented in Figure \ref{multiple-interactions}. The Faddeev component $G^{(i)}$ always starts with an interaction between particles $i$ and $j$ and ends with an interaction between particles $i$ and $j$ or $i$ and $k$. Permutations of this component where the diagrams differ only by which particles are interacting are defined as $G^{(j)}$ and $G^{(k)}$, which leads to the coupling described in Equation \ref{eq:faddeev-components}.

In order to solve the non-relativistic Schr\"{o}dinger equation described in Equation \ref{eq:nonrel-schrodinger}, the wave function is decomposed into three Fadeev calculations. The wave function is thus described by
\begin{equation}
	\ket{\Psi}\equiv \ket{\Phi_0}+\sum_{i=1}^3 \ket{\psi_i}^F,
\end{equation}
where
\begin{equation} 
	\ket{\Phi_0}= \lim_{\epsilon \rightarrow 0} i \epsilon G_0 \ket{\Phi_0} ~~~\mbox{and}~~~ \ket{\psi_i}^F= \lim_{\epsilon \rightarrow 0} i \epsilon G^{(i)} \ket{\Phi_0}.
\end{equation}

\begin{figure}
	\centering
	\includegraphics{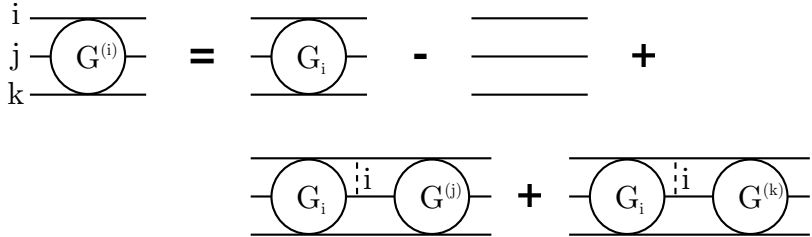}
	\caption [Faddeev Equations] {This set of diagrams is equivalent to the Faddeev equations given in Equation \ref{eq:faddeev-equations}. This diagram is from Reference \cite{Poolman:1999uf}.}
	\label{faddeev-equations}
\end{figure}

Each of the Faddeev components, $\ket{\psi_i}^F$, can be written as the decomposition of the full Green's function, $G(z)$, and the solution $\ket{\phi_i}$ of the channel Hamiltonian, $H_i=H_0+V$, where $\ket{\phi_i}$ is the product of a bound state two-body wave function and a plane wave for a single free particle. The eignenvalue of $H_i\ket{\phi_i}$ is given by
\begin{equation}
	H_i\ket{\phi_i} = \left( \epsilon_i + \frac{3}{4m}q_i^2 \right) \ket{\phi_i} = E_{q_i}\ket{\phi_i},
\end{equation}
where $\epsilon_i$ is the binding energy of the two-body system. It is required to solve the Faddeev equations for both the bound state and the continuum in order to describe electron scattering from a $^3$He nucleus. The Faddeev equations can be described in diagrammatic form as shown in Figure \ref{faddeev-equations} or in matrix notation as
\begin{equation}
	\left[ \begin{array}{c} \ket{\psi_1}^F \\ \ket{\psi_2}^F \\ \ket{\psi_3}^F \end{array} \right] = \left[ \begin{array}{c} \ket{\Phi_0} \\ 0 \\ 0 \end{array} \right] + \begin{bmatrix} 0 & T_1(z) & T_1(z) \\  T_2(z) & 0 & T_2(z) \\ T_3(z) & T_3(z) & 0 \end{bmatrix} G_0(z) \left[ \begin{array}{c} \ket{\psi_1}^F \\ \ket{\psi_2}^F \\ \ket{\psi_3}^F \end{array} \right].
	\label{eq:faddeev-equations}
\end{equation}

The full Faddeev calculations have been solved exactly by the Bochum group for low Q$^2$, where relativistic effects are not taken into account \cite{Golak:2001ge}. They are presented along with the experimental values measured for $A_y^0$ in this dissertation and include contributions from both FSI and MEC. In addition, calculations have been done for both the $^3$He($e,e'd$) and $^3$He($e,e'p$) channels which, when constrained to experimental data, give information into the contributions of the S, S', and D states of the $^3$He wave function \cite{Golak:2001ge}. 

% ^^^^^^^^^^^^^^^^^^^^^^^^^^^^^^^^^^^^^^^^^^^^^^^^^^^^^^^^^^^^^^^^^^^^^^^^^^^^^^^^

% ^^^^^^^^^^^^^^^^^^^^^^^^^^^^^^^^^^^^^^^^^^^^^^^^^^^^^^^^^^^^^^^^^^^^^^^^^^^^^^^^

					% Include the Chapter 2 text (chapter_2.tex)
	% vvvvvvvvvvvvvvvvvvvvvvvvvvvvvvvvvvvvvvvvvvvvvvvvvvvvvvvvvvvvvvvvvvvvvvvvvvvvvvvv
% Chapter 2 (chapter_2.tex)
%
% Chapter 2 of Elena Long's Ph.D. Dissertation
%
% To be completed: March, 2012
%
% ^^^^^^^^^^^^^^^^^^^^^^^^^^^^^^^^^^^^^^^^^^^^^^^^^^^^^^^^^^^^^^^^^^^^^^^^^^^^^^^^
\large
\chapter{Setup of the Experiment}	% Chapter Title
\label{experimentsetup}		% Chapter Label
\normalsize			% Return to Normal font size

\large
\section {Overview of CEBAF and Hall A} 
\label{ch2-hand}
\normalsize
% Target
% vvvvvvvvvvvvvvvvvvvvvvvvvvvvvvvvvvvvvvvvvvvvvvvvvvvvvvvvvvvvvvvvvvvvvvvvvvvvvvvv

\begin{wrapfigure}{r}{0.36\textwidth}
	% \capstart
	\centering
	\includegraphics[width = 0.36\textwidth]{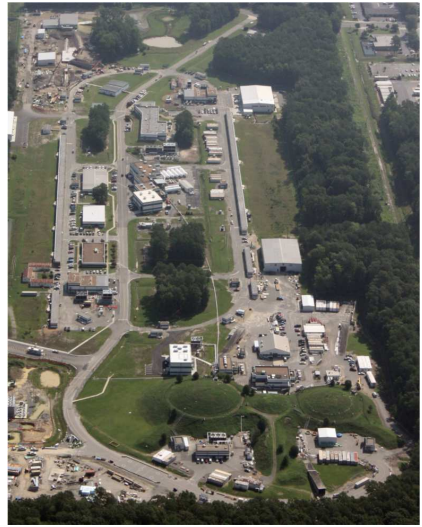}
	\caption [Aerial View of Jefferson Lab] {Aerial view of Jefferson Lab.}
	\label{jlab-aerial}
\end{wrapfigure}

The experiment presented in this dissertation used Jefferson Laboratory's Continuous Electron Beam Accelerator Facility (CEBAF) and was performed in experimental Hall A. CEBAF is a superconducting radio frequency electron accelerator that can provide a beam with polarization greater than 80\% and energies up to 6 GeV \cite{Leemann:2001dg}. An overhead picture of the lab can be seen in Figure \ref{jlab-aerial}. The accelerator is discussed in detail in Section \ref{ch2-beamline}.

%\begin{figure}
%	\centering
%	\includegraphics[width=15 cm, angle=0]{./jlab-aerial.eps}
%	\caption [Aerial View of Jefferson Lab] {Aerial View of Jefferson Lab.}
%	\label{jlab-aerial}
%\end{figure}

\begin{figure}
	\centering
	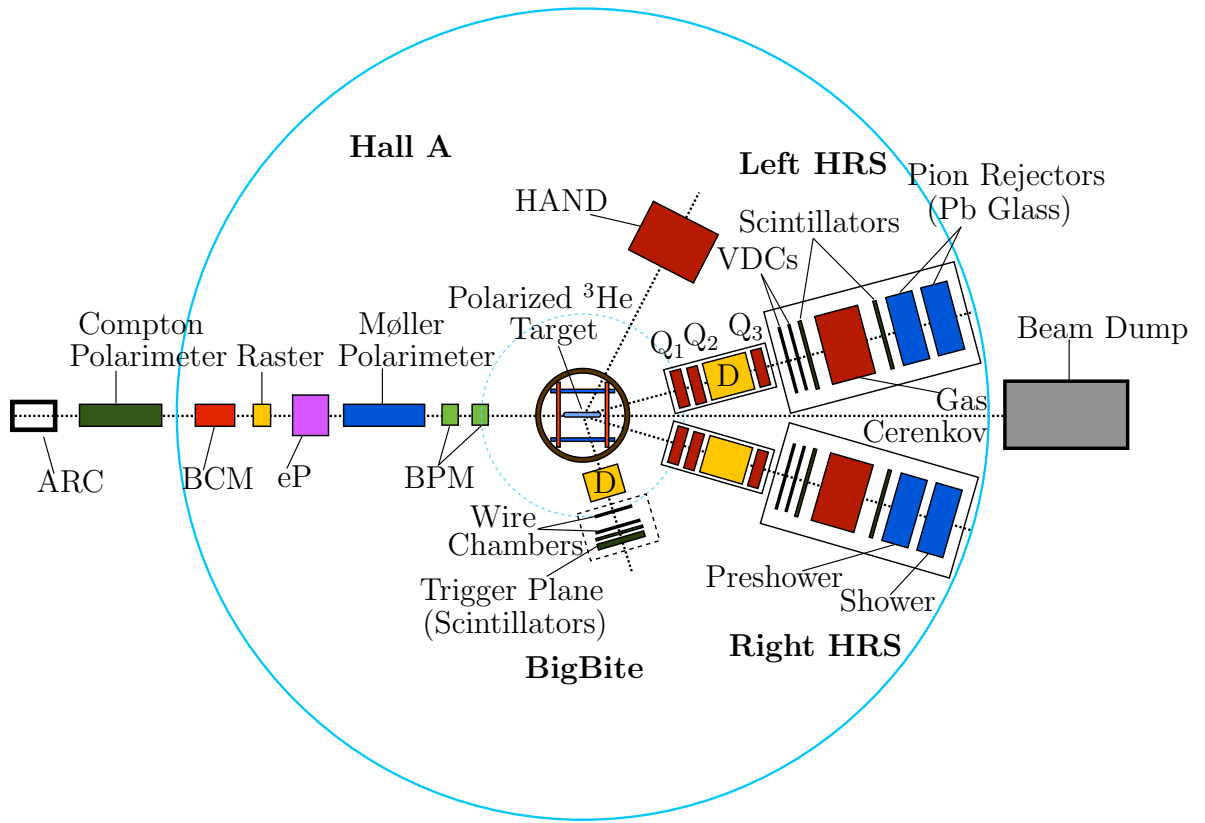
	\caption [Hall A Experimental Setup] {This shows the Hall A equipment that was in place during this experiment. The beam line downstream from the target (towards the right in the schematic) corresponds to a 0$^{\circ}$ angle.}
	\label{halla-layout}
\end{figure}

Hall A contains equipment that includes two High Resolution Spectrometers (HRS), the Hall A Neutron Detector (HAND), and a polarized $^3$He target capable of being polarized in three orthogonal directions. A schematic of the equipment used in Hall A can be seen in Figure \ref{halla-layout}. The equipment in Hall A is discussed in detail in Section \ref{hall-a}.

Due to an improved polarized $^3$He target, this experiment was able to take measurements with the target polarized in each of three orthogonal directions. This is the first time that an experiment has simultaneously measured the $^3$He($e,e'n$) asymmetries with the polarization in three dimensions. Details of the kinematics used during this experiment are discussed in Section \ref{experiment-kinematics}.

% ^^^^^^^^^^^^^^^^^^^^^^^^^^^^^^^^^^^^^^^^^^^^^^^^^^^^^^^^^^^^^^^^^^^^^^^^^^^^^^^

\large
\section {CEBAF and the Electron Beam} 
\label{ch2-beamline}
\normalsize
% Beam Line
% vvvvvvvvvvvvvvvvvvvvvvvvvvvvvvvvvvvvvvvvvvvvvvvvvvvvvvvvvvvvvvvvvvvvvvvvvvvvvvvv

Jefferson Lab's CEBAF is able to produce an 80\%-polarized, continuous-wave electron beam. The beam starts at the polarized electron source, which enters the main accelerator through the injector. It is accelerated up to 6 GeV by two superconducting radio frequency (SRF) linear accelerators and two sets of recirculating arcs. The beam can be circulated up to five times with each pass increasing the energy by up to 1.2 GeV. The final beam is able to be simultaneously sent to three different experimental halls by a beam switchyard. Each experimental hall can receive beam at different energies, as long as they are integer multiples of a single pass. The different components of CEBAF are described in detail below.

\begin{figure}
	\centering
	\includegraphics{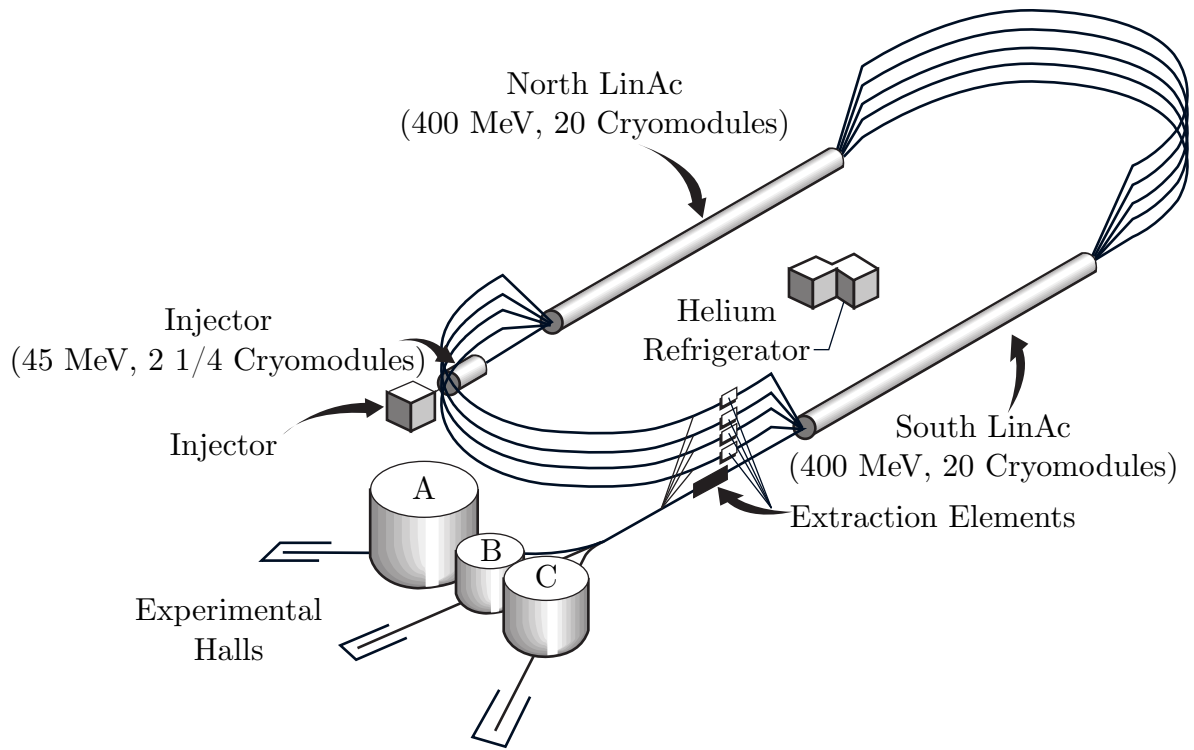}
	\caption [CEBAF Layout] {This figure shows the layout of the Continuous Electron Beam Accelerator Facility. Polarized electrons are produced in the injector and are accelerated in the two LinAcs. There are also two sets of recirculating arcs that allow the beam to go through the LinAcs up to five times. Once the electrons are accelerated, they are sent into one of the three experimental halls, A, B, or C.}
	\label{cebaf-layout}
\end{figure}

\subsection {Injector}
\label{injector}

The polarized electron source is a strained GaAs cathode that is hit by a circularly polarized laser beam. A Pockels cell, which causes bireference induced by magnetic field, causes changes to occur in the laser polarization every 33.3 ms, which in turn causes a flip in the helicity of the electrons every 33.3 ms \cite{Chao:2011zz}. In order to reduce systematic effects dependent on the beam helicity, a half-wave plate can be inserted that reverses the beam's helicity.

These newly polarized electrons are accelerated to 100 keV and injected into the main accelerator through two superconducting radio-frequency (SRF) cavities. These two SRF cavities are referred to as a quarter-cryomodule, since the main accelerator consists of cryomodules that each contain eight SRF cavities \cite{HernandezGarcia:2008zz}.

\subsection {Linear Accelerators}
\label{linacs}
The heart of CEBAF consists of the niobium SRF linear accelerators (LinAcs). There are two sets of these, one towards the north and one towards the south as shown in Figure \ref{cebaf-layout}. Each contains 20 cryomodules, which in turn each contain 8 SRF cavities. Superfluid $^4$He is used to keep the niobium at a superconducting temperature of ~2 K. In the LinAcs, electrons are accelerated up to 600 MeV before entering a recirculating arc, which will allow them to be accelerated again. Due to the unique construction of Jefferson Lab, electrons may pass through the LinAcs up to five times \cite{Leemann:2001dg}. The LinAcs are also used to ensure the highest possible longitudinal electron spin polarization at the experimental halls by adjusting the spin precision through a redistribution of the energy gain \cite{Higinbotham:2009ze}.

\subsection {Recirculating Arcs}
\label{recirc-arcs}
\begin{SCfigure}
	% \capstart
	\centering
	\includegraphics[width=6cm]{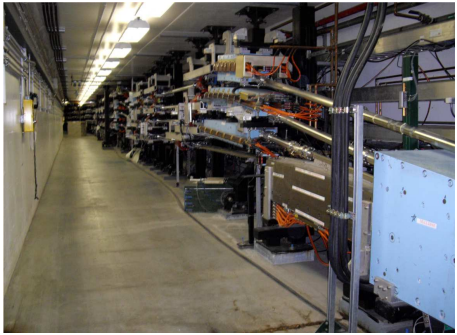}
	\caption [Recirculating Arcs Photograph] {Recirculating Arcs Photograph. This photograph, taken at the end of the west recirculating arc near the injector, shows the recombining dipole magnets in blue. Also visible is each of the four beam pipes, one for each pass.}
	\label{recirc-arcs-photo}
%\end{wrapfigure}
\end{SCfigure}

The recirculating arcs consist of a dipole ``spreading" magnet, followed by a series of dipole magnets that steer the electron beam into a 180$^\circ$ arc, and a final dipole ``recombining" magnet. Each arc contains a beam pipe for electrons at each pass energy. Lower energy electrons, which are easier to steer, are diverted to the higher arcs, while higher energy electrons pass through the lower arcs. The different energy beams are then re-combined at the end of the arc to be put through the LinAcs again \cite{Leemann:2001dg}. A photograph of the arcs is shown in Figure \ref{recirc-arcs-photo}.

%\begin{wrapfigure}{r}{0.4\textwidth}

\subsection {Beam Switchyard}
\label{recirc-arcs}
The beam switchyard is used to send beam to each of the lab's experimental halls. It consists of RF separators, septa, and dipole magnets that separate and divert the beam. The 1/3-harmonic RF separator allows splitting of the three interleaved electron bunch trains into all of the three experimental halls simultaneously. Beam energy can be taken from any of the five passes through the accelerator \cite{Leemann:2001dg}.

% ^^^^^^^^^^^^^^^^^^^^^^^^^^^^^^^^^^^^^^^^^^^^^^^^^^^^^^^^^^^^^^^^^^^^^^^^^^^^^^^^

\section {Hall A}
\label {hall-a}
\normalsize
% Hall A
% vvvvvvvvvvvvvvvvvvvvvvvvvvvvvvvvvvvvvvvvvvvvvvvvvvvvvvvvvvvvvvvvvvvvvvvvvvvvvvvv

\begin{figure}
	\centering
	\import{./}{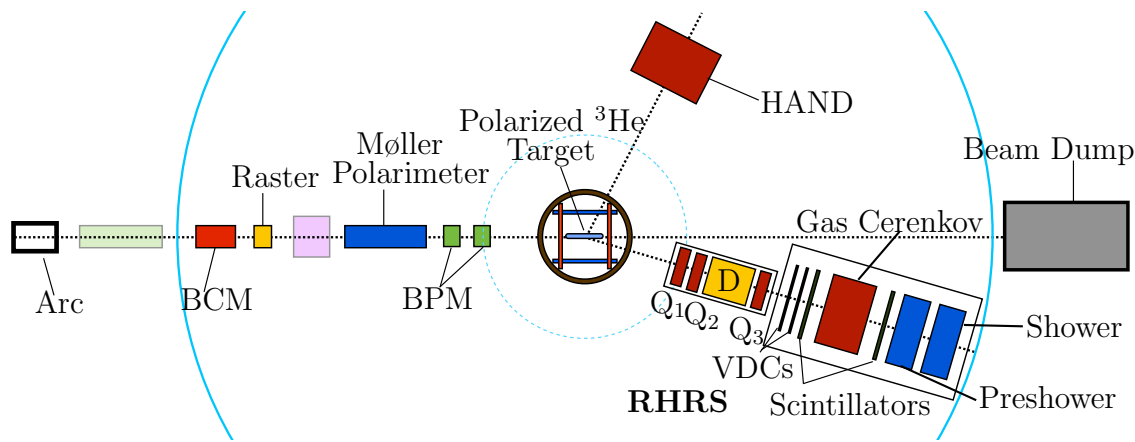}
	\caption [Hall A Equipment used for $^3$He($e,e'n$) Measurements] {This figure is a layout of the Hall A equipment used for the $^3$He($e,e'n$) measurements. Hall A used the arc to measure the beam energy, the BCM and BPM to measure the beam current and position, the raster to spread the beam on the target, the M$\o$ller polarimeter to measure the polarization of the beam, a polarized $^3$He target, the HAND to detect scattered neutrons, the RHRS to detect scattered electrons, and the beam dump to accept electrons from beam which did not scatter.}
	\label{een-hall-layout}
\end{figure}

Experimental Hall A at Jefferson Lab is uniquely suited to measuring $^3$He($e,e'n$) asymmetries due to its high resolution spectrometer (HRS), polarized $^3$He target, and neutron detector. The Hall also contains the Big Bite spectrometer, as well as a second HRS, which were used for the simultaneous measurements of the $^3\vec{\mbox{He}}(\vec{e},e'p$), $^3\vec{\mbox{He}}(\vec{e},e'd$), $^3\vec{\mbox{He}}(\vec{e},e'$),  and $^3$He$^{\uparrow}(e,e'$) asymmetries, which are explored in detail by M. Mihovilovic \cite{Mihovilovic:2012ljubljana}, G. Jin \cite{Jin:2011uva}, and Y.-W. Zhang \cite{Yawei:2013rutgers}. Figure \ref{een-hall-layout} shows the placement of the equipment used for the $^3$He($e,e'n$) asymmetries in Hall A.

\subsection {Beam Measurements}
\label {beam-measurements}
\normalsize

Several pieces of equipment were used to understand the incoming electron beam incident upon the target. A variety of parameters were measured, an overview of which can be seen in Table \ref{tab:beam-params} and details of which are presented in the sections below.

\begin{table}
\begin{center}
\begin{tabular}{c|c|c|c}
Parameter & Method & Accuracy & Comments \\ \hline
Energy & Arc & $2\times 10^{-4}$ & Invasive \\
Energy & Arc & $5\times 10^{-4}$ & Non-invasive \\
% Energy & eP & $2\times 10^{-4}$ & Invasive \\
Energy Width & OTR & $\frac{\Delta E}{E} \approx 1 \times 10^{-5}(\sigma)$ & Non-invasive \\
Current ($\ge 1$ $\mu$A) & 2 RF Cavities & $\le 5 \times 10^{-3}$ & Non-invasive \\
Position (at target) & 2 BPM/Harp & 140 $\mu$m & $x, y$ on line \\
Direction (at target) & 2 BPM/Harp & 30 $\mu$rad & $\theta, \phi$ on line \\
Stability (at target) & Fast Feedback & $\le 720$ Hz motion & \\
Stability (at target) & Position & $\le 20$ $\mu$m ($\sigma$) & \\
Stability (at target) & Energy & $\le 1 \times 10^{-5}$ ($\sigma$) & \\
Polarization & M$\o$ller & $\frac{\Delta P}{P} \approx 2\%$ & Invasive \\
\end{tabular}
\caption[Methods to Determine Beam Parameters]{This table contains an overview of the methods and equipment used to determine beam parameters. The Accuracy column is the width of an assumed Gaussian distribution. Techniques labeled ``Invasive" require dedicated beam time and interrupt the main experiment \cite{Alcorn:2004sb}.}
\label{tab:beam-params}
\end{center}
\end{table}

\subsubsection {Arc Energy Measurements}
\label{arc}

The energy of the beam is determined by measuring the deflection of the beam in the arc section of the beam-line and the field integral of eight dipole magnets. Nominally, the angle of the beam is 34.3$^\circ$. A set of superharp wire scanners are used to determine the position of the incoming and outgoing beam and thus measure the angle. The integrated magnetic field of the eight quadruple magnets that the beam passes through in that bend is also measured and is related to the beam momentum (and thus energy) by
\begin{equation}
	p=k \frac{\int \vec{B}\cdot d\vec{l}} {\theta},
\end{equation}
where $k=0.299792$ $\frac{\mbox{GeV}\cdot \mbox{rad}}{ \mbox{T}\cdot \mbox{m}\cdot \mbox{c}}$ \cite{Alcorn:2004sb}.

\begin{SCfigure}
	\centering
	\includegraphics[width=4.3 in]{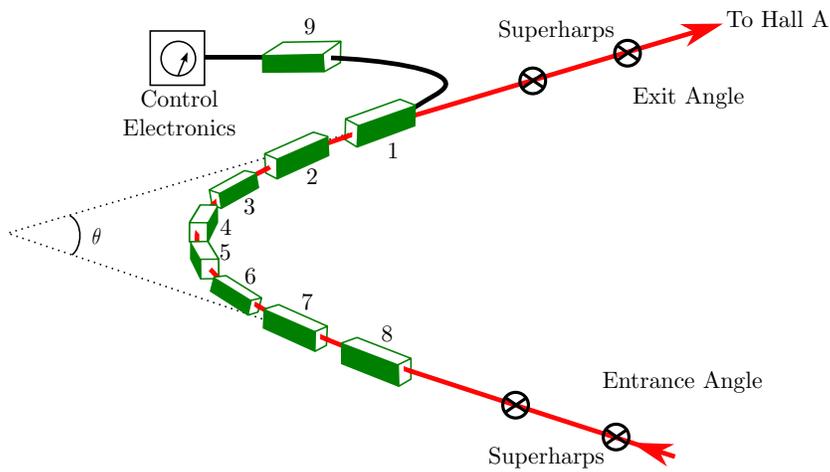}
	\caption [Arc Layout] {Displayed is the layout of the arc energy monitor, which consists of nine quadruple magnets (green), four superharp wire scanners, and control electronics. The measurement is made by finding the angle between the incoming and outgoing beam and comparing it to the integrated magnetic field of the quadruple magnets.}
	\label{arc-layout}
\end{SCfigure}

\subsubsection {Beam Current Monitors}
\label{bcm}

Hall A's Beam Current Monitors (BCMs) consist of an Unser monitor \cite{Unser:1989sj} and two RF cavities. They are located 25 m upstream of the target. They are calibrated against a cavity monitor and a Faraday Cup \cite{brown:696}, which are located at the injector of CEBAF. In order to reduce noise and drift, the Unser monitor must have extensive magnetic shielding and the temperature must be stable. Due to drifts caused by having the beam running through the monitor over a time scale of minutes, it can't be used to measure the beam current continuously.

The RF cavities are stainless steel cylindrical high-Q waveguides tuned to the frequency of the beam (1.497 GHz), which provide output voltages that are proportional to the beam current. The output signals are doubled so that one provides a sampling and the other an integration. The sampling signal is recorded into the data stream approximately every 2-5 s. Each of the integration signals is sent into amplifiers of gains 1, 3, and 10, which extend the non-linear region to lower currents. Each of these integrated signals (three from each BCM) is recorded, which allows for a measurement of the integrated charge during any given run \cite{Alcorn:2004sb}. 

\subsubsection {Beam Raster}
\label{raster}
\begin{figure}
	\centering
	\includegraphics{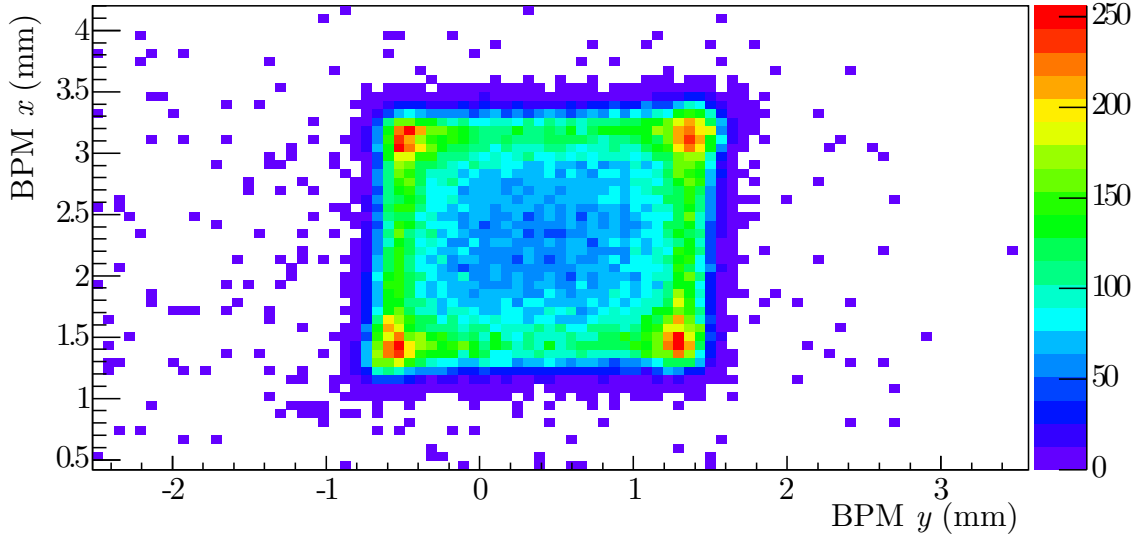}
	\caption [BPM with Beam Raster] {Displayed is a $x$ and $y$ position of the beam from the BPM while the beam raster was on. The beam was spread over the 2 mm $\times$ 2 mm area throughout the experiment. The raster uses a magnetic field that changes rapidly to spread the focused electron beam over a larger area. This allows for smaller target windows which reduces the background contribution due to glass. The data here are from Run 22487.}
	\label{raster-ex}
\end{figure}

In order to prevent damage to the glass target cell, the beam was spread out through the use of quickly changing magnetic fields. This process is called rastering. Rastering also allows for a thinner glass wall on the target, which reduces background scattering. Typically, the raster size is a 2 mm $\times$ 2 mm square, as shown in Figure \ref{raster-ex}. The magnetic fields are provided by dipole magnets that are located 23 m upstream of the target, as shown in Figure \ref{een-hall-layout}. 

\subsubsection {Beam Position Monitors}
\label{bpm}

Two beam position monitors (BPMs) consisting of a 4-wire antenna array tuned to the fundamental RF frequency of the beam were used to determine the position and the direction of the electron beam on the target. They were placed 7.542 m and 1.286 m upstream of the target. The relative position of the beam is determined to within 100 $\mu$m for currents above 1 $\mu$A through the standard difference-over-sum technique \cite{Barry:1990ys}. The absolute position of the BPMs is calibrated through the use of superharp wire scanners located next to the BPMs. The averaged position over 0.3 seconds is logged into the EPICS datastream \cite{Alcorn:2004sb}.

\subsubsection {M$\o$ller Polarimeter}
\label{moller-polarimeter}

\begin{figure}
	\centering
	\includegraphics{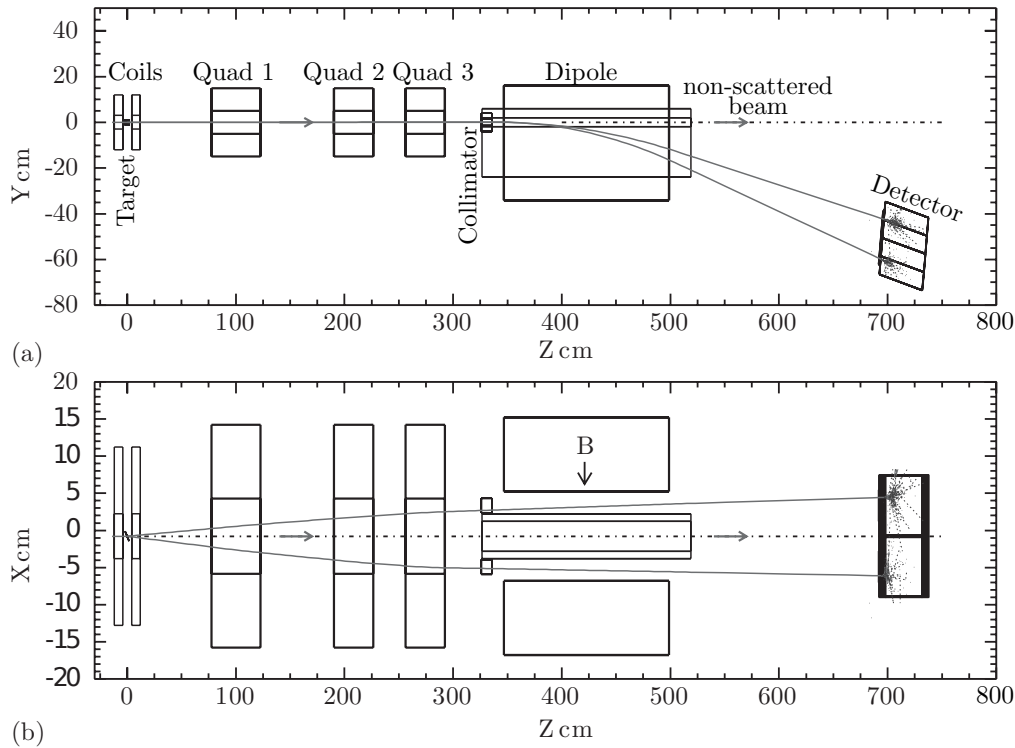}
	\caption [M$\o$ller Polarimeter Layout] {Displayed is the layout of the M$\o$ller polarimeter where (a) is a side view and (b) is a top view. The trajectories dispelled belong to a simulated event of M$\o$ller scattering at $\theta_{CM}=80^{\circ}$ and $\phi_{CM}=0^{\circ}$ at a beam energy of 4 GeV \cite{Alcorn:2004sb}. The polarimeter was used to determine the polarization of the incident electron beam.}
	\label{moller-polarimeter-layout}
\end{figure}

A M$\o$ller polarimeter was used to measure the polarization of the beam in Hall A. It measures a beam-target double-spin asymmetry from M$\o$ller scattering ($\vec{e^-} + \vec{e^-} \rightarrow e^- + e^-$) to extract the beam polarization. The polarimeter consists of a ferromagnetic foil target (which is magnetized in a magnetic field of about 24 mT) as the source of the polarized electrons and a magnetic spectrometer. The spectrometer consists of three quadruple magnets, one dipole magnet, a steel collimator, and two arms of lead-glass calorimeters. The layout of the polarimeter is shown in Figure \ref{moller-polarimeter-layout}.

\begin{SCfigure}
	\centering
	\includegraphics[width=4 in]{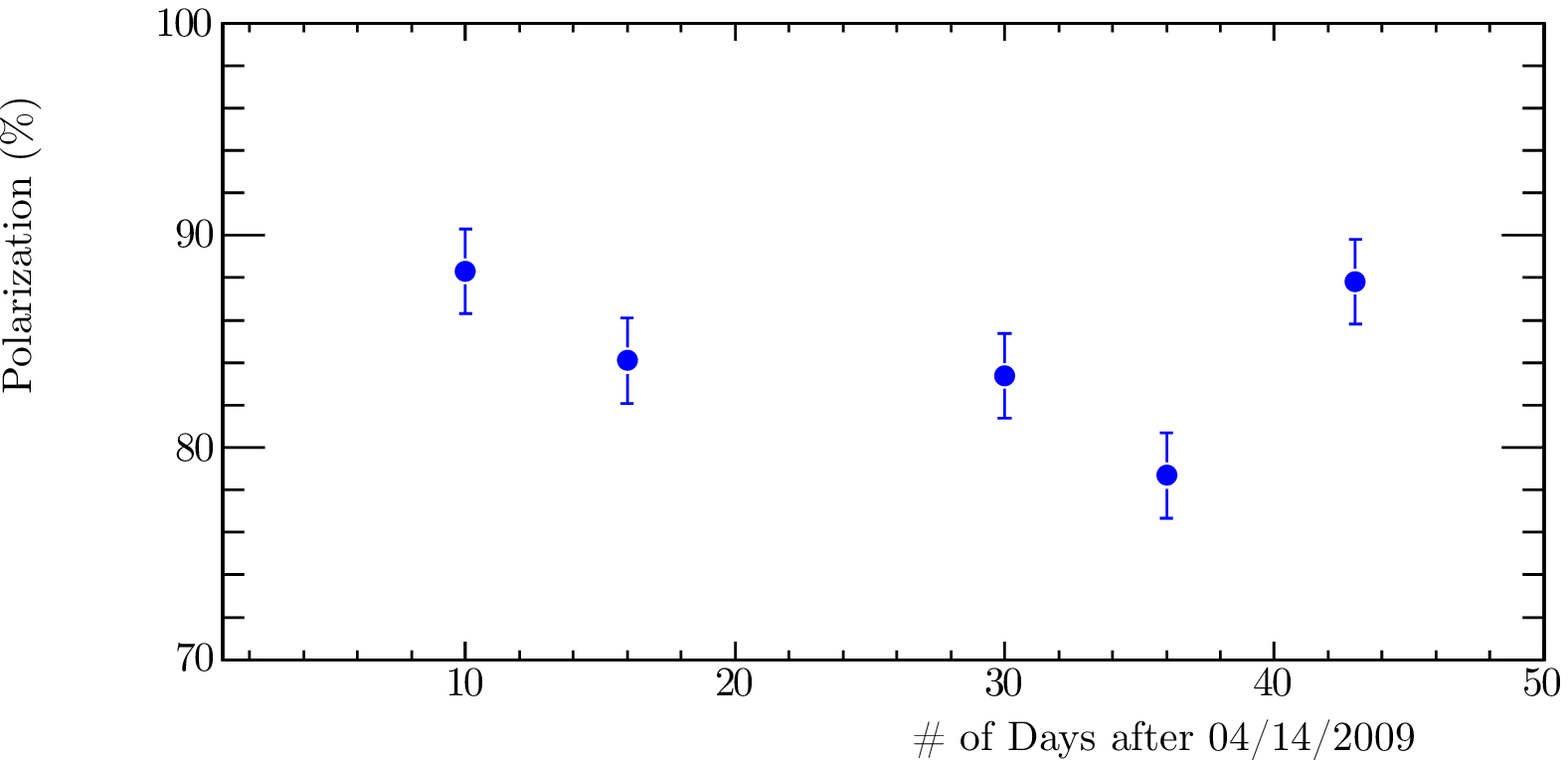}
	\caption [M$\o$ller Measurements] {M$\o$ller Measurements. This plot shows the M$\o$ller measurements taken to determine the beam polarization for the $A_y^0$, $A_T$ and $A_L$ experiments. The average beam polarization during the experiment was 84.5$\%$.}
	\label{moller-meas}
\end{SCfigure}

The steel collimator is 6 cm thick and has a 2 cm radius hole, through which the scattered electrons pass. The spectrometer detects scattered electrons in a kinematic range of $75^{\circ}<\theta_{CM}<105^{\circ}$ and $-5^{\circ}<\phi_{CM}<5^{\circ}$, where $\theta_{CM} ~ (\phi_{CM})$ is the polar (azimuthal) angle \cite{Alcorn:2004sb}. The M$\o$ller polarimeter is an invasive piece of equipment and requires dedicated beam time with runs taking approximately an hour. Measurements of the beam polarization are shown in Figure \ref{moller-meas} are discussed in detail in Section \ref{beam-polarization}.

\large
\subsection {Polarized $^3$He Target} 
\label{ch2-target}
\normalsize
% Target

\begin{SCfigure}
	\centering
	\includegraphics[width=4 in]{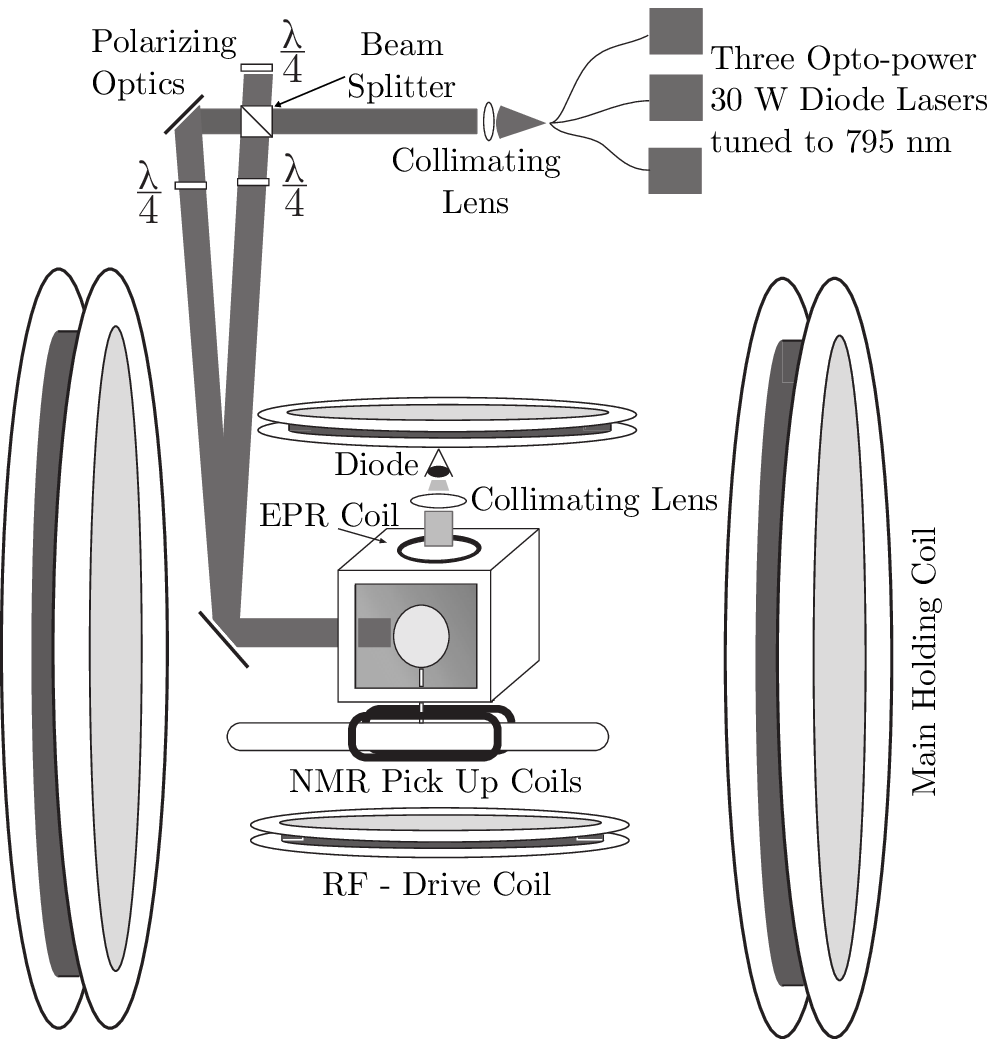}
	\caption [Target System] {This is a schematic of the target system. Not shown is a third set of Helmholtz holding coils that were placed orthogonal to each pair shown. The $^3$He target was polarized $> 50\%$ through SEOP. The polarization was measured every four hours by NMR and was calibrated against invasive EPR measurements, which were taken less frequently.}
	\label{target-system}
\end{SCfigure}

\begin{figure}
	\centering
	\includegraphics{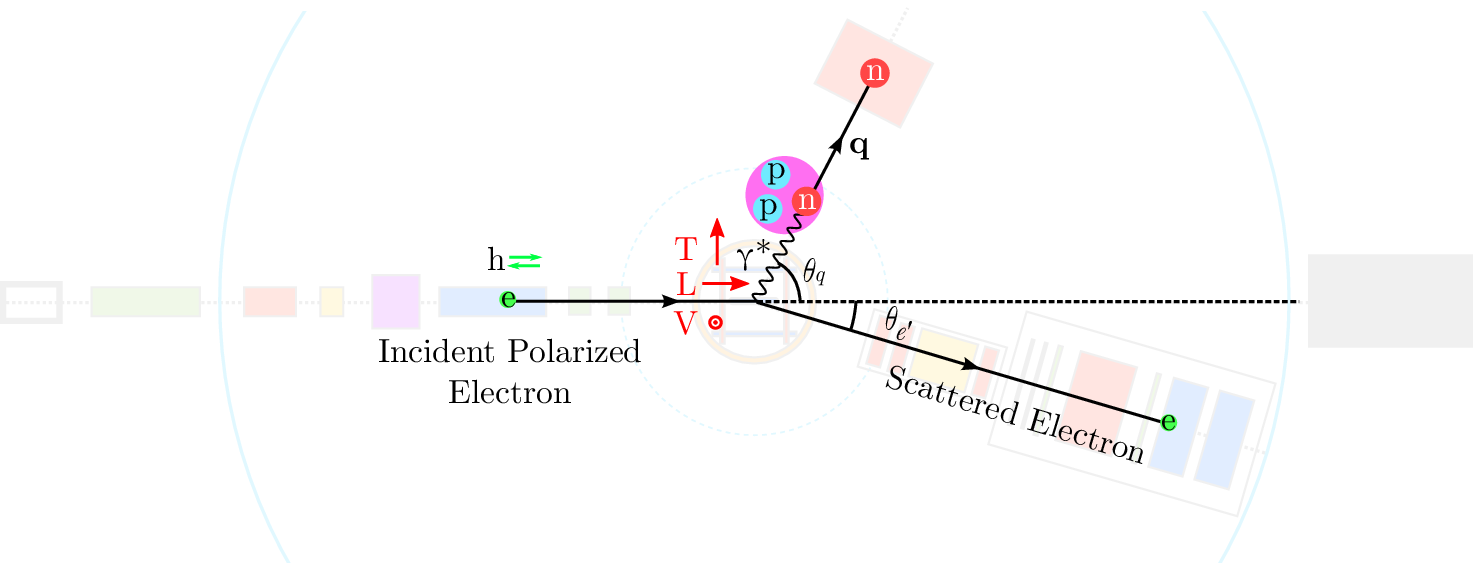}
	\caption [Target Polarization Directions] {The $^3$He target used in this experiment was oriented in three orthogonal directions: transverse to both the beam and q-vector (Vertical, V), transverse to the beam and nearly longitudinal with the q-vector (Transverse, T), and longitudinal with the incident electron (Longitudinal, L).}
	\label{polarization-direction}
\end{figure}

\begin{figure}
	\centering
	\includegraphics[width=15cm]{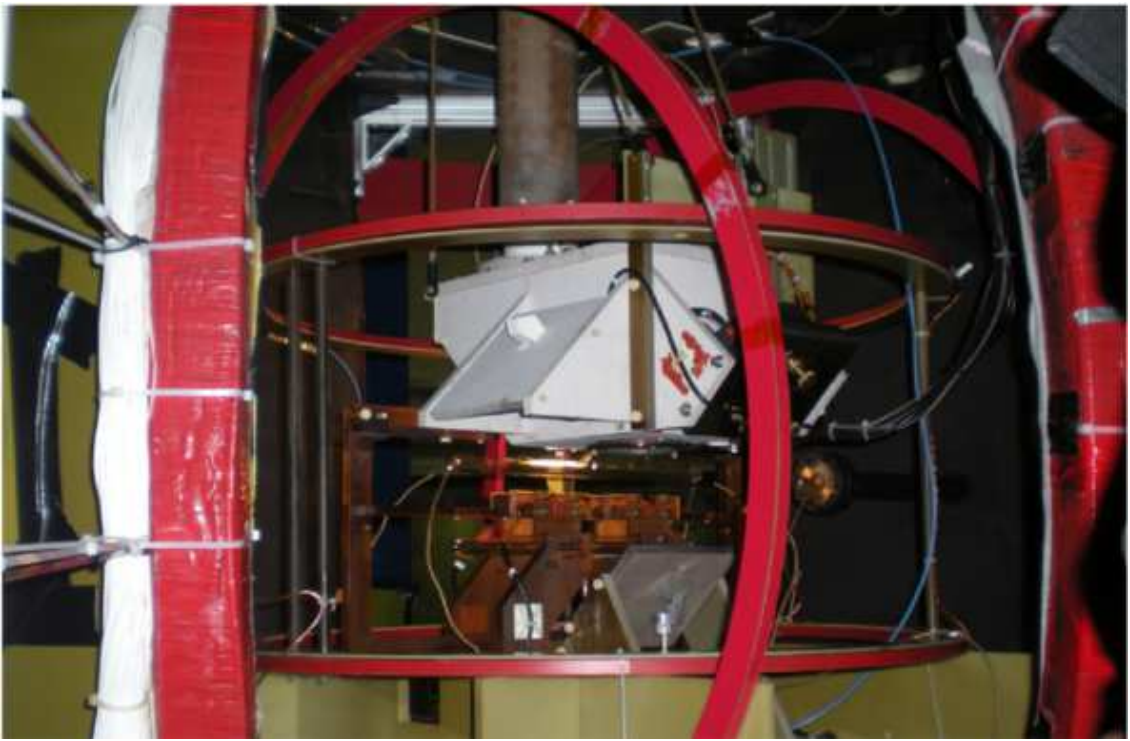}
	\caption [Target Photograph] {This photograph shows the installed target system. The holding coils are painted red, the polarization oven is painted white with the glass target chamber underneath, and the NMR pick-up coils are visible under the target chamber. More detail of the target cell and pick-up coils is shown in Figures \ref{target-cell} and \ref{target-cell-photo}.}
	\label{target-system-photo}
\end{figure}

Experiments E05-102 and E08-005 would not have been possible if not for Hall A's polarized $^3$He target system. Polarization of up to 60\% was obtained through the use of a spin-exchange optical-pumping (SEOP) cell \cite{Walker:1997zzc, PhysRevA.58.1412}. The target consists of a glass cell that holds the He and alkali-metal vapor, a laser-based optical pumping system to polarize the target, three sets of Helmholtz coils to hold the target polarization, an electron paramagnetic resonance (EPR) coil, and nuclear magnetic resonance (NMR) coils. The setup of the target system is shown in Figure \ref{target-system}. Target polarization $>50\%$ was achieved for each of the three polarization directions. During the $A_y^0$ measurement, the direction of the target spin was flipped every 20 minutes. During the $A_T$ and $A_L$ measurements, the direction of the target spin was flipped every few days. Measurements of the polarization are discussed in detail in Section \ref{target-polarization}. The different polarization directions are shown in Figure \ref{polarization-direction}. A photograph of the equipment is presented in Figure \ref{target-system-photo}. 


\subsubsection {Target Cell}
\label{tgt-cell}
\begin{figure}
	\centering
	\includegraphics{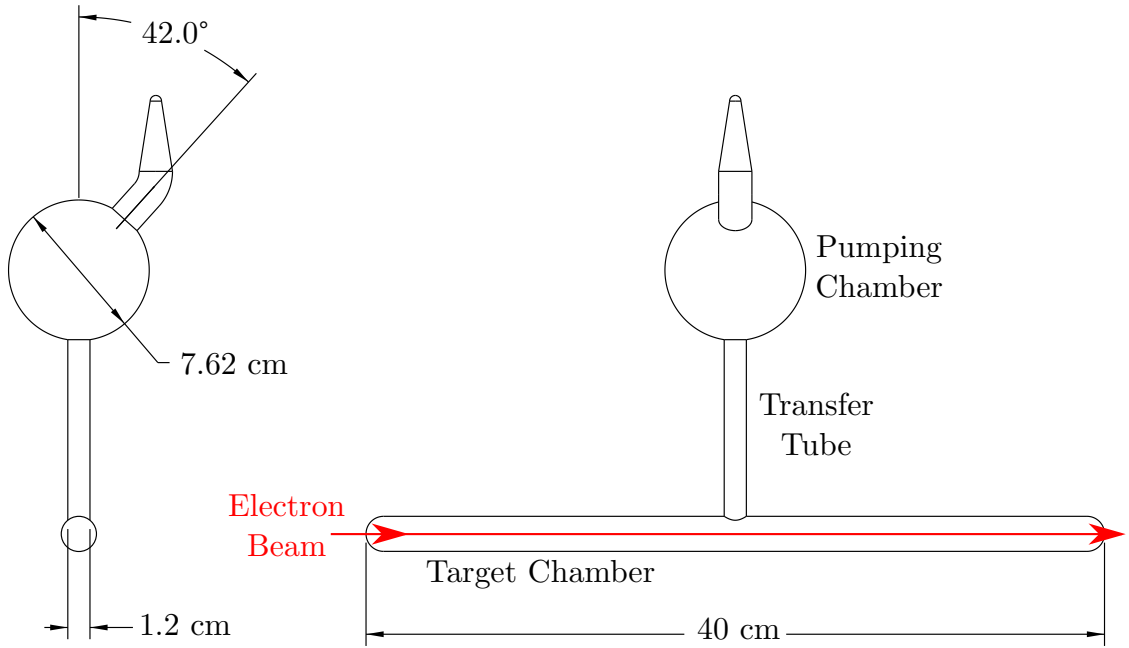}
	\caption [Target Cell] {This schematic of the target cell shows the pumping chamber, transfer tube, and target chamber. All measurements are design specifications and may have varied slightly in production.}
	\label{target-cell}
\end{figure}

\begin{figure}
	\centering
	\includegraphics{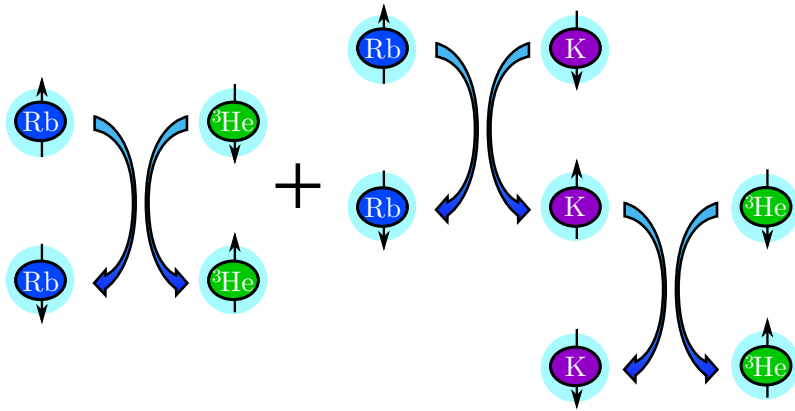}
	\caption [Spin Exchange Process] {This diagram shows the spin-exchange processes of Rb, K, and $^3$He. Circularly polarized laser light excites the Rb atoms to flip their spin which is held in a magnetic field. This spin is transferred, via collisions, to $^3$He nuclei. A secondary process, used to decrease the time needed to reach maximum polarization, occurred with the Rb atoms collided and exchanged spin with K atoms, which in turn collided and exchanged spin with $^3$He nuclei.}
	\label{spin-exchange}
\end{figure}

\begin{figure}
	\centering
	\includegraphics[width=15cm]{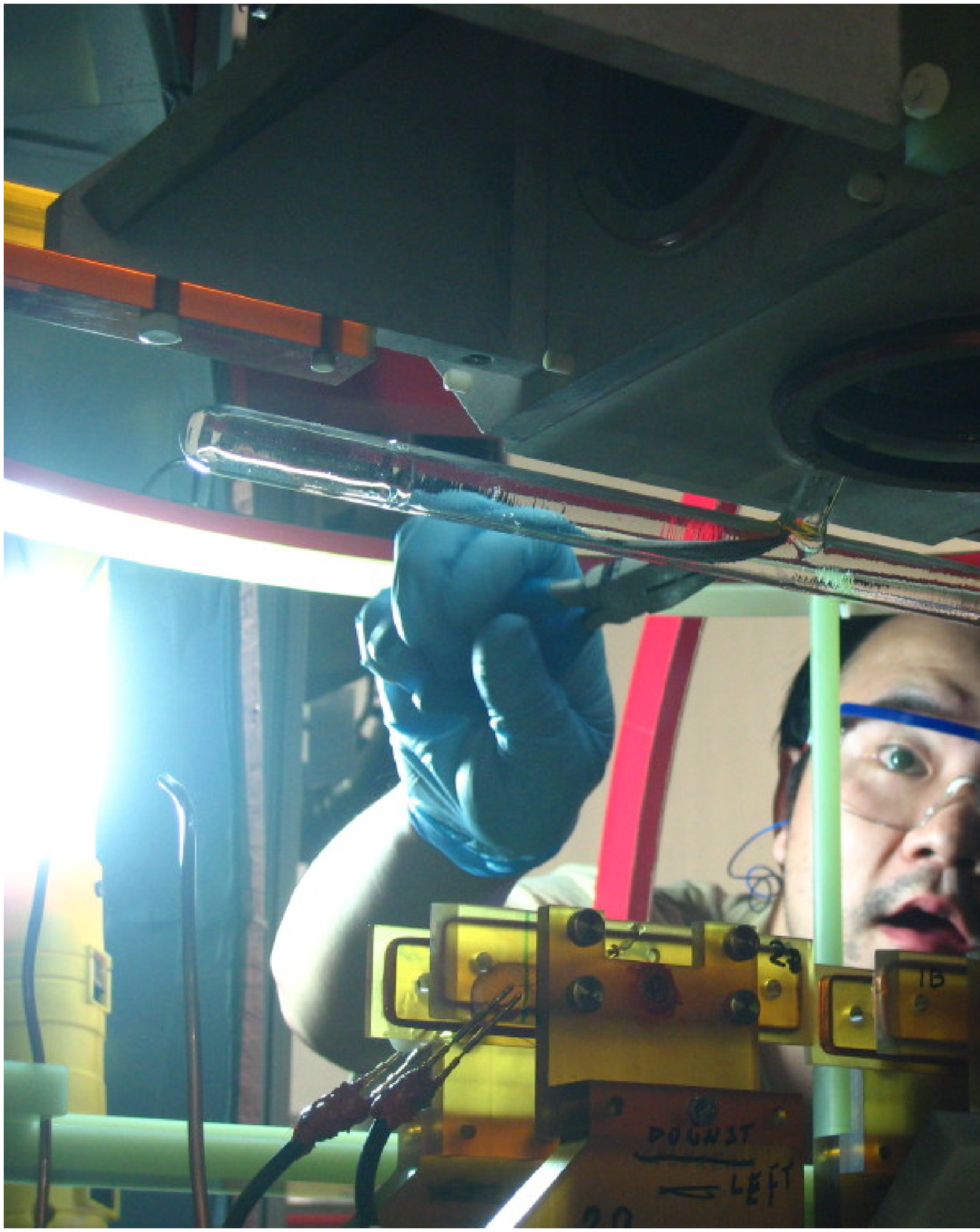}
	\caption [Target Cell Installation] {This photograph shows the polarized $^3$He being put into position by Yi Qiang. Visible are the glass cell, the polarization oven, and the NMR pick-up coils.}
	\label{target-cell-photo}
\end{figure}

The target cell is consists of a pumping chamber, a transfer tube, and a target chamber, as shown in Figure \ref{target-cell}. It contains $^3$He pressurized to about 0.69 MPa. The pumping chamber is where polarization of the $^3$He occurs through SEOP. The particular type of cell used was a RbK hybrid cell. Circularly polarized laser light excited Rb atoms, which collided and exchanged their spins with both $^3$He and K atoms. The now-polarized K atoms also collided with $^3$He nuclei and caused the $^3$He to polarize \cite{PhysRevLett.91.123003}. The second process helped to reduce the time needed for the target to reach maximum polarization \cite{Alcorn:2004sb}. A diagram of these processes is shown in Figure \ref{spin-exchange}. The target chamber is a 40 cm long tube through which the beam passes. Two cells were used for these experiments: ``Dominic" when the target was polarized vertically and ``Moss" when the target was polarized longitudinally and transversely. The glass walls of each cell had a thickness of $<$1.7 mm and a window thickness of $<$0.16 mm. A photograph of the target cell in position is presented in Figure \ref{target-cell-photo}.


\subsubsection {Optical Pumping System}
\label{op-pump}
A laser system consisting of three 30 W diode lasers tuned to 795 nm was used to induce polarization in the Rb mixture. The light was routed via fiber optics and polarizing optics to the pumping chamber of the target cell. The laser light was polarized through the use of a polarizing beam splitter, which polarized the beam linearly, followed by a quarter-wave plate that produced circularly polarized light. The polarization of the light was able to be reversed through through the use of an insertable half-wave plate, which when combined with reversing the direction of the holding field, reversed the direction of $^3$He polarization \cite{Alcorn:2004sb}.

\subsubsection {NMR and EPR}
\label{nmr-epr}
Two systems were used to measure the polarization of the target: nuclear magnetic resonance \cite{Incerti98thenmr} and electron paramagnetic resonance \cite{Liang01theepr}. The $^3$He NMR signal was calibrated against that of a water cell through the technique of adiabatic fast passage (AFP) to find the polarization of protons in water. NMR signals were recorded every four hours throughout the experiment. They were cross checked with EPR measurements of Rb atoms to determine the polarization of the target. There were 15 EPR measurements taken over the entire run period, which are discussed in detail in Section \ref{target-polarization}. The NMR and EPR equipment is shown Figures \ref{target-system} and \ref{target-cell-photo}.

\large
\subsection {High Resolution Spectrometer} 
\label{ch2-hrs}
\begin{figure}
	\centering
%	\import{./}{rhrs-detectors.eps_tex}
	\includegraphics{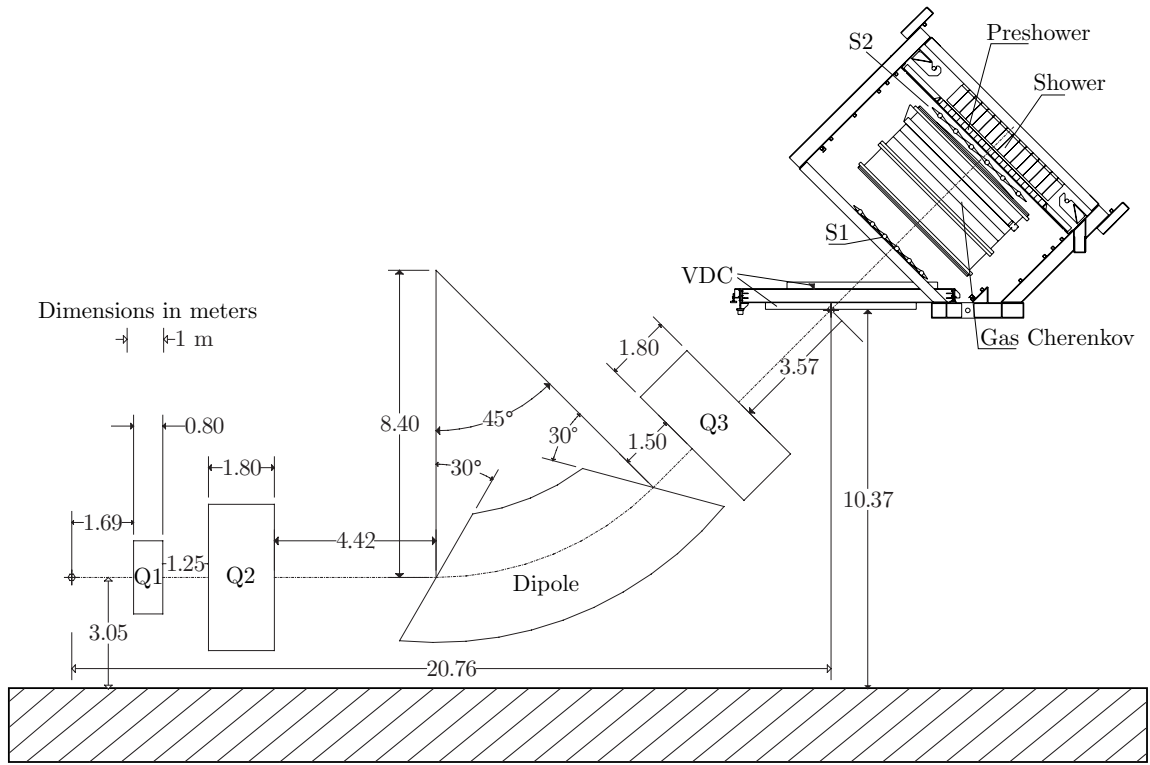}
	\caption [RHRS Layout] {This diagram shows the placement of the detectors inside of the Right High Resolution Spectrometer. The RHRS uses three quadruple magnets, Q1, Q2, and Q3, and one dipole magnet, D, to send particles into the detector package. The detector package for the experiments in this dissertation consisted of two trigger scintillator planes, S1 and S2, a gas \v{C}erenkov detector, and two lead-glass calorimeters called the preshower and shower. This schematic is adapted from \cite{HallAOsp:2011}.}
	\label{rhrs-layout}
\end{figure}

\begin{figure}
	\centering
	\includegraphics[width=15cm]{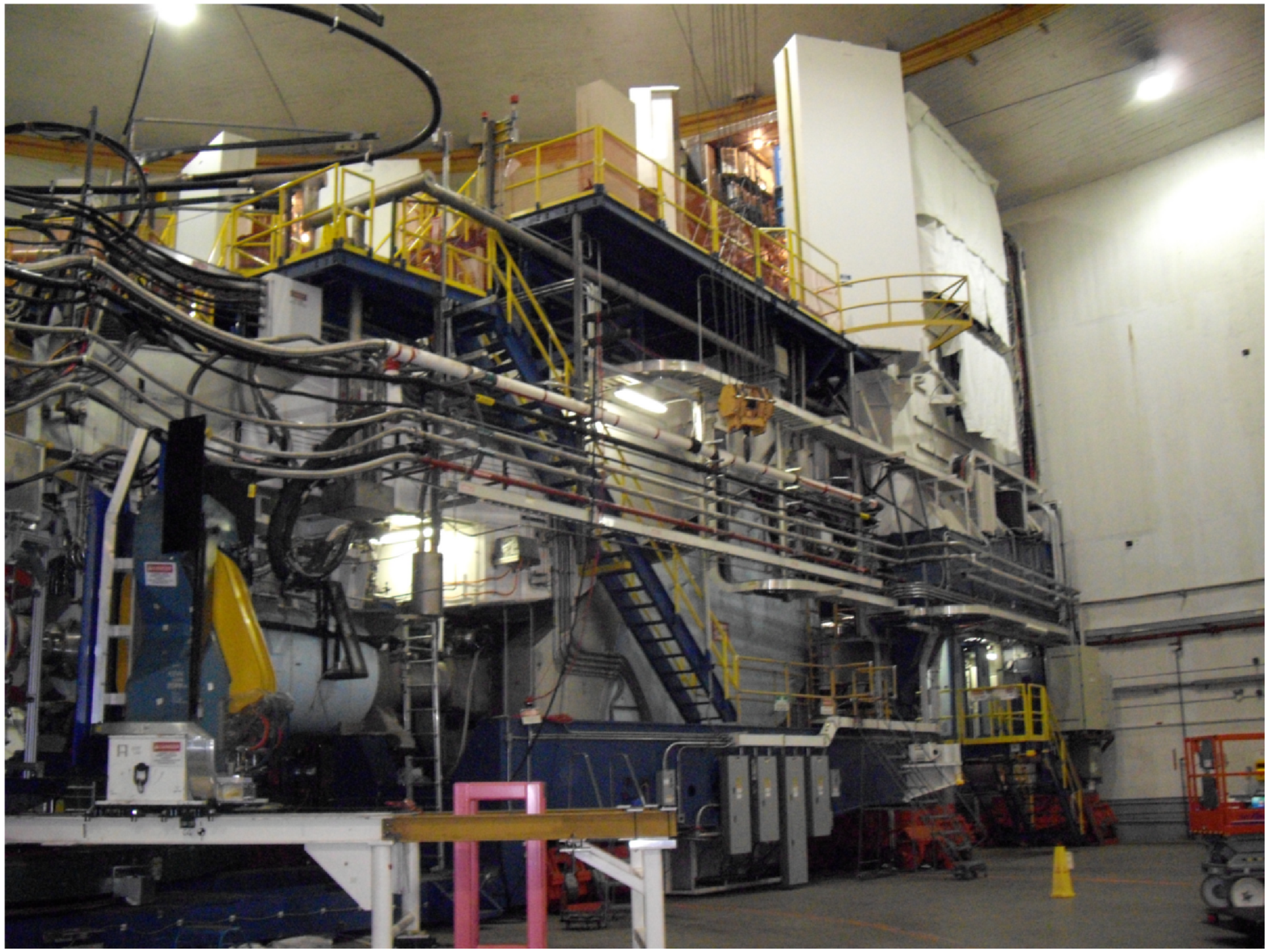}
	\caption [RHRS Photograph] {This photograph shows the Right High Resolution Spectrometer (RHRS) in Hall A. The detector package at the top of the RHRS is outside of the shield hut. During the experiment, the detector package was in place in the hut and the doors were closed. Also visible is the BigBite magnet, painted blue and yellow to the bottom left of the photograph, which was used to detect protons and deuterons.}
	\label{rhrs-photo}
\end{figure}

\normalsize
Jefferson Lab's Hall A has two primary detectors called the Left and Right High Resolution Spectrometers (LHRS and RHRS, respectively) \cite{Alcorn:2004sb}. The $^3$He($e,e'n$) reaction in the measurements presented in this dissertation only used the RHRS. The detector consisted of three quadruple magnets, one dipole magnet, and a detector package. The detector package for this experiment consisted of two multi-wire vertical drift chambers (VDCs), two trigger scintillators (S1 and S2), a gas \v{C}erenkov detector, and two planes of lead glass calorimeters (Preshower and Shower). The layout of each of these is shown in Figure \ref{rhrs-layout}. A photograph of the RHRS in the experimental hall is presented in Figure \ref{rhrs-photo}.


\subsubsection {Vertical Drift Chambers}
\label{ch2-vdc}
\begin{SCfigure}
	\centering
%	\import{./}{rhrs-detectors.eps_tex}
	\includegraphics[width=4 in]{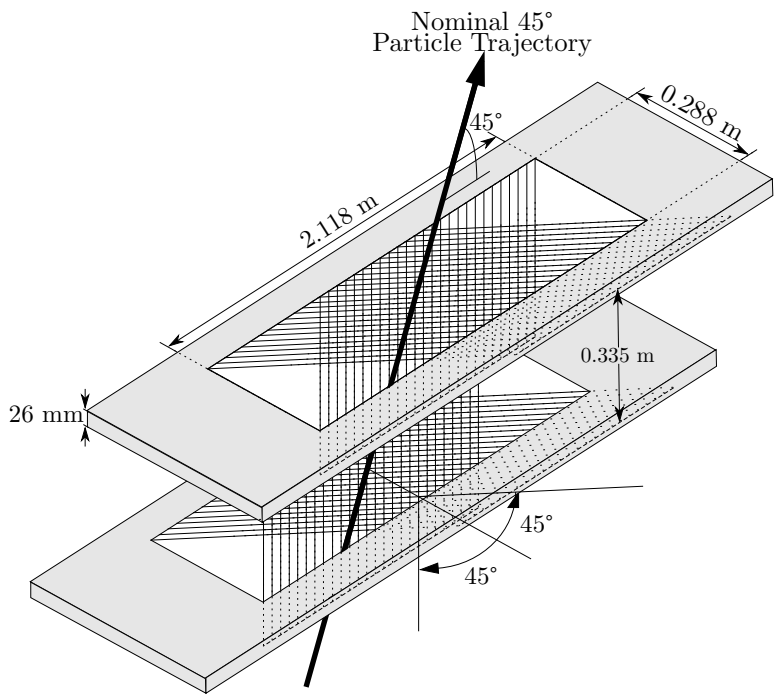}
	\caption [VDC Relative Geometry] {This schematic shows the relative angles and distances of the VDCs with respect to each other. Each VDC contains an upper (V) and lower (U) plane of wires that are orthogonal to each other. The wire planes are separated from their matching plane (U$_{\mathrm{top}}\leftrightarrow$U$_{\mathrm{bottom}}$, V$_{\mathrm{top}}\leftrightarrow$V$_{\mathrm{bottom}}$) by 0.335 m. This figure is adapted from Reference \cite{HallAOsp:2011} and is not to scale.}
	\label{vdc}
\end{SCfigure}

The RHRS has two planes of vertical drift chambers that can measure the position and angle of scattered electrons to within $\pm125$ $\mu$m. Each VDC consists of two orthogonal planes of wires, U and V, held at a high voltage and immersed in a bath of gaseous argon and ethane. As charged particles travel through the gas, they become ionized and are attracted to the wire planes. Upon collision with the wires, a signal pulse is generated that is recorded using multi-hit TDCs. Each VDC has a 2118 mm $\times$ 288 mm active area. The geometry of the VDCs is shown in Figure \ref{vdc} \cite{Fissum:2001st}.

\subsubsection {Trigger Scintillators}
\label{ch2-s1s2}

\begin{figure}
	\centering
	\includegraphics{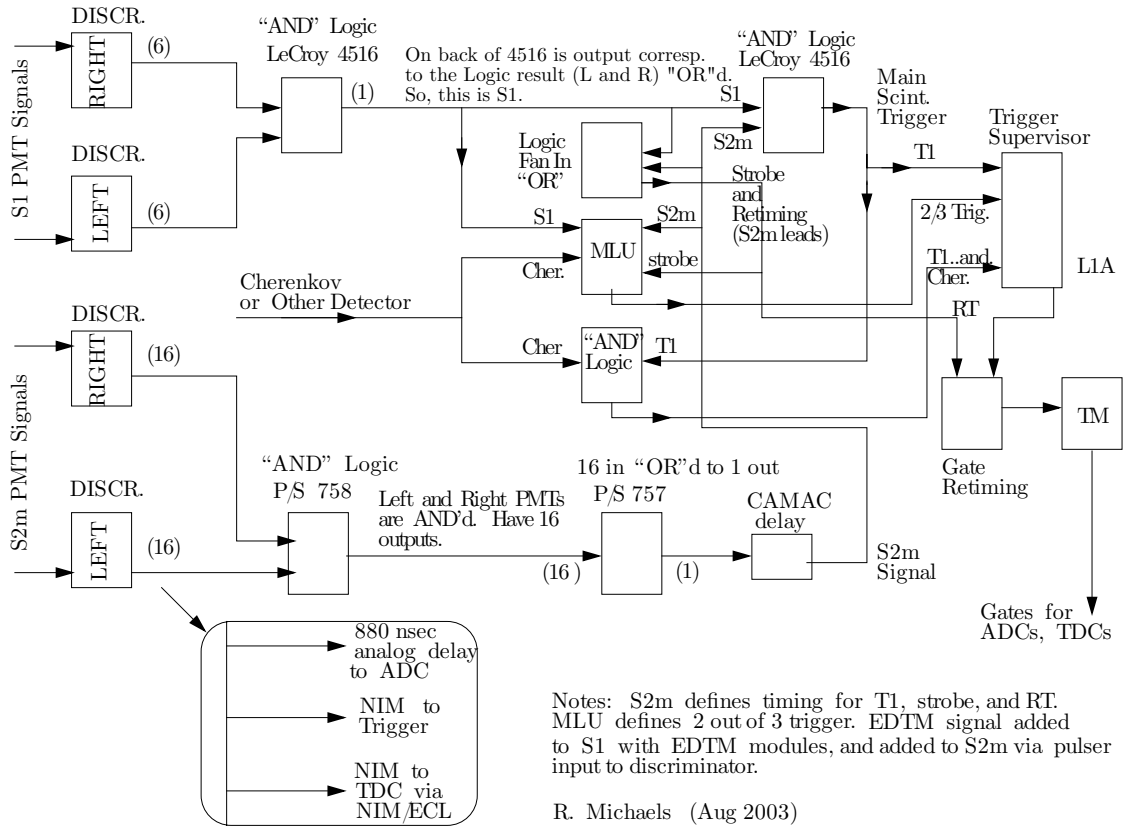}
	\caption [HRS Trigger Scheme] {This diagram maps out the trigger scheme for the RHRS spectrometer. Signals from the trigger scintillators enter on the left, and the output on the right is a gate that the data acquisition uses to open windows in the ADCs and TDCs of the other detectors.}
	\label{trigger-scheme}
\end{figure}

Two planes of thin plastic scintillators were used as triggers in the RHRS. Ionizing radiation deposited in scintillators causes them to fluoresce \cite{Moser199331}. The light given off from this process is collect by photomultiplier tubes (PMTs) \cite{PhysRev.55.966}. Each plane has six overlapping 5 mm-thick paddles. The planes are separated by approximately 2 m and have a time resolution of $\sim0.30$ ns. Every individual paddle records a possible event when there is a coincidence between the two PMTs on that paddle. If the event is picked up in both the front scintillator (S1) and the rear scintillator (S2), then the event is recorded \cite{Alcorn:2004sb}. Details of the trigger electronics are shown in Figure \ref{trigger-scheme}.

\subsubsection {Gas \v{C}erenkov}
\label{ch2-cerenkov}
A CO$_2$-filled gas \v{C}erenkov detector is used to separate pions from electrons. It is positioned between the S1 and S2 scintillator planes, as shown in Figure \ref{rhrs-layout}.. It has a particle path of 130 cm and consists of 10 spherical mirrors of 80 cm focal length that are each focused on a PMT. 

When high-speed particles travel through the gas, they are traveling faster than light can through the CO$_2$. As they progress through the CO$_2$, they excited the atoms in the gas which rapidly go back to the ground state, given off luminous energy in the process \cite{Cerenkov:1937vh, 1063-7869-52-11-R03}. The radiated light, known as \v{C}erenkov radiation, is collected and recorded. Since electrons are lighter than pions, it is easier to accelerate them to speeds required for \v{C}erenkov radiation to occur. A cut made on a small channel of the \v{C}erenkov detector's ADC easily distinguishes between pions and electrons as is discussed in detail in Section \ref{pion-contamination}. 

\subsubsection {Electromagnetic Calorimeters}
\label{ch2-pssh}

\begin{figure}
	\centering
	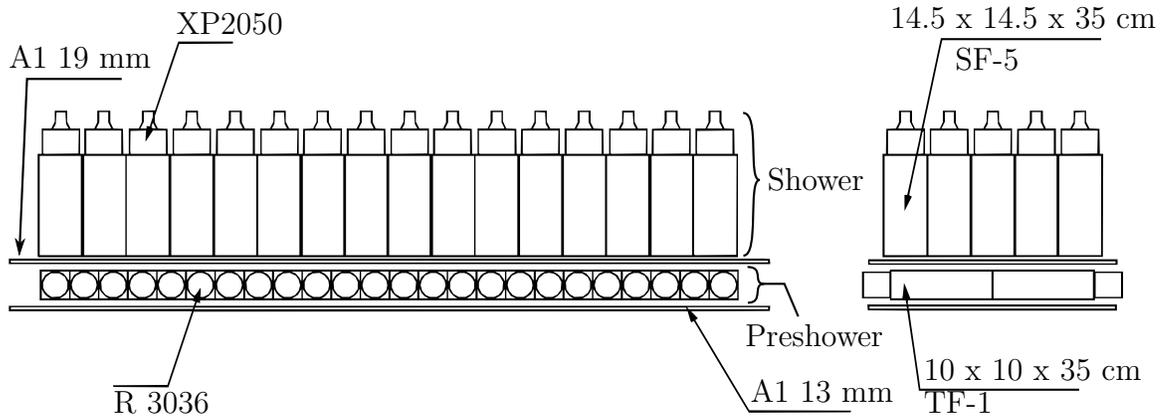
	\caption [Shower and Preshower Layout] {This shows the schematics for the shower and preshower lead glass blocks used in the RHRS \cite{Alcorn:2004sb}. These electromagnetic calorimeters were used to separate electrons from pions.}
	\label{pssh-layout}
\end{figure}

The RHRS contains two layers of electromagnetic calorimeters called the ``preshower" and ``shower" detectors. They are made out of lead glass blocks attached to photomultiplier tubes. The size of the blocks is given in Table \ref{tab:pssh}. Particles can be identified by how much energy they deposit in the preshower compared to the shower \cite{Alcorn:2004sb}, which allows electrons to be separated from hadrons. A schematic of the calorimeters is shown in Figure \ref{pssh-layout}.

\begin{table}
\begin{center}
\begin{tabular}{c|c|c|c|c|c|c}
Name & \# of Blocks & Cols & Rows & X (cm) & Y (cm) & Z (cm) \\ \hline
Preshower & 48 & 2 & 24 & 10.0 & 35.0 & 10.0 \\
Shower & 75 & 5 & 15 & 15.0 & 15.0 & 32.5 \\
\end{tabular}
\caption[Preshower and Shower Properties.]{This table contains the number and dimensions of lead glass blocks used in the Preshower and Shower detectors. ``X" denotes the dispersive plane, ``Z" is along the average particle direction, and ``Y" is parallel to the focal plane.}
\label{tab:pssh}
\end{center}
\end{table}

\newpage
\large
\subsection {Hall A Neutron Detector} 
\label{ch2-hand}
\normalsize

\begin{wrapfigure}{r}{0.4\textwidth}
	% \capstart
	\centering
	\includegraphics[width=0.4\textwidth]{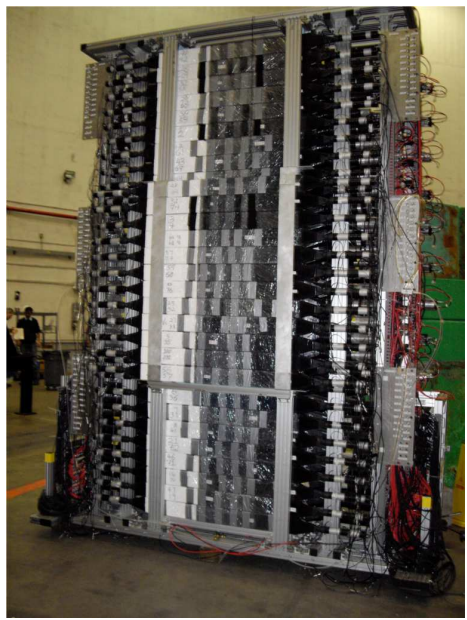}
	\caption [HAND Photograph] {This photograph shows the Hall A Neutron Detector used in this experiment. It is seen from the front view where the veto layer is visible.}
	\label{hand-photo}
\end{wrapfigure}
\begin{figure}
	\centering
	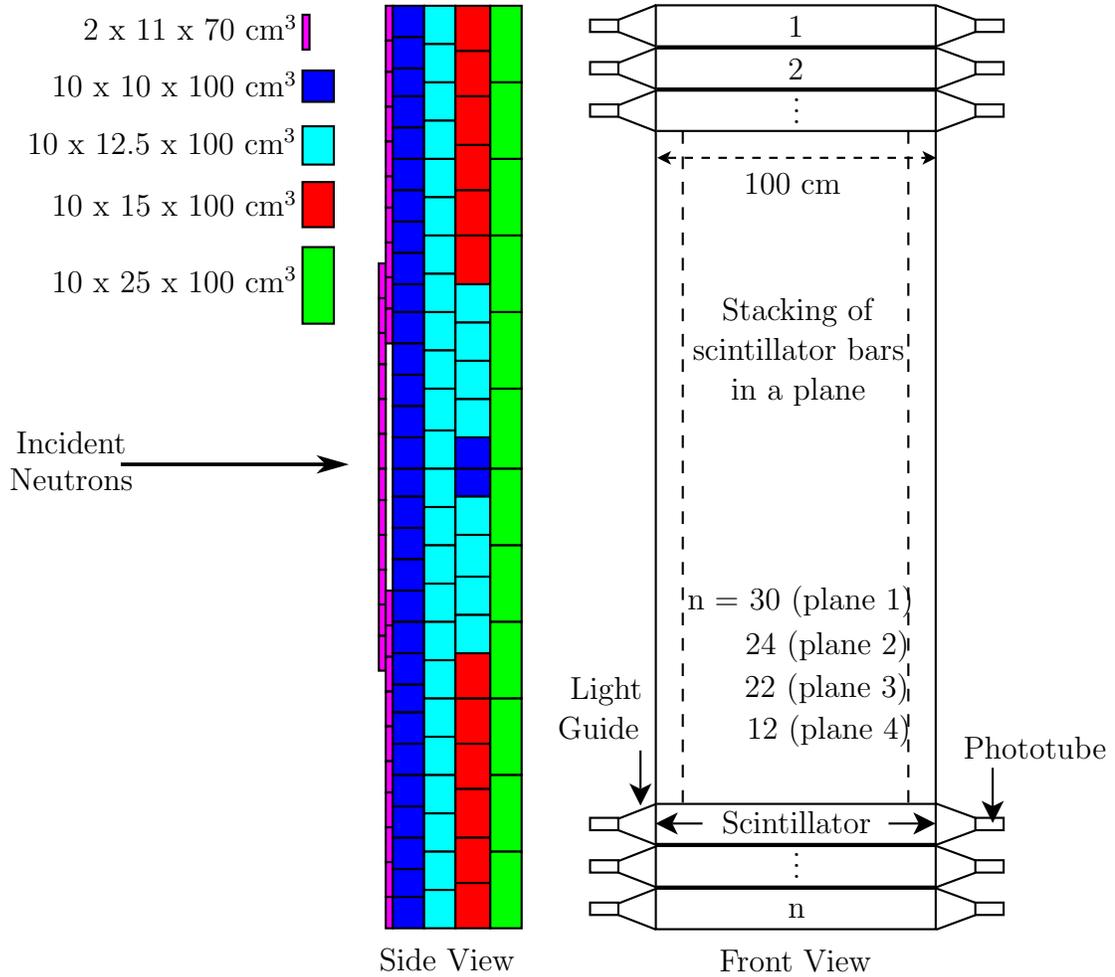
	\caption [HAND Layout] {This shows the arrangement of the scintillator bars in the Hall A Neutron Detector. Incident neutrons collide with protons in the 10 cm thick scintillator bars. The protons cause the material to scintillate, the light from which is sent through a light guide into a PMT. This figure is adapted from \cite{Subedi2007}.}
	\label{hand-layout}
\end{figure}

The Hall A Neutron Detector (HAND) is a non-standard piece of equipment that was used previously in a short-range correlation experiment \cite{Subedi:2008zz}. It consists of an array of plastic scintillators connected to PMTs. Timing information is read through Time-to-Digital Converters (TDCs) for each PMT. HAND is made of 88 main detecting bars arranged in four layers. The thickness of each bar in these layers is 10 cm, the length is 100 cm, and the height varies with the smaller bars placed in front of the larger bars. There is also a thinner ``veto" layer that contains 64 bars with dimensions of 2 x 11 x 70 cm$^3$. The layout of these bars can be seen in Figure \ref{hand-layout}. 

%\begin{figure}
%	\centering
%	\includegraphics[width=15cm]{./hand-photo.eps}
%	\caption [HAND Photograph] {HAND Photograph. This photograph shows the Hall A Neutron Detector used in this experiment. It is seen from the front view where the veto layer is visible.}
%	\label{hand-photo}
%\end{figure}

\begin{SCfigure}
% {l}{0.4\textwidth}
	\centering
	\includegraphics[width=0.4\textwidth]{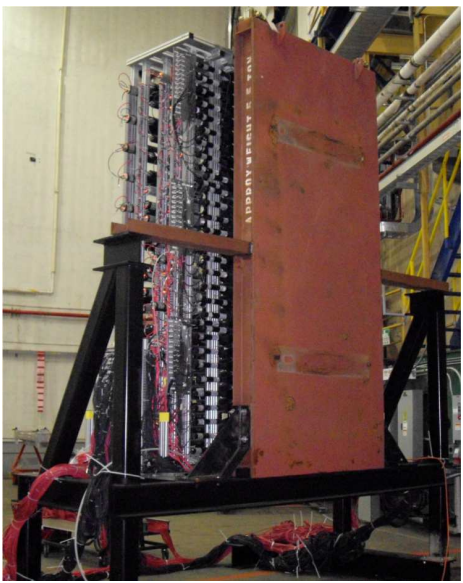}
	\caption [HAND with Pb Wall] {This photograph shows the Hall A Neutron Detector with the lead wall in place. Also visible are the high voltage cables (red) and signal cables (black) that powered and carried the signals from the PMTs. In the background towards the right is the LHRS.}
	\label{hand-pb-wall-photo}
\end{SCfigure}

Since neutrons do not carry charge, they are not directly measured by the scintillator; however, they will elastically knock a proton out of H atoms in the plastic scintillating detectors. The scattered proton then radiates light in the scintillator, which is detected. This process occurs over a distance of approximately 10 cm. Since protons and neutrons are similar in mass, protons scattered from $^3$He will arrive at the detector at approximately the same time as neutrons. In order to separate neutrons from protons, a series of veto cuts, described in detail in Section \ref{ch3-neutronid} and Appendix \ref{hand-vetoes}, are made in post-analysis that exclude events picked up by bars in front of any given bar within the timing window for both protons and neutrons. In particular, a proton should always deposit a signal in the 2 cm thick veto bars whereas a neutron will most likely pass through the thin veto counter without interacting. However, for this experiment the veto layer was often flooded with particles, making neutron detection using only that layer inefficient. In order to accurately determine neutrons, each layer was used as a veto layer for the bars behind it. In addition at higher scattering energies, a 9.08 cm thick wall, made up of 4 cm of iron casing surrounding the 5.08 cm thick lead, was placed in front of HAND that greatly reduced the number of gamma particles, and to a lesser extent the number of protons, that made it to the detector. The lead wall is visible in Figure \ref{hand-pb-wall-photo}.

\begin{SCfigure}
	\centering
	\includegraphics[width=0.5\textwidth]{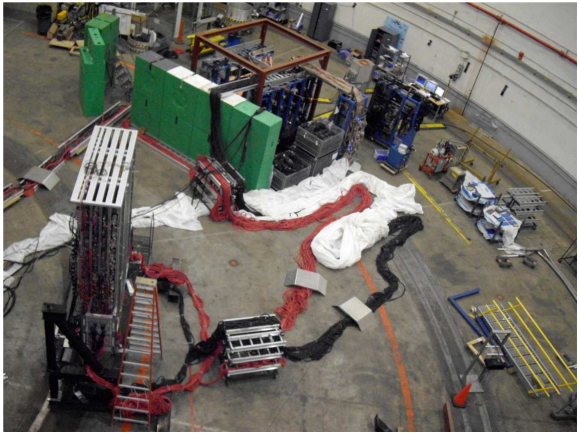}
	\caption [HAND and Electronics Placement] {This photograph shows where both the Hall A Neutron Detector and the electronics hut were located in Hall A during the experiment. Also visible are the high voltage cables (red) and signal cables (black) that powered and carried the signals from the PMTs.}
	\label{hand-in-hall-photo}
\end{SCfigure}

\begin{figure}
	\centering
	\includegraphics[width=15cm]{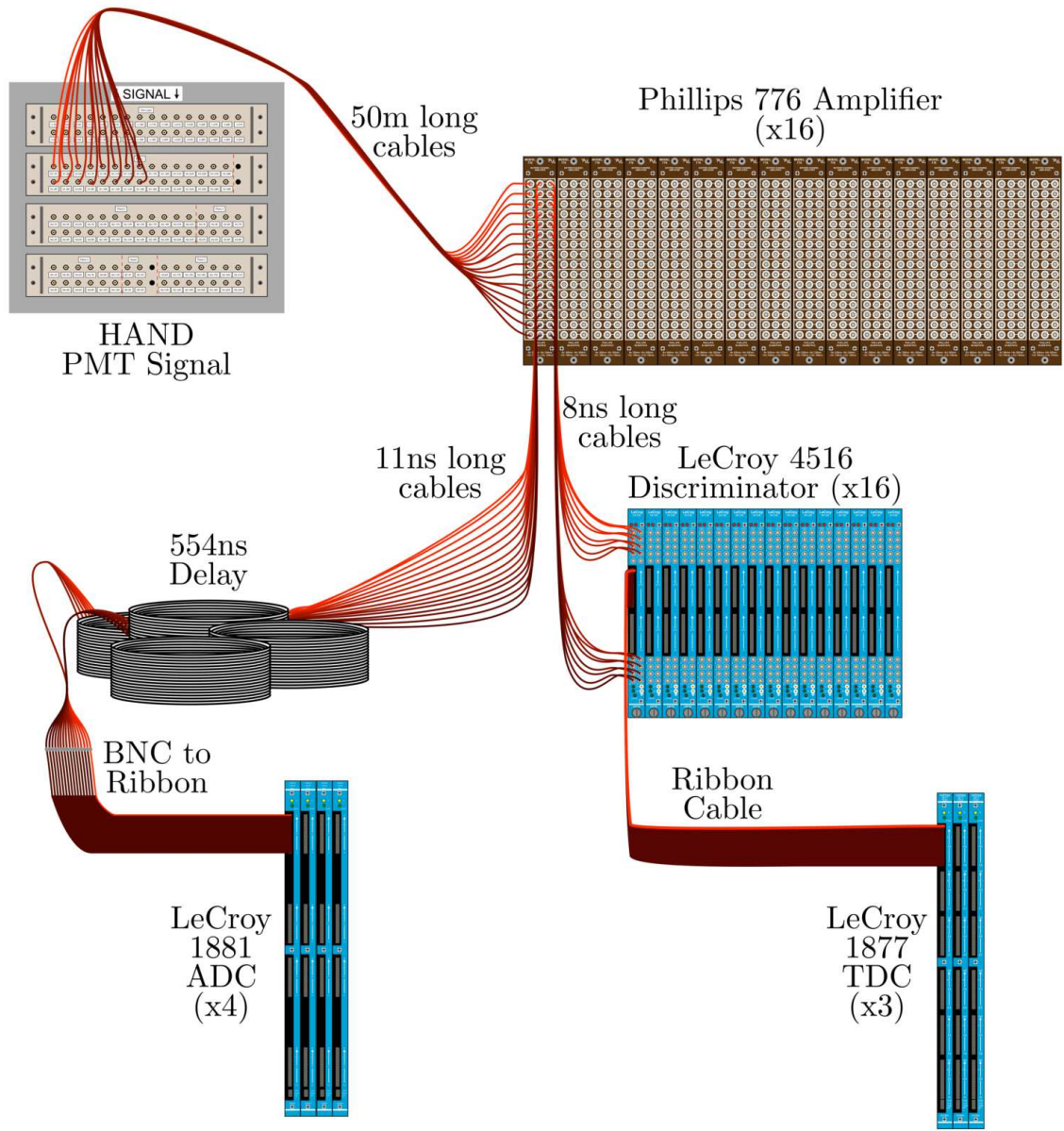}
	\caption [HAND Wiring Diagram] {The electronics for HAND were set up according to this wiring diagram. The incoming signals were from each individual photo-multiplier tube (PMT) shown in Figure \ref{hand-layout}.}
	\label{hand-wiring-diagram}
\end{figure}

\begin{SCfigure}
	\centering
	\includegraphics[width=0.5\textwidth]{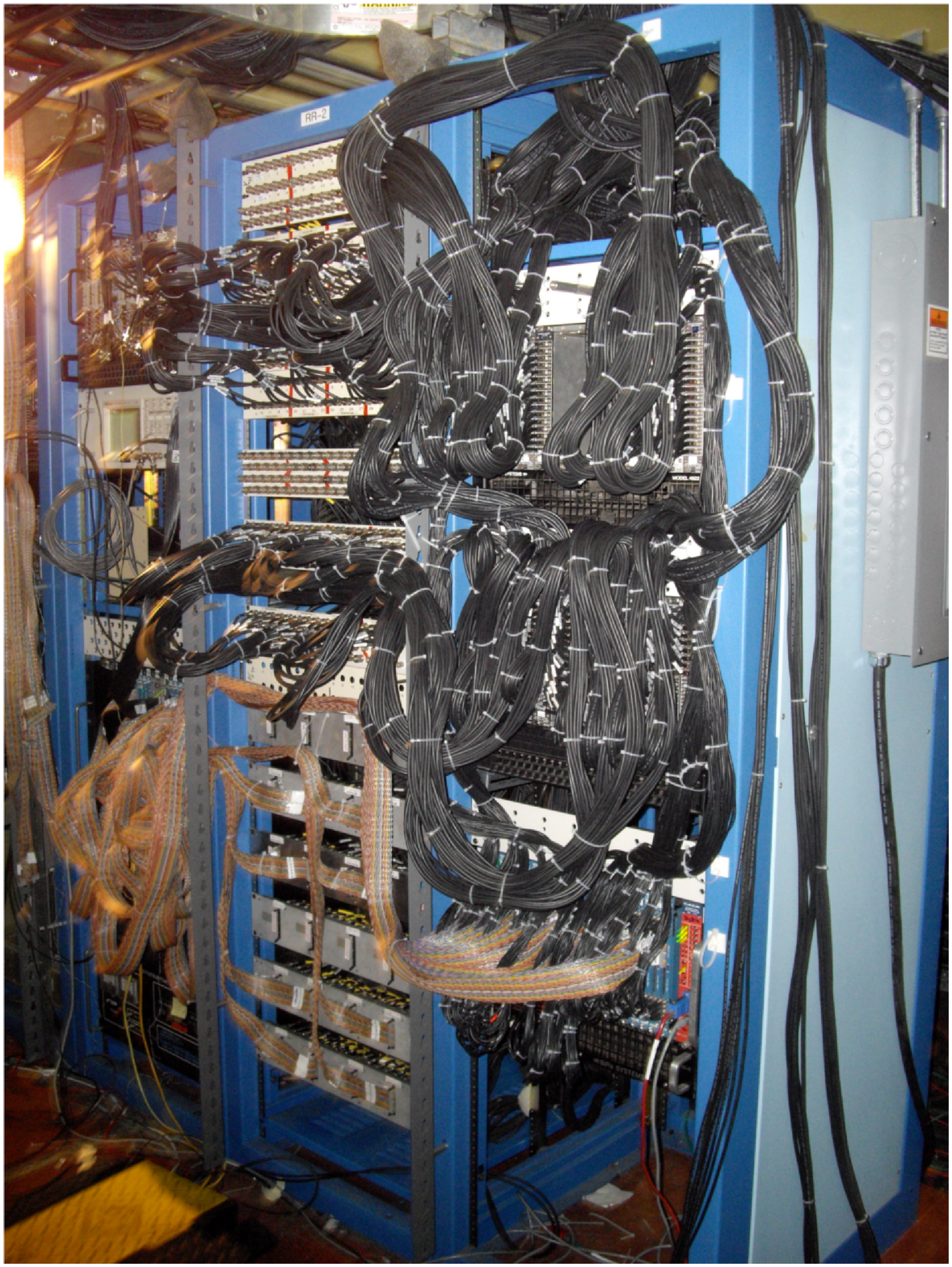}
	\caption [HAND Electronics] {This photograph shows the actual wiring and electronic system for HAND used in the experiment. An overview of the wiring diagram is presented in Figure \ref{hand-wiring-diagram}.}
	\label{hand-electronics-photo}
\end{SCfigure}

The electronics for HAND were used to record timing information. The signal cables from the PMTs were fed into amplifiers where the signal was doubled. One copy of the signal was sent to a discriminator and then a time-to-digital converter (TDC). The other copy was sent through a 554ns delay before being recorded in an analog-to-digital converter (ADC). The wiring diagram for this set-up is presented in Figure \ref{hand-wiring-diagram} and photographs of the electronics are presented in Figures \ref{hand-in-hall-photo} and \ref{hand-electronics-photo}.


% ^^^^^^^^^^^^^^^^^^^^^^^^^^^^^^^^^^^^^^^^^^^^^^^^^^^^^^^^^^^^^^^^^^^^^^^^^^^^^^^
%\pagebreak

\large
\section {Kinematics}
\label {experiment-kinematics}
\normalsize
%Kinematics
% vvvvvvvvvvvvvvvvvvvvvvvvvvvvvvvvvvvvvvvvvvvvvvvvvvvvvvvvvvvvvvvvvvvvvvvvvvvvvvv

In order to map out the quasi-elastic scattering region, the detectors mentioned in the previous sections were placed at different angles, energy settings, and target polarization directions. A listing of each of these kinematics settings is found in Table \ref{tab:qekinematics}. A negative angle corresponds to a detector placed to the right of the beam line whereas a positive angle corresponds to one placed to the left of the beam line. A 0$^{\circ}$ angle corresponds to downstream of the target, along the beamline.

\begin{figure}
	\centering
	\includegraphics[width=15cm]{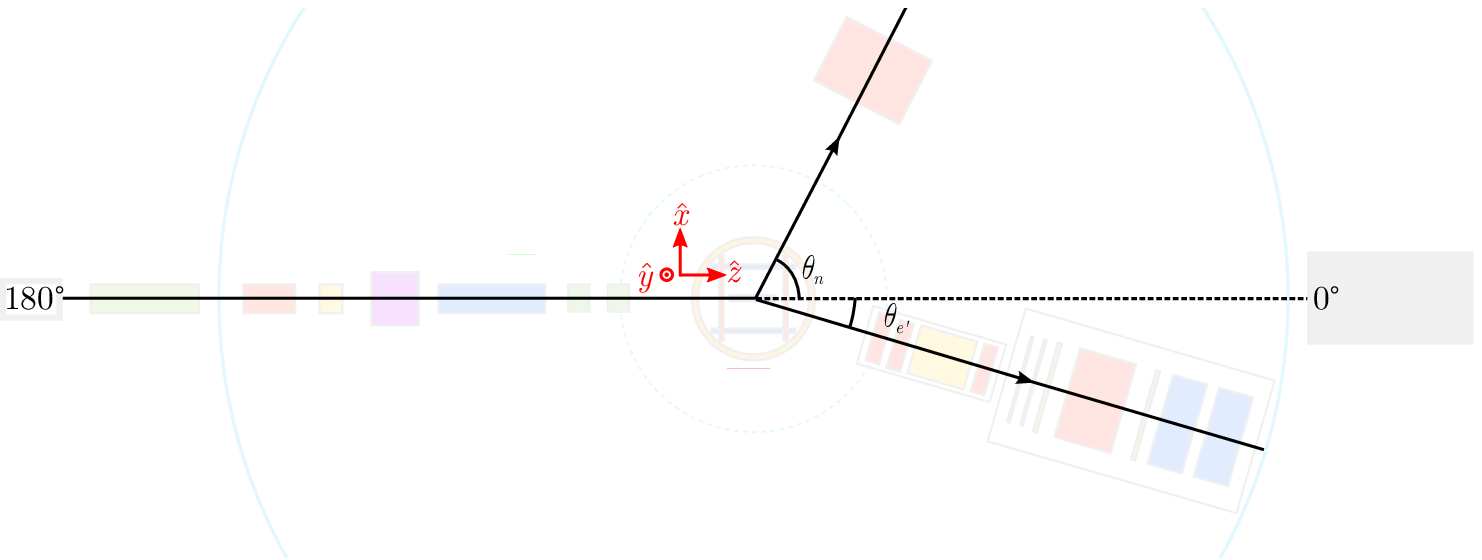}
	\caption [Hall A Angle Definitions] {Hall A Angle Definitions. This figure shows how the angles of HAND ($\theta_n$) and the RHRS ($\theta_{e'}$) are defined, along with the $\hat{x}$, $\hat{y}$, and $\hat{z}$ directions. Downstream of the target, towards the beam dump, is defined as $0^{\circ}$.}
	\label{hall-angles}
\end{figure}
\begin{table}
\begin{center}
\begin{tabular}{c|c|c|c|c}
Target Pol. & E$_0$ (GeV) & RHRS ($^{\circ})$ & RHRS P$_0$ (GeV/$c$) &  HAND ($^{\circ}$) \\ \hline
Vertical & 1.245 & -17.0 & 1.1759 & 71.0 \\
Vertical & 2.425 & -17.0 & 2.1813 & 62.5 \\
Vertical & 3.605 & -17.0 & 3.0855 & 54.0 \\ \hline
%Longitudinal & 2.425 & -16.0 & 2.2500 & 62.5 \\
%Longitudinal & 2.425 & -16.0 & 2.2250 & 62.5 \\
Longitudinal & 2.425 & -18.0 & 2.1750 & 62.5 \\ 
%Longitudinal & 2.425 & -18.0 & 1.8650 & 62.5 \\
%Longitudinal & 2.425 & -18.0 & 0.7000 & 54.0 \\
%Longitudinal & 2.425 & -18.0 & 2.0250 & 54.0 \\
Longitudinal & 3.606 & -17.0 & 3.0855 & 54.0 \\ \hline
%Transverse & 2.425 & -16.0 & 2.2250 & 62.5  \\
Transverse & 2.425 & -18.0 & 2.1750 & 62.5  \\
%Transverse & 2.425 & -18.0 & 1.8500 & 62.5 \\
Transverse & 3.606 & -17.0 & 3.0855 & 54.0 \\
\end{tabular}
\caption[Kinematics for Quasi-Elastic Experiments.]{This table contains the kinematics settings for the Quasi-Elastic family of experiments. Every line corresponds to a change in the kinematics during the experiments. This includes, respectively, the beam energy (E$_0$), the right HRS central angle, the right HRS central momentum (P$_0$), the Hall A Neutron Detector central angle, and the target polarization direction. The angles are defined as in \ref{hall-angles}. Note that the RHRS angle is equal to negative $\theta_{e'}$ in the table above. See Figure \ref{eenDefineDirections} for definitions of the polarization directions.}
\label{tab:qekinematics}
\end{center}
\end{table}

% ^^^^^^^^^^^^^^^^^^^^^^^^^^^^^^^^^^^^^^^^^^^^^^^^^^^^^^^^^^^^^^^^^^^^^^^^^^^^^^^

					% Include the Chapter 3 text (chapter_3.tex)
	% vvvvvvvvvvvvvvvvvvvvvvvvvvvvvvvvvvvvvvvvvvvvvvvvvvvvvvvvvvvvvvvvvvvvvvvvvvvvvvvv
% Chapter 3 (chapter_3.tex)
%
% Chapter 3 of Elena Long's Ph.D. Dissertation
%
% To be completed: March, 2012
%
% ^^^^^^^^^^^^^^^^^^^^^^^^^^^^^^^^^^^^^^^^^^^^^^^^^^^^^^^^^^^^^^^^^^^^^^^^^^^^^^^^

\chapter{Particle Identification}	% Chapter Title
\label{particleid}		% Chapter Label
\normalsize			% Return to Normal font size

As mentioned in Chapter \ref{experimentsetup}, for this experiment the right High Resolution Spectrometer (RHRS) was used to detect electrons scattered from polarized $^3$He and the Hall A Neutron Detector (HAND) was used to detect knocked-out neutrons. This chapter discusses the analysis that went in to identifying these particles.

\large
\section {Electron Identification} 
\label{ch3-electronid}
\normalsize
% Electron Identification
% vvvvvvvvvvvvvvvvvvvvvvvvvvvvvvvvvvvvvvvvvvvvvvvvvvvvvvvvvvvvvvvvvvvvvvvvvvvvvvvv

The RHRS was used to identify electrons that were quasi-elastically scattered from the $^3$He nuclei. The spectrometer contains a gas \v{C}erekov detector and two electromagnetic calorimeters, known as the preshower and shower, that were used to differentiate between pions and electrons. The VDCs provided tracking information that was used to isolate electrons within a certain solid angle. They were also used to isolate events scattered from $^3$He from those scattered from the glass windows of the target cell. Combinations of these detectors were used to find the quasi-elastic scattering peak and to separate it from the elastic scattering peak and background events. Details of the electron cuts are discussed below.

\subsection {HRS Optics}
\label{hrs-optics}

\begin{figure}
	\centering
	\includegraphics[width=6 in]{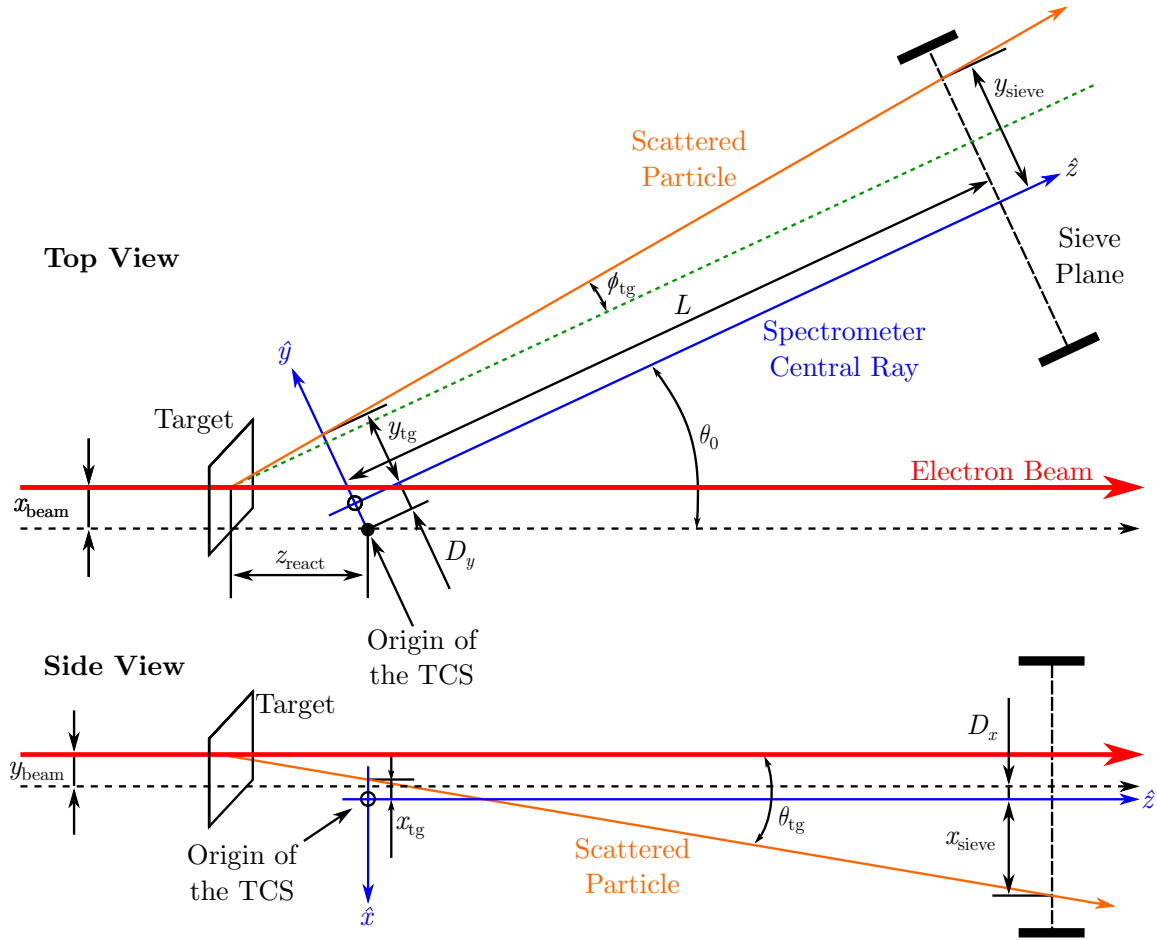}
	\caption [Target Coordinate System] {Displayed here is the target coordinate system used to calibrate the RHRS. The upper image is from a bird's-eye view looking down and the lower image is a side view. This figure is adapted from \cite{Jin:2011uva}.}
	\label{target-coord}
\end{figure}

In order to separate particles for identification within the spectrometer, the optics needed to be calibrated. This was done using a sieve-slit collimator and a multi-foil carbon target. The target coordinate system was used for this calibration, as shown in Figure \ref{target-coord}.

\begin{SCfigure}
	\centering
	\includegraphics[width=4.25 in]{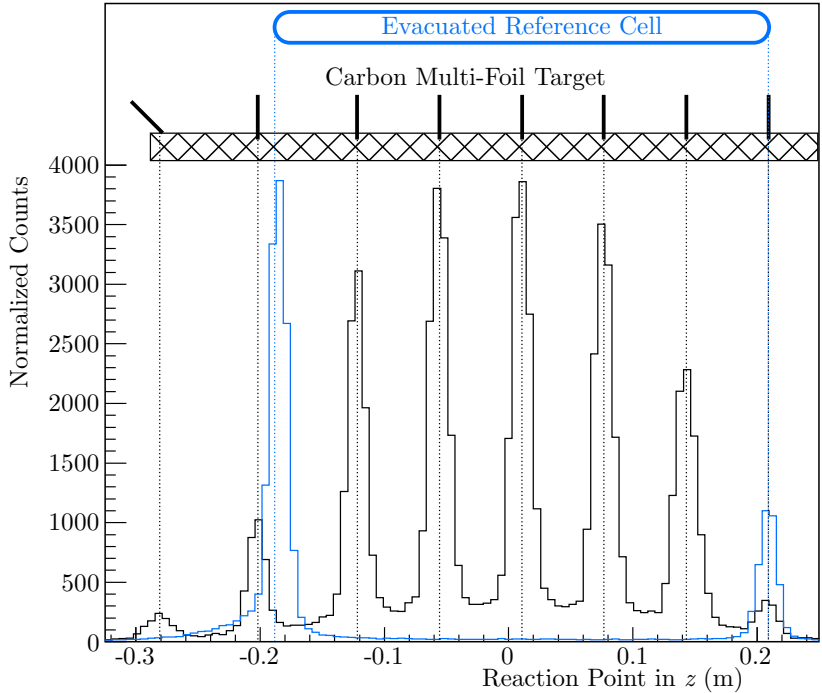}
	\caption [RHRS Extended Target Optics] {This plot shows the calibration of the reaction point in $z$ via the use of a carbon multi-foil target (black) and an evacuated glass reference cell (blue). There was also a small BeO foil shown slanted. The multi-foil target provided small areas of reaction points along $\hat{z}$ which allowed for accurate tracing of particles from the extended, 40 cm $^3$He target.}
	\label{reactptz}
\end{SCfigure}

The multi-foil carbon target was used to trace particles back to the origin of their scattering. It was needed because the $^3$He target consists of an extended, 40 cm long chamber instead of a point target. An identical, but evacuated, 40 cm long glass target was used for calibration. In order to account for the long target, events were traced back to each of the carbon foils in the calibration of the reconstruction matrix. The multi-foil target was made of multiple point targets placed at intervals of approximately 6.67 cm. This is shown in Figure \ref{reactptz}. 
 
\begin{figure}
	\centering
	\includegraphics{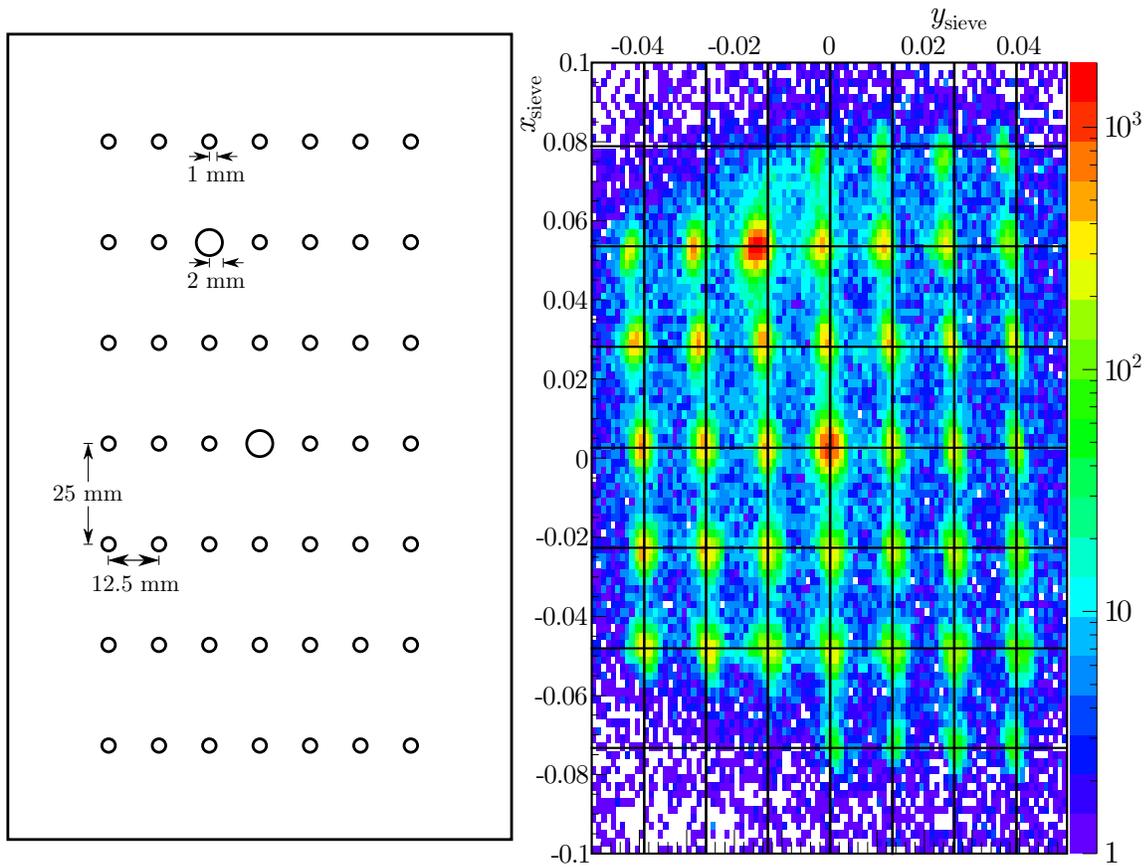}
	\caption [Sieve Pattern] {The schematic on the left shows the orientation of the sieve pattern used to calibrate the RHRS. The plot on the right shows data with the sieve plate in after calibration was completed.}
	\label{sieve-plate}
\end{figure}

A tungsten sieve-slit collimator was used to calibrate the trajectory of a particle ($\theta_{tg}$ and $\phi_{tg}$).  The sieve is a sheet of steel with a pattern of 49 holes (7 x 7) that have a radius of 1 mm and are spaced 25 mm apart vertically and 12.5 mm apart horizontally. Two of the holes have a radius of 2 mm and are placed asymmetrically in the pattern so that its orientation is easily identified. A diagram of the sieve pattern, along with calibration data, is shown in Figure \ref{sieve-plate}.

By using both the sieve pattern and the multi-foil carbon target, events had to pass through multiple known locations. The reconstruction matrix adjusted tracking events to match these locations. The calibration of the reaction point in $z$ , $\theta_{tg}$, and $\phi_{tg}$ was completed by Jin Ge \cite{Jin:2011uva}.

\subsection{Pion Contamination}
\label{pion-contamination}
\begin{figure}
	\centering
	\includegraphics{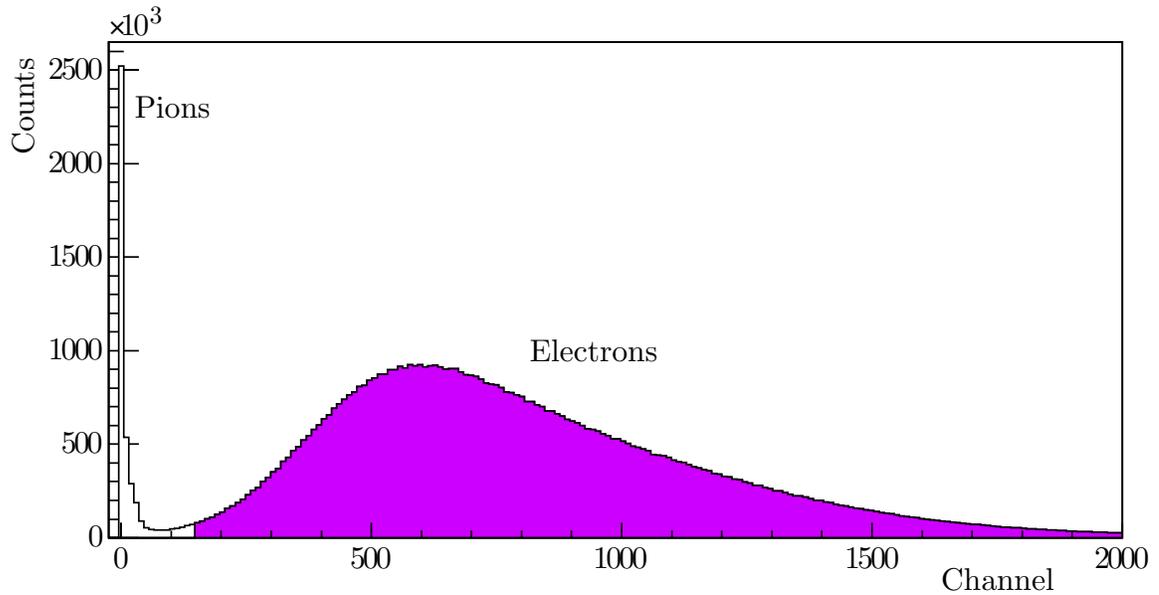}
	\caption [\v{C}erenkov Pion Cut] {The separation of pions from electrons was done through the use of a cut on ADC of the gas \v{C}erenkov detector.}
	\label{ceren-cut}
\end{figure}

In order to separate pions from electrons, the gas \v{C}erenkov, preshower, and shower detectors were used. Pions appear in the output of the \v{C}erenkov detector as a large, sharp peak around channel 0, whereas electrons appear as a much wider peak at channels above 100 as described in Section \ref{ch2-cerenkov}.. This can be seen in Figure \ref{ceren-cut}.

\begin{figure}
	\centering
	\includegraphics{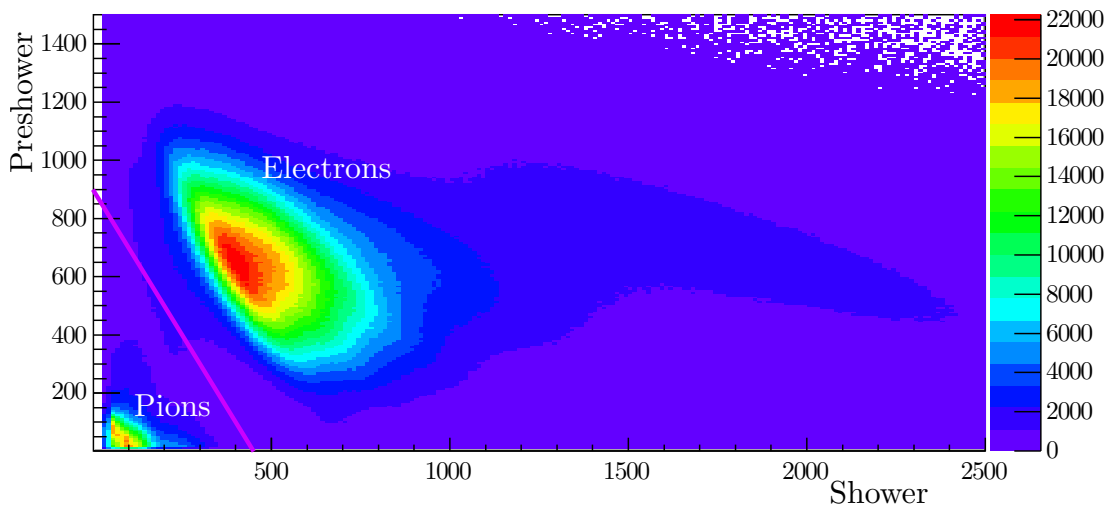}
	\caption [Preshower vs. Shower Pion Cut] {The separation of pions from electrons was done through the use of a cut on the lead glass calorimeters (known as the ``shower" and ``preshower"). This cut shown in pink.}
	\label{pssh-cut}
\end{figure}

In addition, the preshower and shower detectors were used as secondary measurements to separate pions from electrons. There is a clear pion peak and electron peak that can be seen in Figure \ref{pssh-cut}. A linear cut was made between the peaks and only those on the electron side were kept.

\subsection{Glass Window Contamination}
\label{glass-contamination}
\begin{figure}
	\centering
	\includegraphics{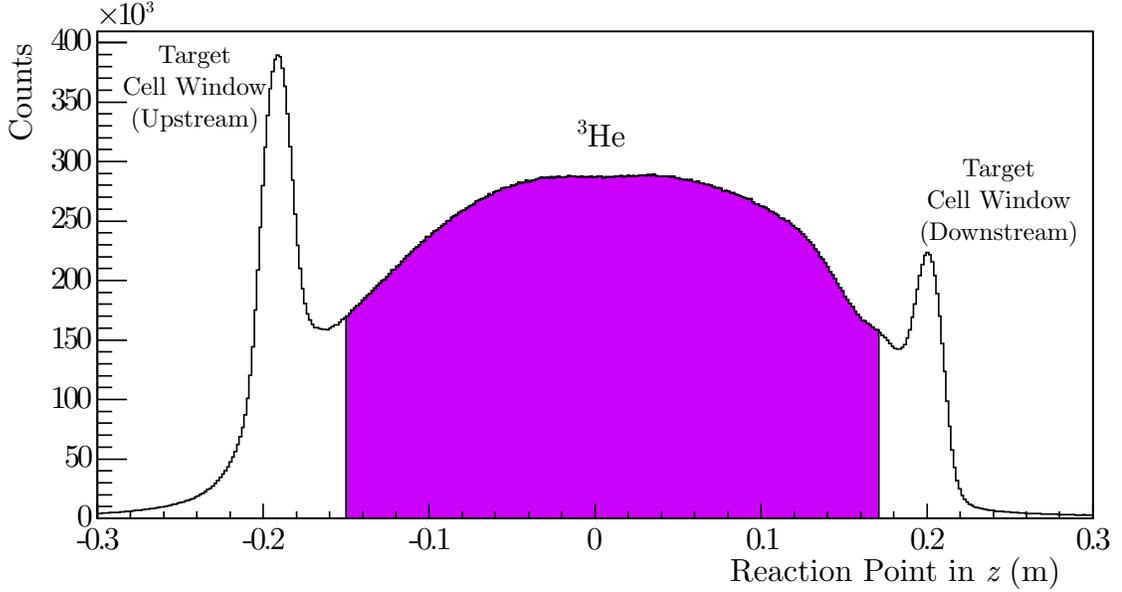}
	\caption [Reaction Point in z Cut] {Separation of $^3$He events from glass window events was achieved by making a cut on the reaction point in $z$.}
	\label{reactz-cut}
\end{figure}

Through the use of tracking variables, the reaction point of scattering along the $\hat{z}$ is able to be determined. From this distribution, it becomes clear that there is a distinction between the $^3$He scattered events and those events that are scattered off of the glass end windows of the target cell. In order to remove events scattered from the windows, a cut was made 3.7 $\sigma$ away from the central value of the upstream window peak and 3 $\sigma$ away from the downstream window peak. The larger cut was made on the upstream side since the magnitude of the peak is much larger than for the downstream side, a trend that becomes more important as the beam energy increased. The cuts used to select $^3$He events is shown in Figure \ref{reactz-cut}.

\subsection{Elastic and Quasi-Elastic Peaks}
\label{elastic-qe-peak}

There are two distinct peaks caused by different types of scattering from $^3$He: the elastic peak, where the whole $^3$He nucleus is scattered by the incoming electron, and the quasi-elastic peak, where a single nucleon is struck by the electron. The contribution of elastic events decreases as the energy of the beam was increased, but it was especially important to take the elastic peak into account for the Q$^2$=0.127 (GeV/$c$)$^2$ data. The $x_{\mathrm{Bjorken}}$ variable was ideal for differentiating between these peaks. It is defined as
\begin{equation}
	x_{\mathrm{Bjorken}} = \frac{\mbox{Q}^2}{2m_N \nu},
\end{equation}
where Q$^2$ is the squared four-momentum transfer, $m_N$ is the average mass of a nucleon, and $\nu$ is the energy transferred.

\begin{figure}
	\centering
	\includegraphics{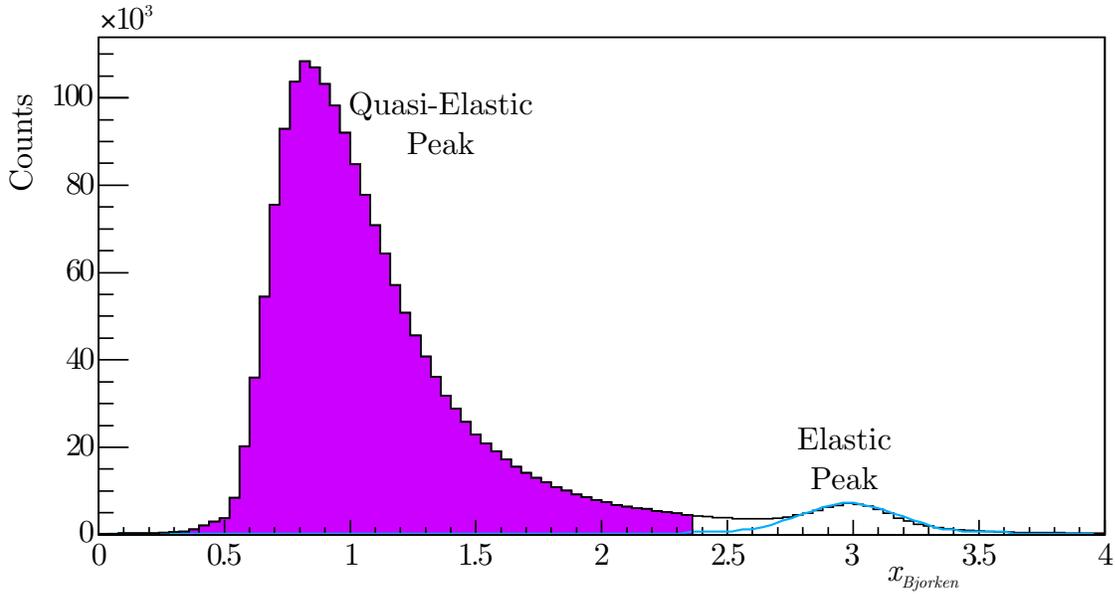}
	\caption [$x_{Bjorken}$ Cut] {The separation of the quasi-elastic from the elastic peak was made by using a cut on $x_{\mathrm{Bjorken}}$. Also shown, in light blue, is a fit on the elastic peak in $x_{\mathrm{Bjorken}}$.}
	\label{xbj-cut}
\end{figure}

\begin{figure}
	\centering
	\includegraphics[width=15 cm, angle=0]{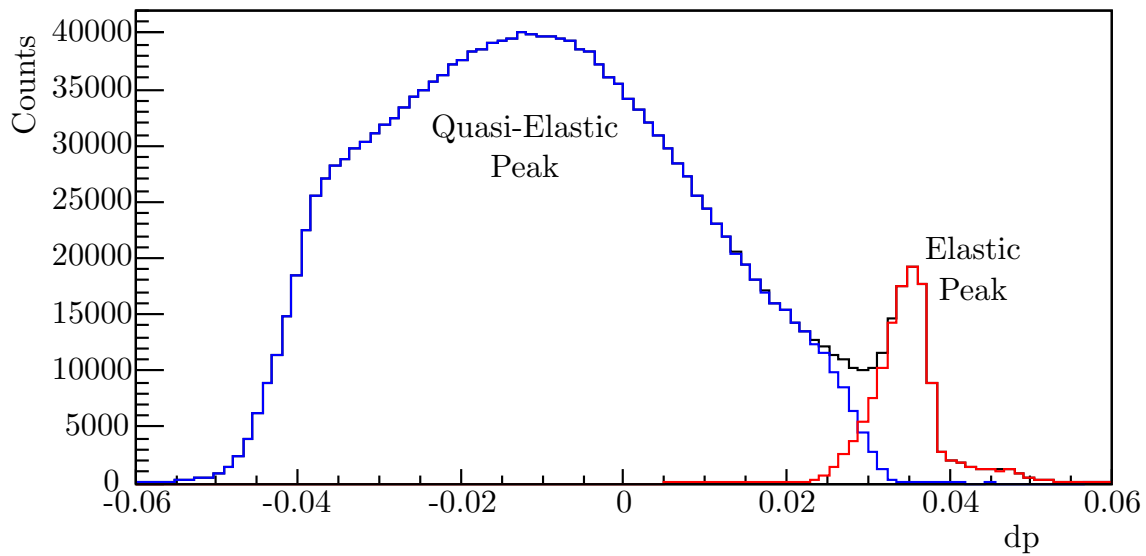}
	\caption [Effect of $x_{Bjorken}$ cut on dp] {This figure shows which events were kept and which were discarded from dp when the cut on $x_{Bjorken}$ was made. The red events were removed while the blue ones were included in the dataset. The black line shows the total events.}
	\label{nu-with-xbj-fit}
\end{figure}

In order to ensure that there was no contamination from the elastic peak, especially for the lowest Q$^2$ point, a fit was made on the elastic peak in $x_{\mathrm{Bjorken}}$. When the elastic peak was removed, as shown in Figure \ref{xbj-cut}, it left behind only those events that were quasi-elastically scattered. This can be further seen in the dp plot shown in Figure \ref{nu-with-xbj-fit}.

% Include dp, Q2, nu here

\subsection{Events in RHRS Acceptance}
\label{qvec-events}

The RHRS has an angular acceptance of approximately 6 msr. Scattered electrons were kept if they fall within this acceptance. A cut was made on $\theta_{tg}$ and $\phi_{tg}$ in the target coordinate system to exclude events outside of the acceptance. Electrons from these events acted as triggers to accept neutrons from HAND. 

\begin{figure}
	\centering
	\includegraphics{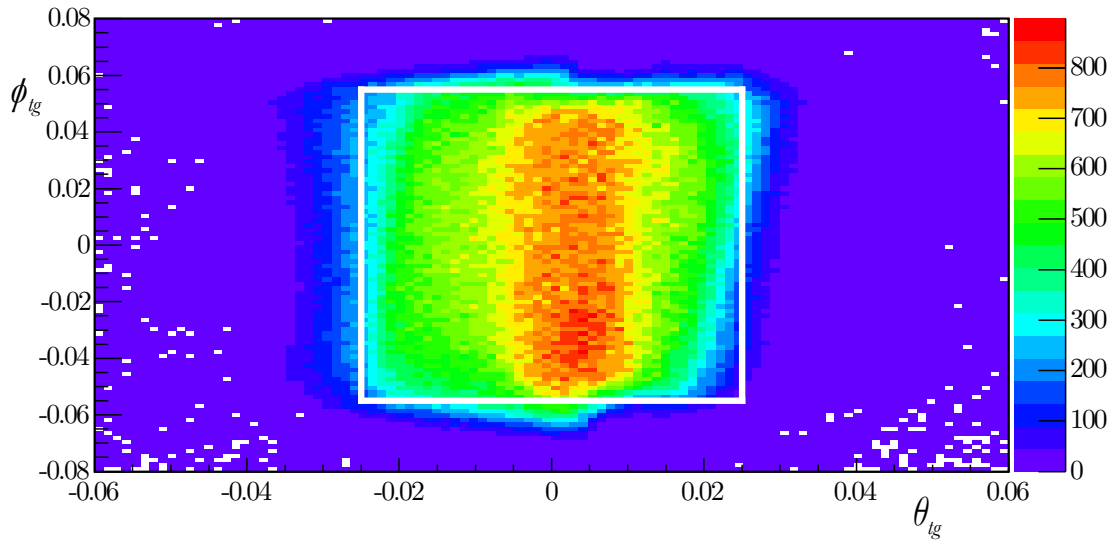}
	\caption [$\theta_{tg} : \phi_{tg}$ Cut] {Events along the q-vector were selected based on $\theta_{tg}$ and $\phi_{tg}$ in the target coordinate system. Selected events are enclosed in the white square.}
	\label{thph-cut}
\end{figure}

\section{Summary of Electron Cuts}
\label{sum-electron-cuts}
Sections \ref{hrs-optics} through \ref{qvec-events} describe in detail each of the cuts made to isolate electrons in the RHRS. Table \ref{tab:ecut-sum} summarizes all of the cuts in a single table. The table includes both the variable name used by the Hall A Analyzer software and a definition of the variable based on the previous discussion in this chapter.

%\begin{table}
%\begin{center}
%\begin{tabular}{c|ccc}
%Description & Variable Name &  & Value \\ \hline
%Q$^2$ (GeV/$c)^2$ & PriKineR.Q2 & $<$ & 10 \\
%$\nu$ (GeV) & PriKineR.nu & $<$ & 10 \\
%dp & ExTgtCor\_R.dp & $>$ & -0.04 \\
%dp & ExTgtCor\_R.dp & $<$ & 0.06 \\
%Reaction Point in $z$ (m) & ReactPt\_R.z & $>$ & -0.15 \\
%Reaction Point in $z$ (m) & ReactPt\_R.z & $<$ & 0.17 \\
%Number of Tracks &  R.tr.n & = & 1 \\
%Number of Hits in VDC Plane U1 & R.vdc.u1.nhits & $>$ & 3 \\
%Number of Hits in VDC Plane U1 & R.vdc.u1.nhits & $<$ & 7 \\
%Number of Hits in VDC Plane U2 & R.vdc.u2.nhits & $>$ & 3 \\
%Number of Hits in VDC Plane U2 & R.vdc.u2.nhits & $<$ & 7 \\
%Number of Hits in VDC Plane V1 & R.vdc.v1.nhits & $>$ & 3 \\
%Number of Hits in VDC Plane V1 & R.vdc.v1.nhits & $<$ & 7 \\
%Number of Hits in VDC Plane V2 & R.vdc.v2.nhits & $>$ & 3 \\
%Number of Hits in VDC Plane V2 & R.vdc.v2.nhits & $<$ & 7 \\
%$\phi_{tg}$ (Radians) & ExTgtCor\_R.ph & $<$ & 0.025 \\
%$\phi_{tg}$ (Radians) & ExTgtCor\_R.ph & $>$ & -0.025 \\
%$\theta_{tg}$ (Radians) & ExTgtCor\_R.th & $<$ & 0.055 \\
%$\theta_{tg}$ (Radians) & ExTgtCor\_R.th & $>$ & -0.055 \\
%$x_{Bjorken}$ & PriKineRHe3.x\_bj & $<$ & 2.353 \\
%Preshower ADC Channel & R.ps.e & $>$ & 1\\
%Shower ADC Channel & R.sh.e & $>$ & 1\\
%Preshower and Shower ADC Channels & R.ps.e  + 2*R.sh.e & $>$ & 900 \\
%\v{C}erenkov ADC Channel & R.cer.asum\_c & $>$ & 150 \\
%\end{tabular}
%\caption[Summary of Electron Cuts]{Summary of Electron Cuts. This table each of the cuts used to select electrons in the RHRS.}
%\label{tab:ecut-sum}
%\end{center}
%\end{table}

\begin{table}
\begin{center}
\begin{tabular}{c|c}
Description & Cut Definition \\ \hline
Q$^2$ (GeV/$c)^2$ & PriKineR.Q2 $<$ 10 \\
$\nu$ (GeV) & PriKineR.nu $<$ 10 \\
dp & ExTgtCor\_R.dp $>$ -0.04 \\
dp & ExTgtCor\_R.dp $<$ 0.06 \\
Reaction Point in $z$ (m) & ReactPt\_R.z $>$ -0.15 \\
Reaction Point in $z$ (m) & ReactPt\_R.z $<$ 0.17 \\
Number of Tracks &  R.tr.n = 1 \\
Number of Hits in VDC Plane U1 & R.vdc.u1.nhits $>$ 3 \\
Number of Hits in VDC Plane U1 & R.vdc.u1.nhits $<$ 7 \\
Number of Hits in VDC Plane U2 & R.vdc.u2.nhits $>$ 3 \\
Number of Hits in VDC Plane U2 & R.vdc.u2.nhits $<$ 7 \\
Number of Hits in VDC Plane V1 & R.vdc.v1.nhits $>$ 3 \\
Number of Hits in VDC Plane V1 & R.vdc.v1.nhits $<$ 7 \\
Number of Hits in VDC Plane V2 & R.vdc.v2.nhits $>$ 3 \\
Number of Hits in VDC Plane V2 & R.vdc.v2.nhits $<$ 7 \\
$\phi_{tg}$ (Radians) & ExTgtCor\_R.ph $<$ 0.025 \\
$\phi_{tg}$ (Radians) & ExTgtCor\_R.ph $>$ -0.025 \\
$\theta_{tg}$ (Radians) & ExTgtCor\_R.th $<$ 0.055 \\
$\theta_{tg}$ (Radians) & ExTgtCor\_R.th $>$ -0.055 \\
$x_{Bjorken}$ & PriKineRHe3.x\_bj $<$ 2.353 \\
Preshower ADC Channel & R.ps.e $>$ 1\\
Shower ADC Channel & R.sh.e $>$ 1\\
Preshower and Shower ADC Channels & R.ps.e  + 2*R.sh.e $>$ 900 \\
\v{C}erenkov ADC Channel & R.cer.asum\_c $>$ 150 \\
\end{tabular}
\caption[Summary of Electron Cuts]{Summary of Electron Cuts. This table lists each of the cuts used to select electrons in the RHRS. The left column is a description of the variable being cut and the right column is the definition of the cut using the variable name as defined in the Hall A Analyzer software.}
\label{tab:ecut-sum}
\end{center}
\end{table}

% ^^^^^^^^^^^^^^^^^^^^^^^^^^^^^^^^^^^^^^^^^^^^^^^^^^^^^^^^^^^^^^^^^^^^^^^^^^^^^^^

\newpage

\large
\section {Neutron Identification} 
\label{ch3-neutronid}
\normalsize
% Neutron Identification
% vvvvvvvvvvvvvvvvvvvvvvvvvvvvvvvvvvvvvvvvvvvvvvvvvvvvvvvvvvvvvvvvvvvvvvvvvvvvvvvv

\subsection{Neutron Selection via Veto bars}
\label{veto-bars}
\begin{figure}
	\centering
	\includegraphics{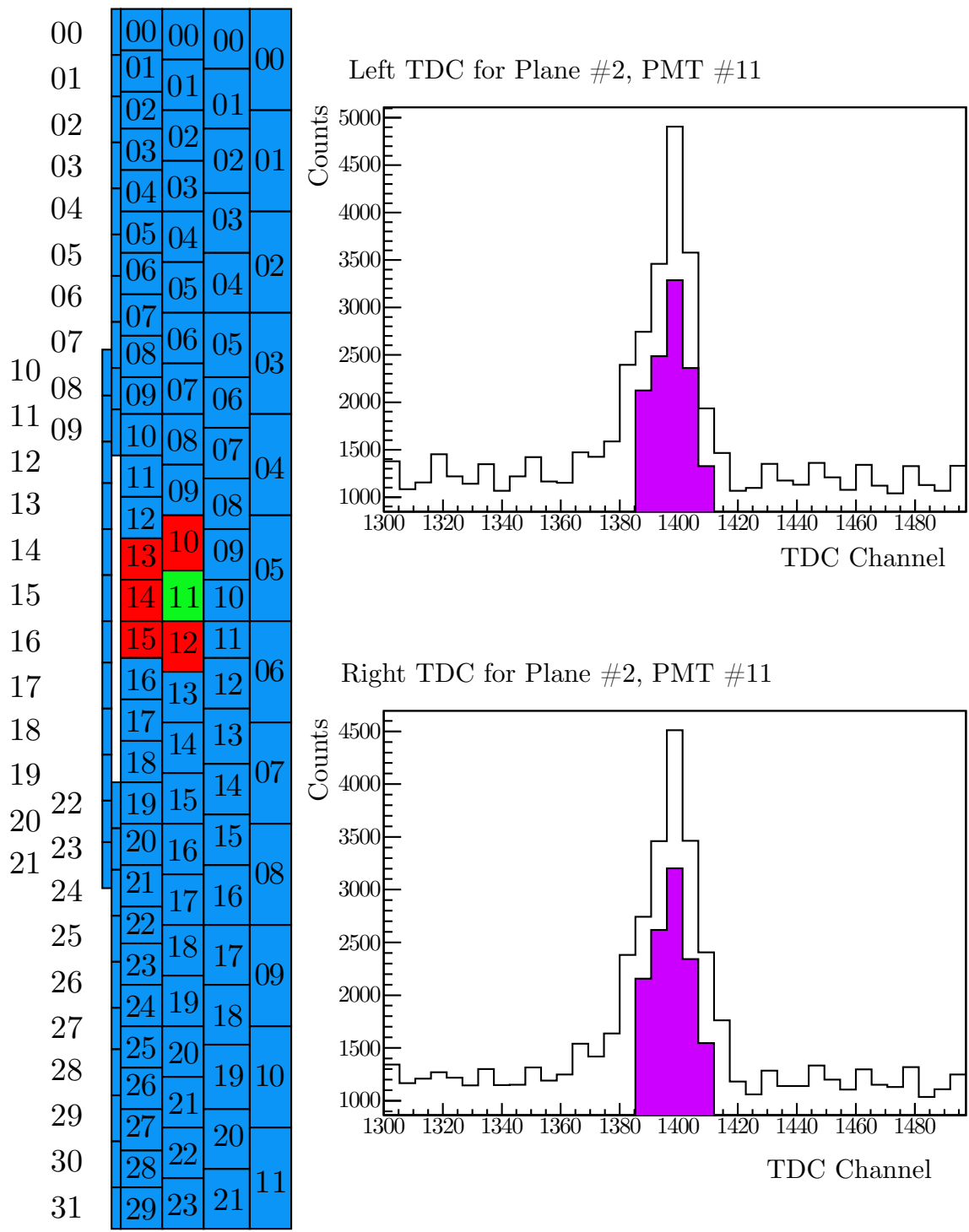}
	\caption [HAND Veto Example] {For each bar in HAND, the bars in front of, above, and below were identified as ``veto" bars used to isolate neutrons from protons. For example, Plane 2 Bar 11 uses Bars 13, 14, and 15 in the first plane and bars 10 and 12 in the second plane as veto bars. The larger, black peak shows the TDCs before the veto cuts are made, and the smaller purple peaks show the TDC after the veto cut is made. The larger peak is protons and neutrons, whereas the smaller purple peak is only neutrons after the veto cut has been made.}
	\label{hand-veto-example}
\end{figure}

As discussed in Section \ref{ch2-hand}, the scintillators that make up HAND cannot directly detect neutrons. However, they easily detect struck protons. Another complication is that the knocked-out protons from electron scattering, having approximately the same mass as neutrons, will reach HAND at approximately the same time. In order to differentiate between the two, a series of ``veto" bars was used. When a neutron enters HAND, it is not detected in a scintillator bar until it knocks out a proton in one of the bars. This means that if there is a signal located in one bar, but not in any of the bars in front of it, then the signal comes from a neutron. If instead, there is a signal in a bar and in the veto bars in front of it, then it is a proton. Although protons and neutrons come in the same timing peak, the TDCs of HAND were used to discriminate between protons an neutrons by excluding events appearing in the timing window of the veto bars that correspond to the timing of neutrons or protons. An example is shown in Figure \ref{hand-veto-example}.

\subsection{TDC Calibration}
\label{tdc-calibration}
In order for the veto cuts described in Section \ref{veto-bars} to work, the TDCs had to be calibrated. This was done using $^2$H runs where the timing peak consisted of protons. For each TDC channel, the calibration matrix was adjusted so that the mean TDC peak for each bar as aligned to TDC channel 1400. This resulted in all of the proton and neutron peaks being centered around channel 1400 as shown in Figure \ref{hand-veto-example}.

\subsection{Time of Flight}
\label{tof}
Through the use of veto bars, it is possible to separate protons from neutrons in the TDC timing peaks. However, there are a number of other background events that also need to be removed to select only neutrons. These events appear as a broad background in time and come from processes such as $^3$He($e,e'$), dark noise in the PMTs, and other sources. In order to separate them out, the time of flight (ToF) was used. 

\begin{figure}
	\centering
	\includegraphics{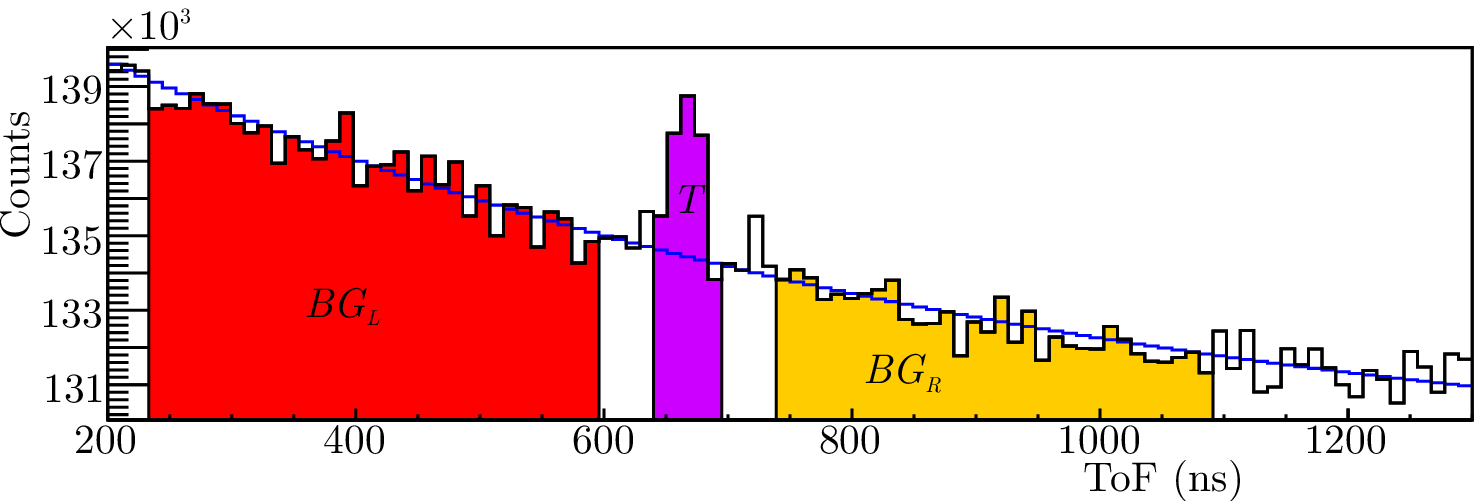}
	\includegraphics{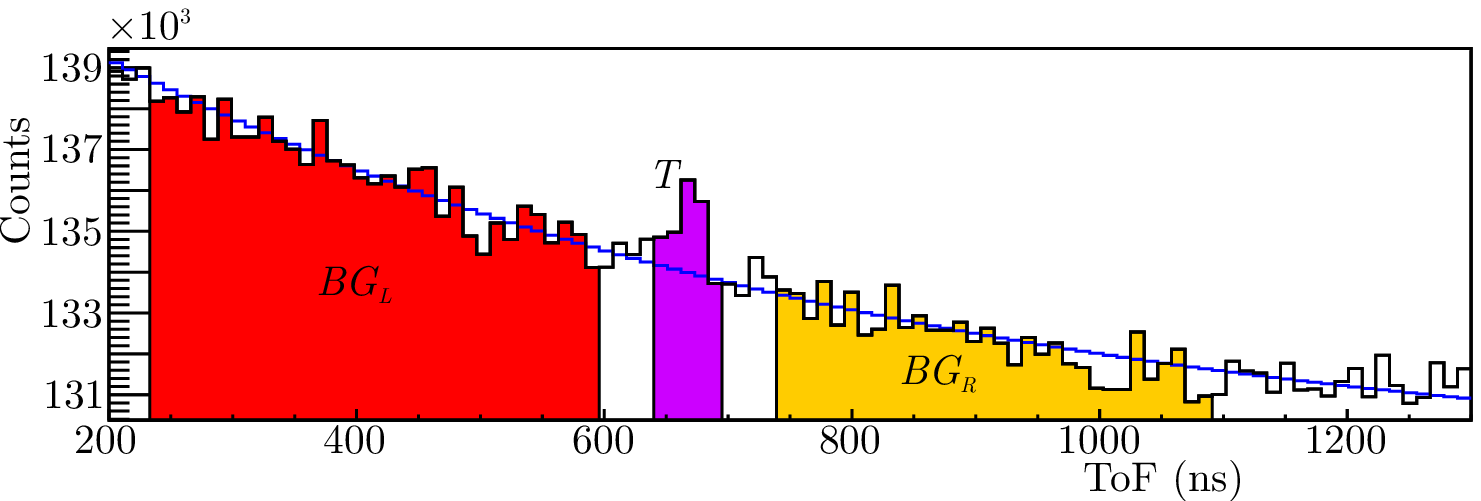}
	\caption [ToF for Q$^2$=0.127 (GeV/$c)^2$] {ToF for Q$^2$=0.127 (GeV/$c)^2$. The upper plot is the ToF for target spin-up events and the lower plot is the ToF for target spin-down events. $BG_{L}$, $BG_{R}$, and $T$ are used in the uncertainty analysis as described in Section \ref{ch3-error}. Events highlighted in purple above the blue fit line were considered ``good" events. There is also a small $\gamma$ peak to the right of the main neutron peak.}
	\label{q2-01-tof}
\end{figure}

\begin{figure}
	\centering
	\includegraphics{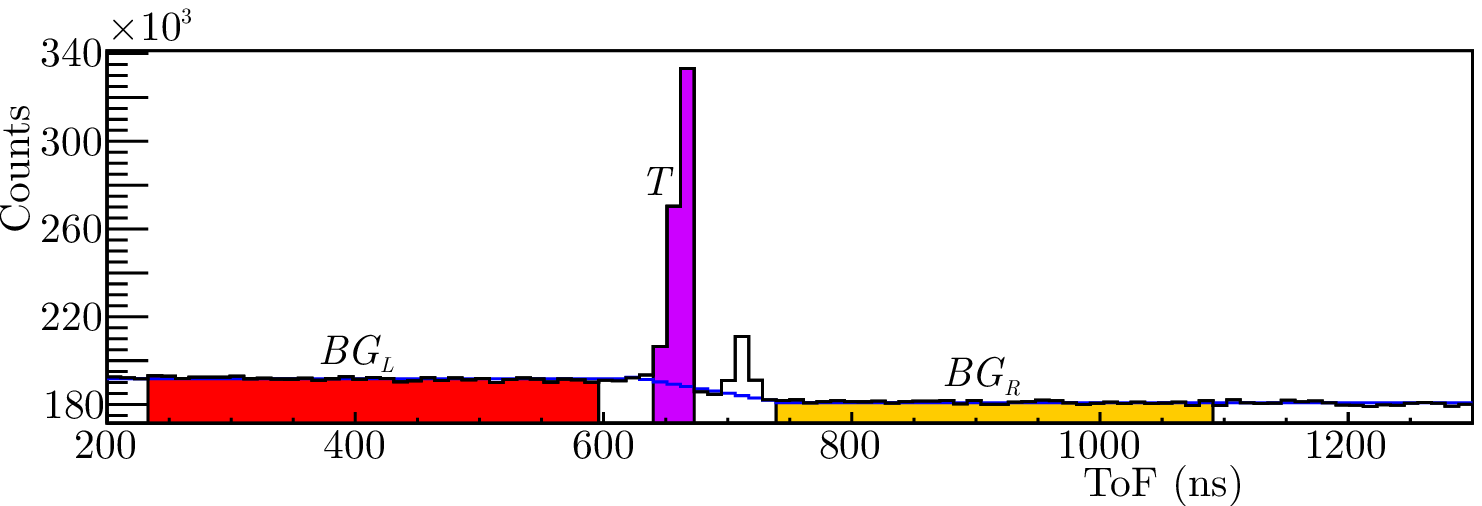}
	\includegraphics{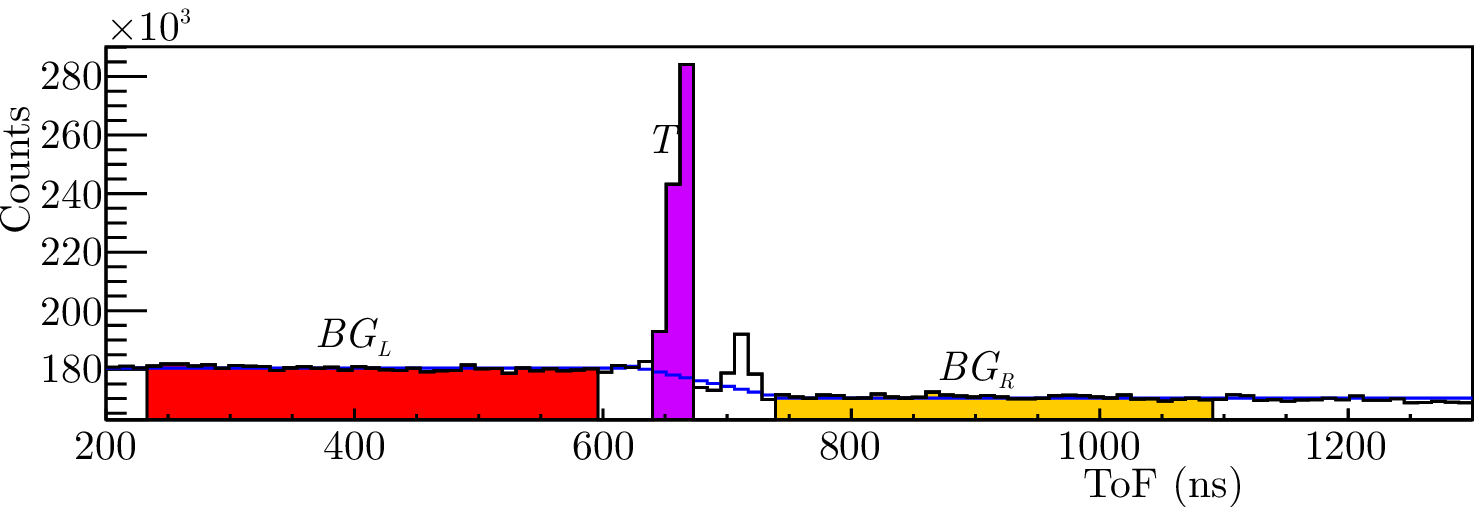}
	\caption [ToF for Q$^2$=0.456 (GeV/$c)^2$] {ToF for Q$^2$=0.456 (GeV/$c)^2$. The upper plot is the ToF for target spin-up events and the lower plot is the ToF for target spin-down events. $BG_{L}$, $BG_{R}$, and T are used in the uncertainty analysis as described in Section \ref{ch3-error}. Events highlighted in purple above the blue fit line were considered ``good" events. There is also a small $\gamma$ peak to the right of the main neutron peak.}
	\label{q2-05-tof}
\end{figure}

\begin{figure}
	\centering
	\includegraphics{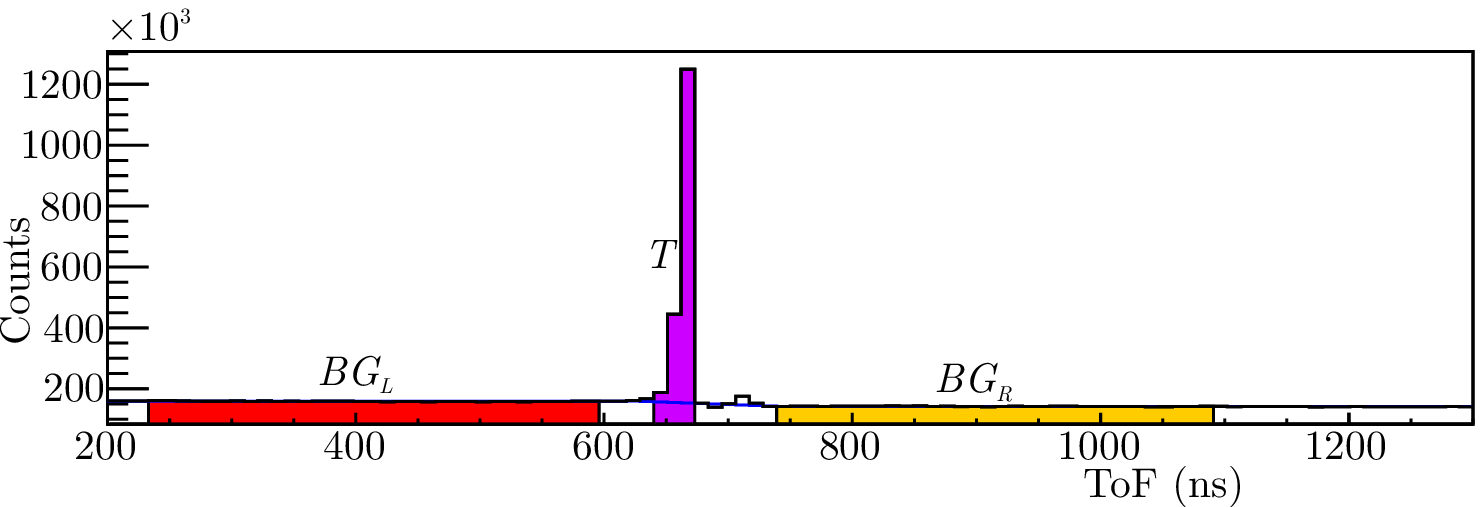}
	\includegraphics{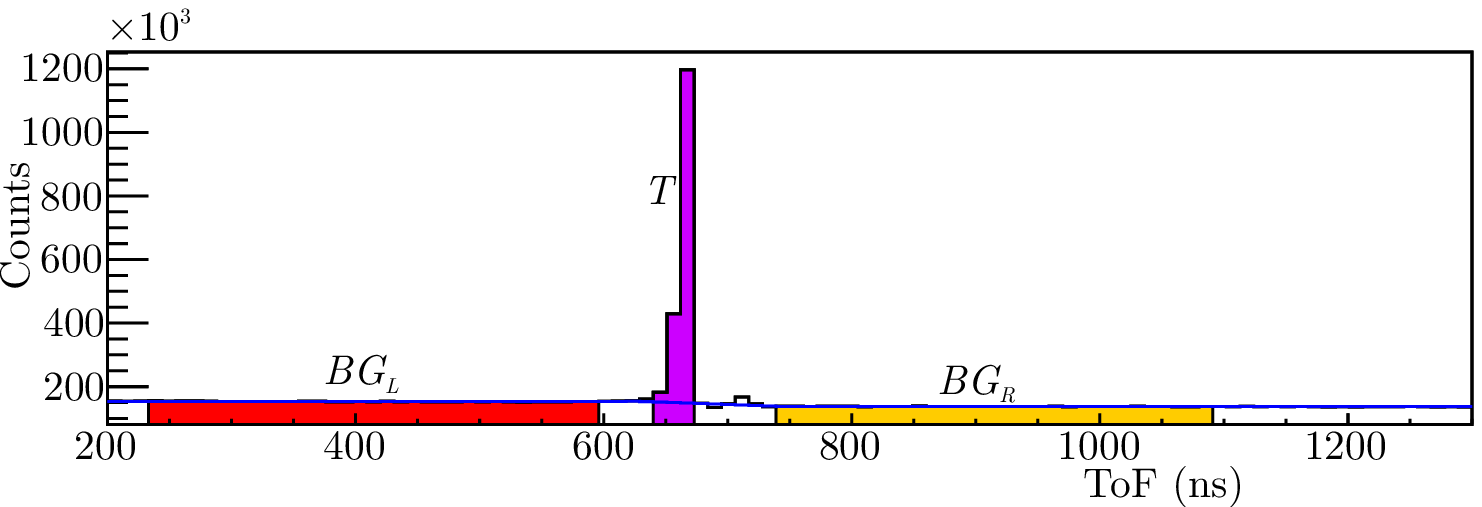}
	\caption [ToF for Q$^2$=0.953 (GeV/$c)^2$] {ToF for Q$^2$=0.953 (GeV/$c)^2$. The upper plot is the ToF for target spin-up events and the lower plot is the ToF for target spin-down events. $BG_{L}$, $BG_{R}$, and T are used in the uncertainty analysis as described in Section \ref{ch3-error}. Events highlighted in purple above the blue fit line were considered ``good" events. There is also a small $\gamma$ peak to the right of the main neutron peak.}
	\label{q2-1-tof}
\end{figure}

In the case of Q$^2$=0.127 (GeV/$c)^2$, this was accomplished by a simple exponential fit on the background. This allowed the neutron peak to be isolated and the number of events in it to be counted. This is shown in Figure \ref{q2-01-tof}. The higher Q$^2$ points were slightly more complicated. The background for those points was constant, however, there is a difference in magnitude on either side of the neutron peak. In order to account for this, a linear fit was made under the neutron peak to bridge the gap between the constant background on either side. This is shown in Figures \ref{q2-05-tof} and \ref{q2-1-tof} for Q$^2$ = 0.505 (GeV/$c)^2$ and 0.953 (GeV/$c)^2$, respectively. The uncertainty due to background subtraction is discussed in Section \ref{ch3-error}.


% ^^^^^^^^^^^^^^^^^^^^^^^^^^^^^^^^^^^^^^^^^^^^^^^^^^^^^^^^^^^^^^^^^^^^^^^^^^^^^^^

					% Include the Chapter 4 text (chapter_4.tex)
	% vvvvvvvvvvvvvvvvvvvvvvvvvvvvvvvvvvvvvvvvvvvvvvvvvvvvvvvvvvvvvvvvvvvvvvvvvvvvvvvv
% Chapter 4 (chapter_4.tex)
%
% Chapter 4 of Elena Long's Ph.D. Dissertation
%
% To be completed: March, 2012
%
% ^^^^^^^^^^^^^^^^^^^^^^^^^^^^^^^^^^^^^^^^^^^^^^^^^^^^^^^^^^^^^^^^^^^^^^^^^^^^^^^^

\chapter{Dilutions and Uncertainties}	% Chapter Title
\label{dilution-uncertainties}					% Chapter Label
\normalsize					% Return to Normal font size

\large
\section {Polarization of Target and Beam} 
\label{ch3-polarization}
\normalsize
% Polarization of Target and Beam
% vvvvvvvvvvvvvvvvvvvvvvvvvvvvvvvvvvvvvvvvvvvvvvvvvvvvvvvvvvvvvvvvvvvvvvvvvvvvvvvv

\subsection {Target Polarization}
\label{target-polarization}

\begin{figure}
	\centering
	\includegraphics[width=6 in]{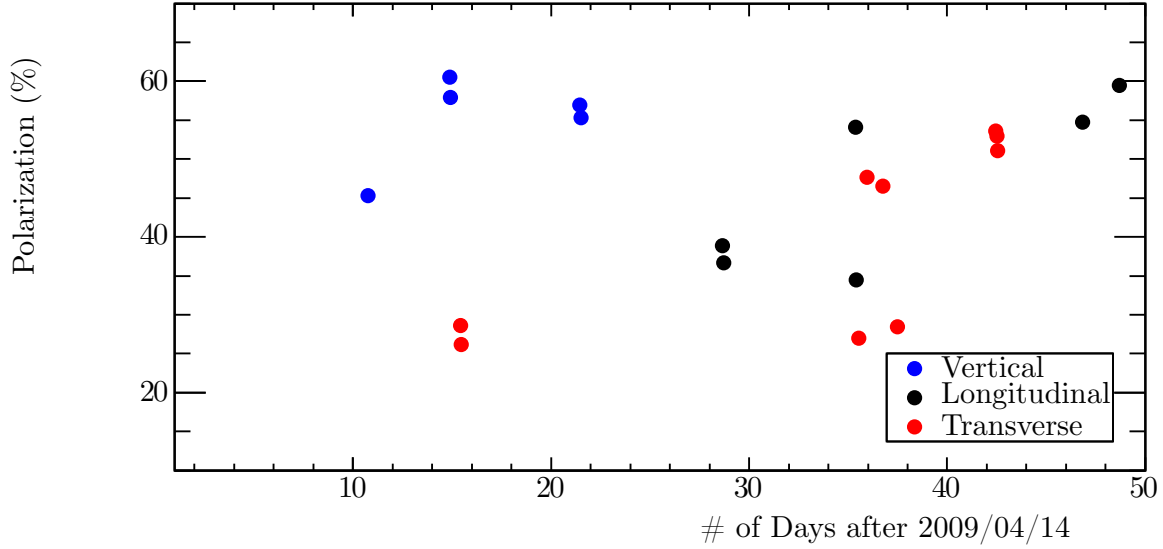}
	\caption [EPR Measurements] {EPR Measurements. This plot shows the target polarization measured by EPR that occurred for the $A_y^0$, $A_T$ and $A_L$ measurements.}
	\label{epr}
\end{figure}

The target polarization was measured by two independent methods: nuclear magnetic resonance (NMR) at the target chamber and electron paramagnetic resonance (EPR) in the pumping chamber, as described in Section \ref{nmr-epr}. For the $A_y^0$ experiment, there were five EPR measurements taken and NMR measurements were taken every 20 minutes after the spin was flipped. For the $A_T$ measurement, there were nine EPR measurements and for the $A_L$ measurement there were six EPR measurements. NMR measurements were taken at intervals of approximately four hours for both $A_T$ and $A_L$. Each of the EPR measurements are shown in Figure \ref{epr}. The EPR measurements allow for a measurement of a calibration constant that can be used with the NMR measurements to find the target polarization. This is necessary since non-invasive NMR only measures the relative polarization, whereas the invasive EPR measures the absolute polarization.

\begin{figure}
	\centering
	\includegraphics{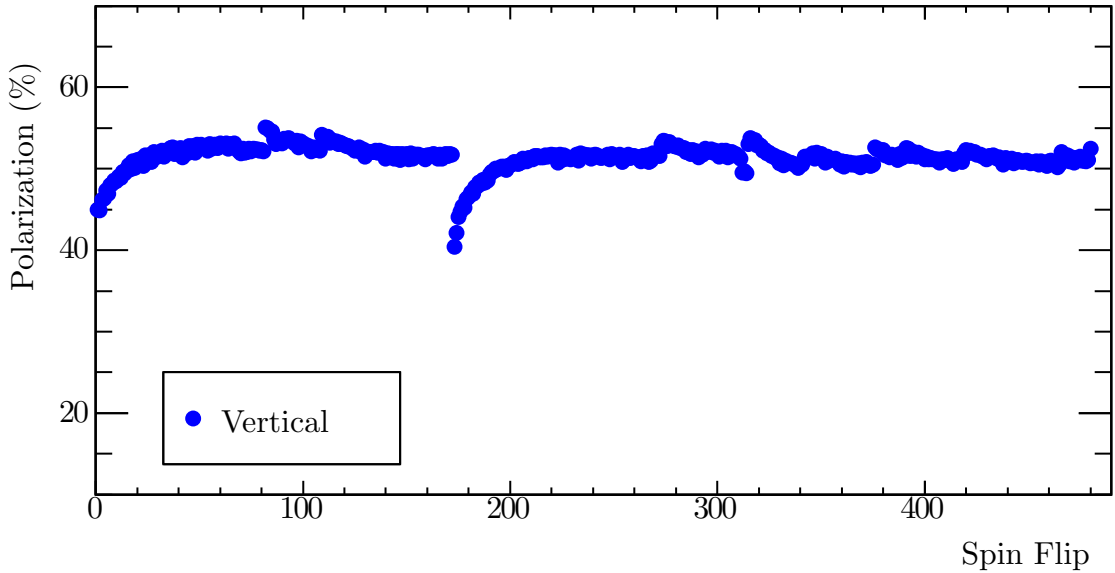}
	\caption [Target Polarization Measurements for $A_y^0$] {This plot shows the target polarization measurements that occurred for the $A_y^0$ experiment. These are NMR measurements, taken approximately every four hours, calibrated against EPR measurements.}
	\label{tgt-pol-ay}
\end{figure}

\begin{figure}
	\centering
	\includegraphics{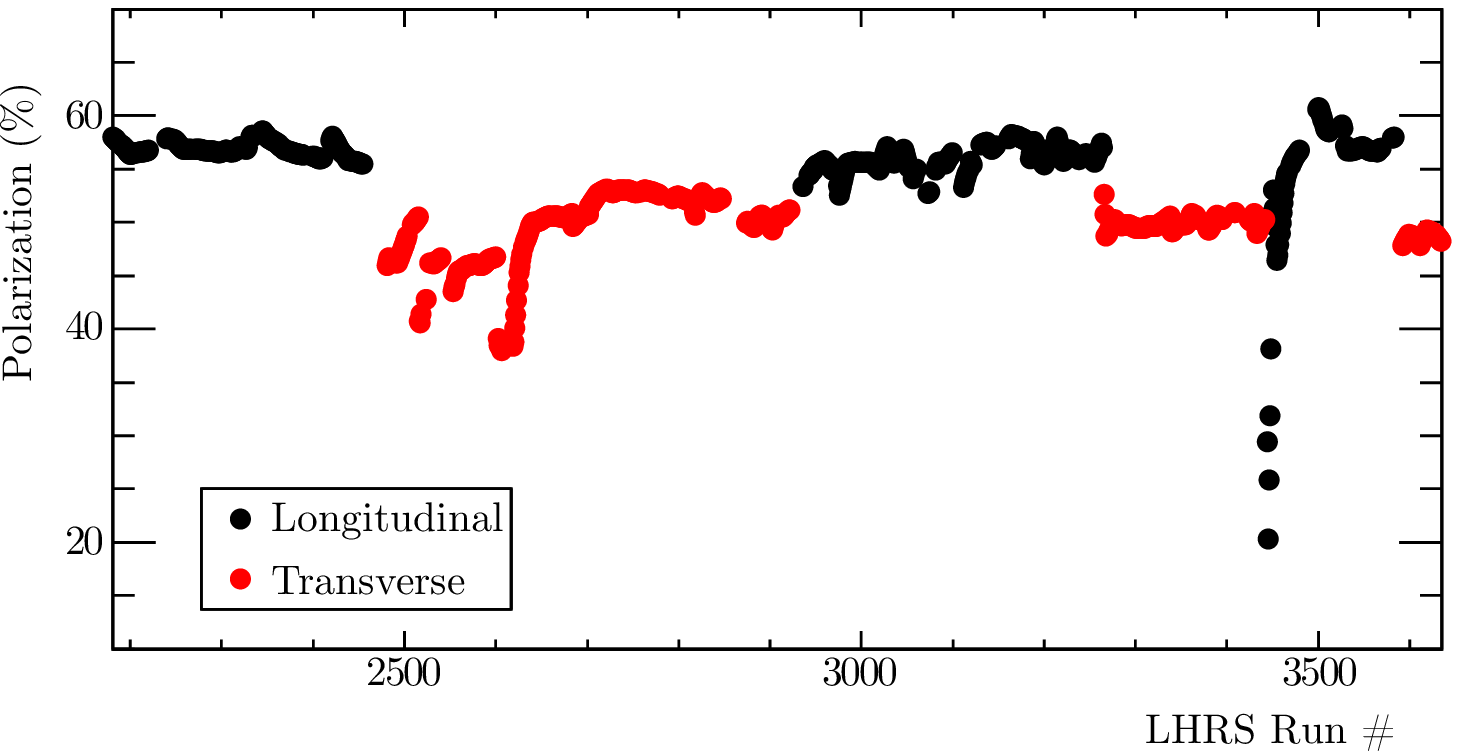}
	\caption [Target Polarization Measurements for $A_T$ and $A_L$] {This plot shows the target polarization measurements that occurred for the $A_T$ and $A_L$ experiments. These are NMR measurements, taken approximately every four hours, calibrated against EPR measurements.}
	\label{tgt-pol}
\end{figure}

The description of the polarization used with a correction factor is
\begin{equation}
	P_{tc} = \frac{d_{tc}}{d_{tc}+\Gamma_{tc}}P_p,
\end{equation}
where $P_{tc}$ is the polarization in the target chamber, $d_{tc}$ is the reduced diffusion constant, $\Gamma_{tc}$ is the depolarization rate in the target chamber, and $P_p$ is the polarization in the pumping chamber. The reduced diffusion constant for the target chamber is defined as
\begin{equation}
	d_{tc} = \frac{A_{tt}D_{tc}}{V_{tc}L_{tt}} K,
\end{equation}
where
\begin{equation}
	D_{tc} = D_{T_0} \left( \frac{T_{tc}}{T_0} \right) ^{(m-1)} \frac{n_0}{n_{tc}},
\end{equation}
\begin{equation}
	K = \frac{(2-m)(t-1)}{t^{2-m}-1},
\end{equation}
\begin{equation}
	t = T_{pc}/T_{tc}.
\end{equation}
Here $A_{tt}$ is the transfer tube cross section, $V_{tc}$ is the volume of the target chamber, $T_0$ is the equilibrium temperature, $T_tc$ is the target chamber temperature, $T_pc$ is the pumping chamber temperature,  $n_0$ is the equilibrium density, $n_tc$ is the target chamber density, $L_{tt}$ is the length of the transfer tube, and $D_{tc}$ is the diffusion constant. The depolarization rate in the target chamber is defined as
\begin{equation}
	\Gamma_{tc} = \Gamma^{\mathrm{He}} + \Gamma^{\mathrm{Wall}}_{tc} + \Gamma^{\mathrm{Beam}} + \Gamma^{\mathrm{AFP}} + \Gamma^{\Delta B},  
\end{equation}
where $\Gamma^{\mathrm{He}}$ is the nuclear dipole interaction, $\Gamma^{\mathrm{Wall}}$ is the relaxation of polarization due to the glass walls, $\Gamma^{\mathrm{Beam}}$ is the depolarization due to the beam, $\Gamma^{\mathrm{AFP}}$ is the loss from the adiabatic fast passage, and $\Gamma^{\Delta B}$ is the relaxation from the magnetic field gradient. Taking all of these into account, the beam polarization was measured and can be seen in Figures \ref{tgt-pol-ay} and \ref{tgt-pol}. This work was done by Yawei Zhang \cite{Yawei:2013rutgers}. The polarization dilution factors used in this experiment are shown in Table \ref{tab:targ-pol-values} and the systematic error budget is shown in Table \ref{tab:targ-pol-err-eed}.

\begin{table}
\begin{center}
\begin{tabular}{c|c|c}
\multicolumn{2}{c|}{Items} & Rel. Pol. Error \\ \hline
 & K-$^3$He EPR $\kappa_0$ & 2.8$\%$ \\
 Pumping Chamber & Pumping Chamber Density & 1.5$\%$ \\
  & EPR Signal Fit & 0.6$\%$ \\
  & NMR Signal Fit & 0.6$\%$ \\ \hline
 & Diffusion Rate & 1.2$\%$ \\
 Target Chamber & Target Chamber Intrinsic Life-Time & 1.4$\%$ \\
  & Beam Depolarization & 2.6$\%$ \\
   & Spin Flip Loss & 0.1$\%$ \\ \hline
\multicolumn{2}{c|}{Total} & 4.6$\%$ \\
\end{tabular}
\caption[Target Polarization Systematic Uncertainty Budget]{This table shows the uncertainties involved in obtaining the target polarization for the $A_T$ and $A_L$ experiments.}
\label{tab:targ-pol-err-eed}
\end{center}
\end{table}

\begin{table}
\begin{center}
\begin{tabular}{c|c|c|c}
Experiment & Tgt. Pol. ($\%$) & Stat. Err. (Abs. $\%$) & Sys. Err. (Abs. $\%$)\\ \hline
$A_y^0$ & 51.4 & 0.4 & 2.8 \\
$A_T$ & 49.6 & 0.4 & 2.3 \\
$A_L$ & 54.7 & 0.4 & 2.5 \\
\end{tabular}
\caption[Target Polarization Dilution]{This table shows the target polarization and uncertainty that was used as a dilution factor for the $A_y^0$, $A_T$, and $A_L$ experiments.}
\label{tab:targ-pol-values}
\end{center}
\end{table}

\newpage

\subsection {Beam Polarization}
\label{beam-polarization}

\begin{figure}
	\centering
	\includegraphics[width=6 in]{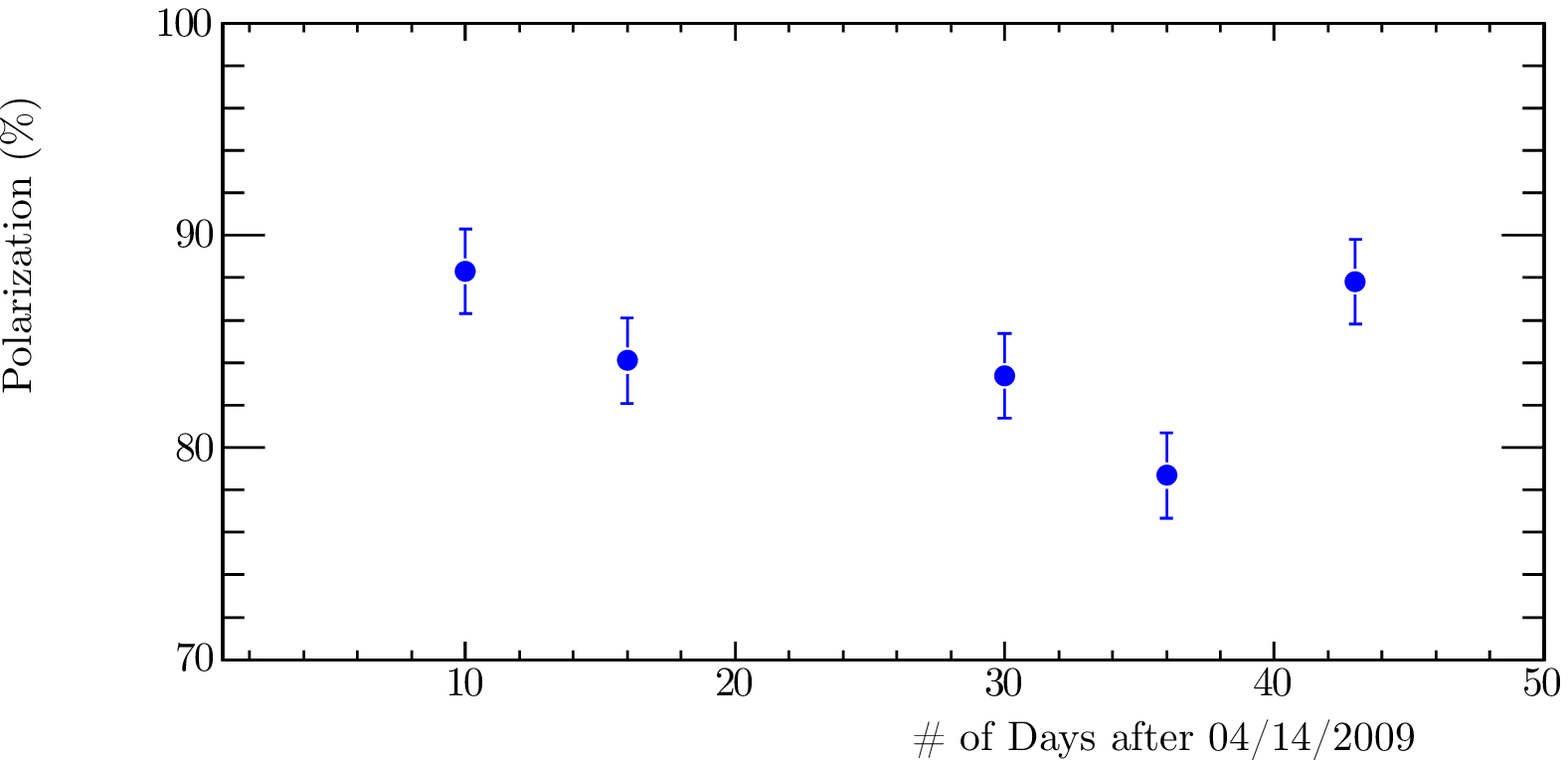}
	\caption [Beam Polarization Measurements] {This plot shows the M$\o$ller measurements taken to determine the beam polarization for the $A_y^0$, $A_T$, and $A_L$ experiments.}
	\label{beam-pol}
\end{figure}

The beam polarization was measured with a M$\o$ller polarimeter, which is described in Section \ref{moller-polarimeter}. The M$\o$ller measurements are invasive and require beam time separate from production running. The polarimeter utilizes the fact that the M$\o$ller scattering ($\vec{e^-} + \vec{e^-} \rightarrow e^- + e^-$) cross section depends on the beam and target polarizations. A thin, magnetically saturated ferromagnetic foil is used as a target. The saturation leads to an electron polarization of approximately 8$\%$ in the target. The foil can be rotated to $\pm20^{\circ}$ with respect to the beam, which causes the effective target polarization to be $P_{\mathrm{target}} = P_{\mathrm{foil}}\cdot \cos \theta_{\mathrm{target}}$. Since the target polarization is known, a beam-target double-spin asymmetry measurement is taken that allows the beam polarization to be determined by
\begin{equation}
P^{\mathrm{beam}}_Z = \frac{N_+ - N_-} {N_+ + N_-} \cdot \frac{1}{P^{\mathrm{foil}}}\cdot \cos \theta_{\mathrm{target}} \cdot \left< A_{ZZ} \right>,
\end{equation}
where $\left< A_{ZZ} \right>$ is the average analyzing power \cite{kelleher:1997cwm}. $\left< A_{ZZ} \right>$ is dependent only on the center-of-mass-angle scattering and was determined via a Monte Carlo calculation of the spectrometer acceptance. Five M$\o$ller measurements were taken over the course of the entire run period, which can be seen in Figure \ref{beam-pol}. Individually, each run has a statistical uncertainty of $0.2 \%$ and a systematic uncertainty of $2.0 \%$. The average polarization was $84.5 \pm 3.9 \%$, which is used as a dilution factor for the $A_T$ and $A_L$ double-spin asymmetries. Since $A_y^0$ is a target single-spin asymmetry, the beam was treated as being unpolarized.

% ^^^^^^^^^^^^^^^^^^^^^^^^^^^^^^^^^^^^^^^^^^^^^^^^^^^^^^^^^^^^^^^^^^^^^^^^^^^^^^^
\newpage

\large
\section {Proton Contamination} 
\label{ch3-protoncontamination}
\normalsize
% Proton Contamination
% vvvvvvvvvvvvvvvvvvvvvvvvvvvvvvvvvvvvvvvvvvvvvvvvvvvvvvvvvvvvvvvvvvvvvvvvvvvvvvvv
Since neutrons and protons have roughly the same mass, if an electron knocks out either one, it can be difficult to tell them apart based only on the timing information. Although veto cuts were applied to identify neutrons, as discussed in Section \ref{ch3-neutronid}, protons will occasionally make it past those cuts. This can be due to a number of reasons. The largest contributor is charge-conversion, where the proton knocks out a neutron along the flight path toward HAND, which is then detected by HAND. The significance of this problem increases with Q$^2$ as it becomes more likely that protons make it to the detector. At the highest Q$^2$ points, a lead wall was placed in front of the neutron detector to reduce the number of protons reaching the detector; however, it also acts as a converter for protons to knockout neutrons.

\subsection {Nucleons Along the q-vector}
\label{qvector-nucleons}

In order to estimate the number of protons that make it to HAND, we first need to get an estimate of the number of protons being emitted along the q-vector. For hydrogen data, this is simply the number of particles detected. For $^3$He, however, it becomes a bit more complicated. If it is assumed that the $^3$He nucleus is made up of two free protons and one free neutron, we can use the Rosenbluth equation \cite{Rosenbluth:1950yq} to estimate the cross section:

\begin{equation}
\left(\frac{d\sigma}{d\Omega}\right) = \left(\frac{d\sigma}{d\Omega}\right) _{\mbox{Mott}} \left[ \frac{G^{2}_{E}(Q^2) + \tau G^{2}_{M}(Q^2)}{1+\tau} + 2 \tau G^2_M(Q^2)\tan^2\frac{\theta}{2} \right].
\end{equation}

For the above equation, $\theta_{e'}$ is the electron scattering angle, $G_E$ is the nucleon electric form factor \cite{Sachs:1962zzc}, $G_M$ is the nucleon magnetic form factor \cite{Sachs:1962zzc},
 \begin{equation}
\left(\frac{d\sigma}{d\Omega}\right)_{\mbox{Mott}} = \left( \frac{E'}{E} \right)\left( \frac{4 Z^2 \alpha^2(\hbar c)^2 E'^2}{|\vec{q}c|^4} \cos^2 \frac{\theta_{e'}}{2}\right) \mbox{ in the limit $\beta\rightarrow1$} \cite{Mott01011930},
\end{equation}
\begin{equation}
\tau = \left(\frac{Q^2}{4M^2c^2}\right),
\end{equation}
\begin{equation}
E' = \left(\frac{E}{1 + \frac{E}{Mc^2} (1-\cos\theta)}\right),
\end{equation}
$E$ is the incoming electron energy, $E'$ is the outgoing electron energy, $M$ is the nucleon mass, $Z$ = 1, $|Q|^2$ = $\nu^2 - |\vec{q}|^2$ is the four-momentum transfer squared, $\beta$ is the speed of the electron divided by the speed of light, and $\alpha$ is the fine structure constant. The Kelly parametrization \cite{Kelly:2004hm} was used to find the values of the form factors at various Q$^2$. Using the calculated cross sections, the ratio of protons to neutrons at any value of Q$^2$ can be calculated by taking the ratio 
\begin{equation}
r_{p:n} = \left(\frac{2\left(\frac{d\sigma}{d\Omega}\right)_p}{\left(\frac{d\sigma}{d\Omega}\right)_n}\right).
\end{equation}
This ratio was calculated for Q$^2$=0.127 (GeV/$c)^2$, 0.456 (GeV/$c)^2$, 0.505 (GeV/$c)^2$, and 0.953 (GeV/$c)^2$ and can be found in Table \ref{tab:cross-sections}.

\begin{table}
\begin{center}
\begin{tabular}{c|c|c|c|c|c}
Q$^2$ (GeV$^2$/c$^2$) & E (GeV) & $\theta_{e'}$ ($^\circ$) & $\frac{d\sigma}{d\Omega}|_{p}$ (m$^2$) & $ \frac{d\sigma}{d\Omega}|_{n}$ (m$^2$) & $r_{p:n}$\\ \hline
0.127 & 1.245 & 17.0 & 4.060$\times 10^{-34}$ & 4.302$\times 10^{-35}$ & 18.87:1\\
0.456 & 2.425 & 17.0 & 4.050$\times 10^{-35}$ & 1.066$\times 10^{-35}$ & 7.599:1\\
0.505 & 2.425 & 18.0 & 2.835$\times 10^{-35}$ & 7.944$\times 10^{-36}$ & 7.138:1\\
0.953 & 3.606 & 17.0 & 6.299$\times 10^{-36}$ & 2.363$\times 10^{-36}$ & 5.331:1\\
\end{tabular}
\caption[Rosenbluth Cross Sections for Nucleons.]{Rosenbluth Cross Sections for Nucleons. This table shows the Rosenbluth cross sections for each of the nucleons at various Q$^2$, electron energies (E), and scattered electron angles ($\theta_{e'}$). It also shows the estimated ratio of protons:neutrons if it is assumed that $^3$He consists of three free nucleons.}
\label{tab:cross-sections}
\end{center}
\end{table}

If every detected scattered electron came from a nucleon, then we can calculate how many of each particle was sent towards HAND along the q-vector by taking those scattered electrons that make it past the acceptance cuts (see Sections \ref{ch3-electronid} and \ref{ch3-neutronid}) and multiplying it by the ratio $r_{p:n}$ such that 

\begin{equation}
N_e = N_p + N_n,
\end{equation}
\begin{equation}
N_p = r_{p:n}N_n,
\end{equation}
\begin{equation}
N_e = (1+r_{p:n})N_n,
\end{equation}
\begin{equation}
N_n = \frac{N_e}{r_{p:n}+1},
\end{equation}
where $N_n$ is the number of knocked-out neutrons, $N_p$ is the number of knocked-out protons, and $N_e$ is the number of scattered electrons. Results of this with the data taken are found in Table \ref{tab:qvector-nucleons}. In general, as Q$^2$ increases, the ratio of protons to neutron decreases. This is expected because the $G_E$ contribution of the cross section of the protons drops off at higher Q$^2$, whereas the for neutrons it levels out \cite{Kelly:2004hm}.

\begin{table}
\begin{center}
\begin{tabular}{c|c|c|c|c|c}
Experiment & Q$^2$ (GeV$^2$/c$^2$) & $N_e$ & $r_{p:n}$ & $N_p$ & $N_n$ \\ \hline
		&	0.127	&	35,496,060	&	18.87:1	& $	3.371\times 10^{7}$ & $	1.786\times 10^{6}$ \\
$A_y^0$	&	0.456	&	52,758,650	&	7.599:1	& $	4.662\times 10^{7}$ & $	6.135\times 10^{6}$ \\
		&	0.953	&	55,623,240	&	5.331:1	& $	4.684\times 10^{7}$ & $	8.786\times 10^{6}$ \\ \hline
$A_T$	&	0.505	&	51,550,460	&	7.138:1	& $	4.522\times 10^{7}$ & $	6.335\times 10^{6}$ \\
		&	0.953	&	13,416,160	&	5.331:1	& $	1.130\times 10^{7}$ & $	2.119\times 10^{6}$ \\ \hline
$A_L$	&	0.505	&	22,130,450	&	7.138:1	& $	1.941\times 10^{7}$ & $	2.719\times 10^{6}$ \\
		&	0.953	&	10,910,390	&	5.331:1	& $	0.9187\times 10^{7}$ & $	1.723\times 10^{6}$ \\
		
% & 0.1 & $35,496,060$ & 22.91:1 & 3.401$\times 10^{7}$ & 1.485$\times 10^{6}$\\
% $A_y^0$ & 0.5 & $52,758,650$& 7.196:1 & 4.632$\times 10^{7}$ & 6.437$\times 10^{6}$ \\
% & 1.0 & $55,623,240$ & 5.252:1 & 4.673$\times 10^{7}$ & 8.897$\times 10^{6}$ \\ \hline
% $A_T$ & 0.5 & $???$& 7.196:1 & ??? & ??? \\
% & 1.0 & $12,416,160$ & 5.252:1 & 1.127$\times 10^{7}$ & 2.146$\times 10^{6}$ \\  \hline
% $A_L$ & 0.5 & $22,130,450$& 7.196:1 & 1.943$\times 10^{7}$ & 2.700$\times 10^{6}$ \\
% & 1.0 & $ 10,910,300 $ & 5.252:1 & 9.165$\times 10^{6}$ & 1.745$\times 10^{6}$ \\ 
\end{tabular}
\caption[Estimated Number of Nucleons Along q-vector.]{This table shows the estimated number of protons and neutrons that were scattered along the q-vector towards HAND for each of the kinematic settings used in this experiment.}
\label{tab:qvector-nucleons}
\end{center}
\end{table}

\subsection {Protons Detected by HAND}
\label{handprotons}

From Section \ref{qvector-nucleons}, we know how many protons and neutrons were headed towards HAND. Protons were scattered towards HAND large part by the scattering of the knocked-out protons on the target glass windows, from the protons in the $^3$He nuclei, the plastic around the target enclosure, the air between the target and HAND, and the lead wall when it was installed. However, only a fraction of these were actually detected. In order to calculate the proton dilution in $^3$He($e,e'n$), a calibration was done using $^1$H. Using the hydrogen data, all of the particles detected in HAND are protons, so it can be used to find how many protons make it to HAND, and how many are converted into neutrons along the way.

Three values are necessary to calculate how many protons are diluting the neutron data: the number of protons that make it past the neutron cuts from the hydrogen data ($P_n$), the total number of protons headed along the q-vector from the hydrogen data ($T_P$), the charge accumulation of the hydrogen data ($C_P$), the estimated number of protons along the q-vector for $^3$He data ($N_p$), and the charge accumulation of the $^3$He data ($C_{^3\mathrm{He}}$). For any given Q$^2$, the number of protons misidentified as neutrons is defined as
\begin{equation}
E_p = \frac{P_n}{T_P} \cdot N_p \cdot \frac{C_P}{C_{^3\mathrm{He}}}.
\end{equation}
From this, we can find the percentage of misidentified protons ($\% P$) by
\begin{equation}
\%P = \frac{E_p}{E_{n}}\cdot100\% ,
\end{equation}
where $E_n$ is the number of $^3$He scattered events that are identified as neutrons using the cuts described in Section \ref{ch3-neutronid}. The calculated percentage of protons and neutrons is found in Table \ref{tab:proton-dilution-factors}, where $\% N = 100 - \% P$. The percentage of neutrons in the $^3$He data is used as the proton dilution factor.

\begin{table}
\begin{center}
\begin{tabular}{c|c|c|c|c}
Experiment	&	Q$^2$ (GeV$^2$/$c^2$) 	&	$\%P$	&	$\%N$	&	Uncertainty ($\%$)	\\ \hline
	&	0.127	&	6.963	&	93.04	&	0.19	\\
$A_y^0$	&	0.456	&	7.252	&	92.75	&	0.06	\\
	&	0.953	&	0.9411	&	99.06	&	0.01	\\ \hline
$A_T$	&	0.505	&	4.140	&	95.86	&	0.03	\\
	&	0.953	&	2.158	&	97.84	&	0.02	\\ \hline
$A_L$	&	0.505	&	9.024	&	90.98	&	0.08	\\
	&	0.953	&	2.504	&	97.50	&	0.03	\\% & 0.1 & 1.175 & 98.83\\
% $A_y^0$& 0.5 & 7.205 & 92.80 \\
% & 1.0 & 0.9388 & 99.06 \\ \hline
% $A_T$ & 0.5 & ??? & ??? \\
% & 1.0 & 4.369 & 95.63 \\ \hline
% $A_L$ & 0.5 & 18.34 & 81.66 \\
% & 1.0 & 5.072 & 94.93 \\ 
\end{tabular}
\caption[Proton Contamination.]{This table shows the dilution factor of protons for all of the asymmetry measurements taken.}
\label{tab:proton-dilution-factors}
\end{center}
\end{table}

% In order to tag protons, three different cuts were made. The first is only a cut on the time of flight peak with no cuts on the individual TDCs of HAND. For the hydrogen data, this is referred to as $P_{np}$ and for the $^3$He data as $E_{np}$. The second is a cut identical to that used to find neutrons (see Section \ref{ch3-neutronid}) which is referred to as $P_n$ for the hydrogen data and $E_n$ for the $^3$He data. The final cut is a reverse of the neutron veto cuts. It requires an event to be fired in one of the previous bars so that a basic track is identified. This cut is referred to as $P_p$ for hydrogen and $E_p$ for $^3$He.

% ^^^^^^^^^^^^^^^^^^^^^^^^^^^^^^^^^^^^^^^^^^^^^^^^^^^^^^^^^^^^^^^^^^^^^^^^^^^^^^^

\newpage

\large
\section {Nitrogen Contamination} 
\label{ch3-nitrogen-contamination}
\normalsize
% Nitrogen Contamination
% vvvvvvvvvvvvvvvvvvvvvvvvvvvvvvvvvvvvvvvvvvvvvvvvvvvvvvvvvvvvvvvvvvvvvvvvvvvvvvvv
The $^3$He target was polarized due to spin-exchanges processes between Rb, K, and $^3$He, as discussed in Section \ref{ch2-target}. Unfortunately, the excited Rb and K atoms will also give off photons that can depolarize the $^3$He within minutes. In order to combat this effect, a small amount of nitrogen was added to the target cell to absorb these photons. Contamination due to events scattered from this N$_2$ must be taken into account. 

\begin{figure}
	\centering
	\includegraphics[width=15cm]{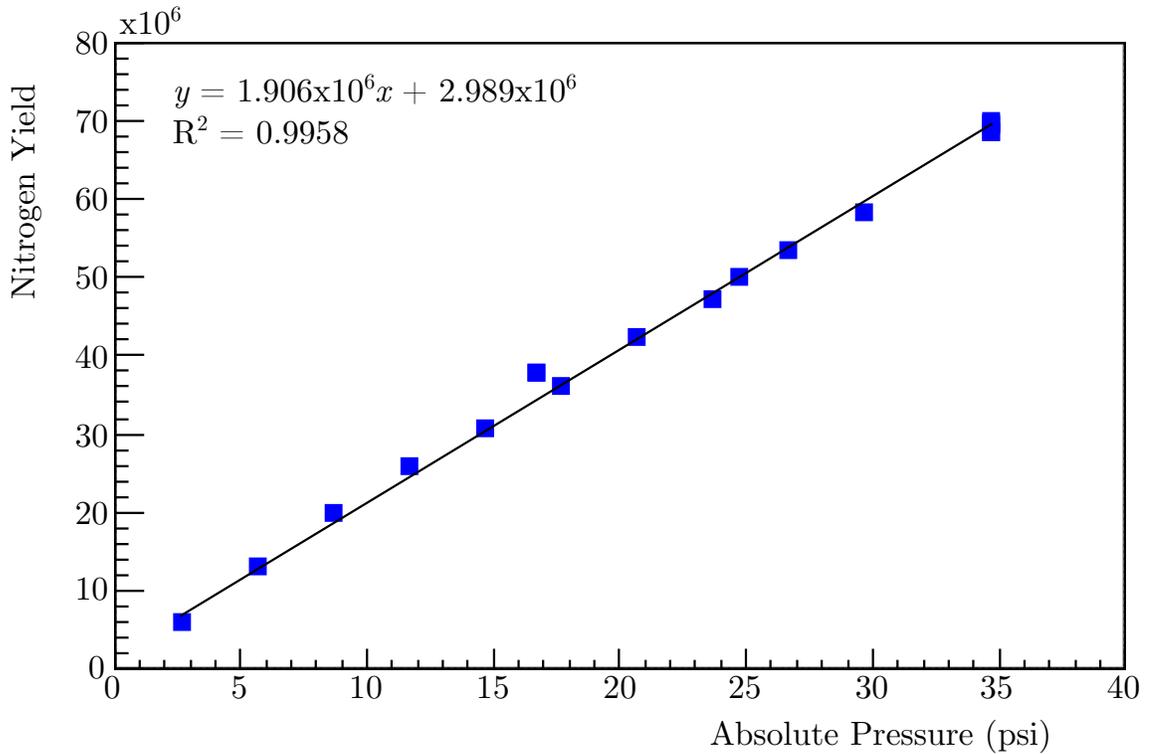}
	\caption [Pressure Curve Fit Example] {For each Q$^2$ value, a pressure curve was fitted to determine the slope, $k$, in the analysis of the nitrogen dilution factor. Presented is a pressure scan with Q$^2$=0.505 (GeV/$c)^2$.}
	\label{k-fit}
\end{figure}

Dilution from N$_2$ was calculated using the pressure curve method. Using a reference cell filled with N$_2$ and comparing it to the production $^3$He cell, a dilution can be found. The relationship between the two cells can be described as
\begin{equation}
	Y^{\mathrm{prod}}_{N_2} = k \cdot P^{\mathrm{prod}}_{N_2},
\end{equation}
where $Y^{\mathrm{prod}}_{N_2}$ is the charge and live-time normalized nitrogen yield of the $^3$He production cell, $P^{\mathrm{prod}}_{N_2}$ is the nitrogen pressure in the $^3$He production cell, and
\begin{equation}
	k = \frac{Y^{\mathrm{ref}}_{N_2}}{P^{\mathrm{ref}}_{N_2}},
\end{equation}
where $Y^{\mathrm{ref}}_{N_2}$ is the charge and live-time normalized yield of a N$_2$-filled reference cell and $P^{\mathrm{ref}}_{N_2}$ is the pressure in that cell. The value of $k$ was determined by taking the slope of a linear fit of $Y^{\mathrm{ref}}_{N_2}$ against $P^{\mathrm{ref}}_{N_2}$ for each Q$^2$ value. An example of this fit is shown in Figure \ref{k-fit}.

The unit density of the $^3$He production cells was measured in amagats and is defined as
\begin{equation}
	\eta = \left( \frac{p}{p_0} \right) \left( \frac{T_0}{T} \right) \mbox{amg},
\end{equation}
where $\eta$ is the number density, $p$ is the pressure of the cell, $p_0$ is 1 atm (or 14.7 psi), $T$ is the temperature of the target cell, and $T_0$ is 273.15 K. The $^3$He production cell for $A_y^0$ has a N$_2$ density of 0.0783 amg and the cell for both $A_T$ and $A_L$ has a N$_2$ density of 0.1132 amg. The $^3$He production cell was held at a temperature of 46 $^{\circ}$C for $A_y^0$ and 45 $^{\circ}$C for $A_T$ and $A_L$. This leads to $P^{\mathrm{prod}}_{N_2}=1.345$ psi for $A_y^0$, and 1.938 psi for $A_T$ and $A_L$.  This analysis leads to the dilution values shown in Table \ref{tab:n2-dilution}.

\begin{table}
\begin{center}
\begin{tabular}{c|c|c}
Asymmetry Type	&	Q$^2$ (GeV/$c)^2$	&	N$_2$ Dilution Factor	\\ \hline
		&	0.127	&	$0.9468	\pm	0.0077$	\\
$A_y^0$	&	0.456	&	$0.9788	\pm	0.0029$	\\
		&	0.953	&	$0.9721	\pm	0.0120$	\\ \hline
$A_T$	&	0.505	&	$0.9454	\pm	0.0075$	\\
		&	0.953	&	$0.9390	\pm	0.0262$	\\ \hline
$A_L$	&	0.505	&	$0.9711	\pm	0.0040$	\\
		&	0.953	&	$0.9380	\pm	0.0267$	\\
\end{tabular}
\caption[N$_2$ Dilution Factors.]{This table shows the dilution factors due to nitrogen contamination that were used for each of the asymmetries measured.}
\label{tab:n2-dilution}
\end{center}
\end{table}

% ^^^^^^^^^^^^^^^^^^^^^^^^^^^^^^^^^^^^^^^^^^^^^^^^^^^^^^^^^^^^^^^^^^^^^^^^^^^^^^^

\large
\section {Uncertainty Analysis} 
\label{ch3-error}
\normalsize
% Error Analysis
% vvvvvvvvvvvvvvvvvvvvvvvvvvvvvvvvvvvvvvvvvvvvvvvvvvvvvvvvvvvvvvvvvvvvvvvvvvvvvvvv
Since this dissertation examines two different types of asymmetries, the target single-spin asymmetry in the case of $A_y^0$ and the beam-target double-spin asymmetry in the case of $A_L$ and $A_T$, the uncertainty analysis is handled differently for each. The single-spin asymmetry uncertainty analysis is discussed in Section \ref{ay-error} while that for the double-spin asymmetry is presented in Section \ref{axaz-error}. 

\subsection{$A_y^0$ Uncertainty Analysis}
\label{ay-error}
The measured target single-spin asymmetry,  $A_y^0$, is defined as
\begin{equation}
	A_y^0 = \frac{1}{|P_y|}\left( \frac{Y_{\uparrow}-Y_{\downarrow}}{Y_{\uparrow}+Y_{\downarrow}} \right),
\end{equation}
where 
\begin{equation}
	Y_{\uparrow (\downarrow)} = \frac{S_{\uparrow (\downarrow)}}{C_{\uparrow (\downarrow)} \cdot LT_{\uparrow (\downarrow)}},
	\label{yield}
\end{equation}
\begin{equation}
	S_{\uparrow (\downarrow)} = T_{\uparrow (\downarrow)} - B_{\uparrow (\downarrow)} = \mbox{\# of Signal Events}_{\uparrow (\downarrow)},
\end{equation}
\begin{equation}
	T_{\uparrow (\downarrow)}  = \mbox{Total \# of Events Under Peak}_{\uparrow (\downarrow)},
\end{equation}
and
\begin{equation}
	B_{\uparrow (\downarrow)} = \mbox{Background Fit}_{\uparrow (\downarrow)}.
\end{equation}
If a new variable, $r$, is defined as 
\begin{equation}
	 r = \frac{Y_\uparrow}{Y_\downarrow}, 
\end{equation}
then
\begin{equation}
	A_y^0 = \frac{1}{|P_y|}\left( \frac{r-1}{r+1} \right).
\end{equation}
Propagating the uncertainties in $P_y$ and $r$ in quadrature, we find
\begin{equation}
	\delta r = r \left[ \left(\ \frac{\delta Y_\uparrow}{Y_\uparrow} \right) ^2 + \left(\ \frac{\delta Y_\downarrow}{Y_\downarrow} \right) ^2 \right]^{\frac{1}{2}}
\end{equation}
and 
\begin{equation}
	\delta A_y^0 = \left( \frac{A_{y}^{2} \delta P _{y}^2}{P_y^2} + \frac{1}{P_y^2}\frac{4}{\left(r+1\right)^4} \delta r ^2 ,\right)^{\frac{1}{2}}.
\end{equation}
If we replace $r$ with the yields, we find
\begin{equation}
	\delta A_y^0 = \left( \frac{A_{y}^{2} \delta P _{y}^2}{P_y^2} + \frac{1}{P_y^2}\frac{4}{\left(\frac{Y_\uparrow}{Y_\downarrow}+1\right)^4} \cdot \left( \frac{Y_\uparrow}{Y_\downarrow} \right)^2 \cdot \left[ \left( \frac{\delta Y_\uparrow}{Y_\uparrow} \right) ^2 + \left( \frac{\delta Y_\downarrow}{Y_\downarrow} \right)^2 \right]  \right) ^{\frac{1}{2}},
	\label{full-err}
\end{equation}
or, more simply, 
\begin{equation}
	\delta A_y^0 = \left( \epsilon_{P_T}^2 + \epsilon_S^2  \right) ^{\frac{1}{2}},
\end{equation}
where $\epsilon_{P_T}$ and $\epsilon_S$ are defined as in Table \ref{tab:ay-error-eq}.

\begin{table}
\begin{center}
\begin{tabular}{c|c}
Uncertainty Type & Equation \\ \hline
Statistical & $\epsilon_S=\frac{1}{P_y} \cdot \frac{2}{\left(\frac{Y_\uparrow}{Y_\downarrow}+1\right)^2} \cdot \left( \frac{Y_\uparrow}{Y_\downarrow} \right) \cdot \left[ \left( \frac{\delta Y_\uparrow}{Y_\uparrow} \right) ^2 + \left( \frac{\delta Y_\downarrow}{Y_\downarrow} \right)^2 \right]^\frac{1}{2} $\\
Target Polarization & $\epsilon_{P_T}=\frac{A_y^0 \delta P_y}{P_y}$ \\ \\ \hline \\
Total & $\delta A_y^0=\sqrt{\epsilon_{P_T}^2 + \epsilon_S^2}$ \\
\end{tabular}
\caption[$A_y^0$ Uncertainties.]{This table shows the equations used to calculate the uncertainties for $A_y^0$.}
\label{tab:ay-error-eq}
\end{center}
\end{table}

In order to use Eq. \ref{full-err}, we need to look at the uncertainty in the yields ($Y_{\uparrow(\downarrow)}$, defined in Eq. \ref{yield}). Since the uncertainty in the charge and live-time are negligible, this leads to
\begin{equation}
	\delta Y = \frac{\delta S}{C\cdot LT},
\end{equation}
where
\begin{equation}
	\delta S = \sqrt{\delta T^2 + \delta B^2}.
\end{equation}
Since $T$ deals with the statistical fluctuations of the signal and background,
\begin{equation}
	\delta T = \sqrt{T}.
\end{equation}
\begin{figure}
	\centering
	\includegraphics{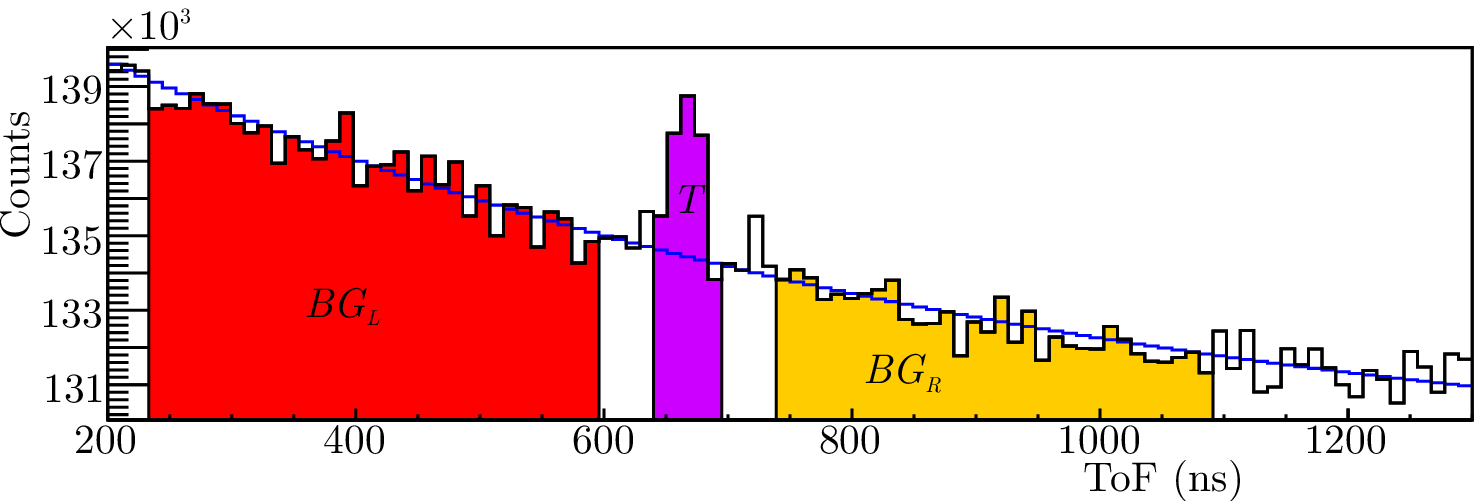}
	\caption [ToF Background Example] {For each ToF plot, the background was fitted to the left and right of the peak. In order to estimate the uncertainty from background contributions, three sections were used. $BG_{L}$ and $BG_{R}$ consist of the number of events within the same number of bins on the left and right side of the peak. $T$ consists of the total number of events under the peak, which includes both the signal and background. Signal events are those events in $T$ that are above the blue fit line.}
	\label{bg-exp}
\end{figure}

Things are more complicated with the uncertainty in the background fit. Since the time-of-flight background was measured over a large range, a fit was made for each Q$^2$ value as defined in Section \ref{tof}. An equal range of bins, $R_{BG}$/2, was integrated on the background left of the signal, $BG_L$, and right of the signal, $BG_R$. An example of this is shown in Figure \ref{bg-exp} and is shown for each Q$^2$ for $A_y^0$ in Figures \ref{q2-01-tof} through \ref{q2-1-tof}. In order to find $\delta B$, the fractional uncertainty was multiplied by the range of bins used to define the signal ($R_{S}$),
\begin{equation}
	\delta B = \left(\frac{1}{\sqrt{BG_L+BG_R}}\right) \cdot \left(\frac{R_S}{R_{BG}}\right).
\end{equation}
Taken together, we obtain
\begin{equation}
	\delta Y = \frac{1}{C\cdot LT} \cdot \sqrt{ \left( \sqrt{T}\right)^2  + \left[ \left(\frac{1}{\sqrt{BG_L+BG_R}}\right) \cdot \left(\frac{R_S}{R_{BG}}\right)\right]^2},
	\label{bg-eq}
\end{equation}
which can then be used in Eq. \ref{full-err} to complete the full error analysis. The total uncertainty budget is shown in Table \ref{tab:ay-error-budget}.

\begin{table}
\centering
\begin{tabular}{c|c|c}
Q$^2$  (GeV/$c)^2$& Uncertainty Type & Amount \\ \hline
0.127 & Statistical ($\epsilon_S$) & 0.12617 \\
0.127 & Target Polarization ($\epsilon_T$) & 0.03042 \\ \hline
0.127 & Total ($\delta A_y^0$) &  0.12979 \\ \hline \hline
0.456 & Statistical ($\epsilon_S$) & 0.00121\\
0.456 & Target Polarization ($\epsilon_T$) & 0.01087 \\ \hline
0.456 & Total ($\delta A_y^0$) & 0.01093 \\ \hline \hline
0.953 & Statistical ($\epsilon_S$) & 0.001321   \\
0.953 & Target Polarization ($\epsilon_T$) & 0.000298 \\ \hline
0.953 & Total ($\delta A_y^0$) & 0.001354\\ \hline \hline
\end{tabular}
\caption[$A_y^0$ Uncertainties.]{This table shows the magnitude of the uncertainties for $A_y^0$.}
\label{tab:ay-error-budget}
\end{table}

\subsection{$A_L$ and $A_T$ Uncertainty Analysis}
\label{axaz-error}
The measured target single-spin asymmetry,  $A_L$, is defined as
\begin{equation}
	A_{x(z)} = \frac{1}{|P_T \cdot P_B|}\left( \frac{Y_{\uparrow}-Y_{\downarrow}}{Y_{\uparrow}+Y_{\downarrow}} \right),
\end{equation}
where $P_T$ is the target polarization, $P_B$ is the beam polarization, and $Y$, $S$, $T$, and $B$ are defined as in Section \ref{ay-error}. Following the discussion in Section \ref{ay-error}, we obtain 
\begin{equation}
	\delta A_{x(z)} = \left( \epsilon_{P_B}^2 + \epsilon_{P_T}^2 + \epsilon_{S}^2 \right)^\frac{1}{2},
\end{equation}
where $\epsilon_{P_B}$, $\epsilon_{P_T}$, and $\epsilon_{S}$ are defined as in Table \ref{tab:axaz-error-eq}. The background fluctuations are included in the statistical uncertainty as in Section \ref{ay-error} and in particular as in Eq. \ref{bg-eq}. The uncertainty from the beam and target polarizations for $A_T$ and $A_L$ is shown in Table \ref{tab:ax-az-error-budget}. The full uncertainties, which include terms based on the asymmetries, are discussed in Sections \ref{ch4-trans} and \ref{ch4-long}.

\begin{table}
\begin{center}
\begin{tabular}{c|c}
Uncertainty Type & Equation \\ \hline
Statistical & $\epsilon_S=\frac{1}{P_T P_B} \cdot \frac{2}{\left(\frac{Y_\uparrow}{Y_\downarrow}+1\right)^2} \cdot \left( \frac{Y_\uparrow}{Y_\downarrow} \right) \cdot \left[ \left( \frac{\delta Y_\uparrow}{Y_\uparrow} \right) ^2 + \left( \frac{\delta Y_\downarrow}{Y_\downarrow} \right)^2 \right]^\frac{1}{2} $\\
Target Polarization & $\epsilon_{P_T}=\frac{A \delta P_T}{P_T}$ \\ \\
Beam Polarization & $\epsilon_{P_B}=\frac{A \delta P_B}{P_B}$ \\ \\ \hline \\
Total & $\delta A=\sqrt{\epsilon_{P_B}^2 + \epsilon_{P_T}^2 + \epsilon_S^2}$ \\
\end{tabular}
\caption[$A_L$  and $A_T$ Uncertainties.]{This table shows the equations used to calculate the uncertainties for $A_L$ and $A_T$.}
\label{tab:axaz-error-eq}
\end{center}
\end{table}

\begin{table}
\begin{center}
\begin{tabular}{c|c|c}
Experiment  & Uncertainty Type & Amount (Abs. $\%$)\\ \hline
$A_T$  & Target Polarization ($\delta P_T$) & 2.33 \\ 
 & Beam Polarization ($\delta P_B$) & 3.9 \\ \hline
$A_L$  & Target Polarization ($\delta P_T$) &  2.53  \\
& Beam Polarization ($\delta P_B$) &  3.9  \\ 
\end{tabular}
\caption[Polarization Uncertainties in $A_T$ and $A_L$.]{Uncertainties in $A_T$ and $A_L$. This table shows the magnitude of the uncertainties in the beam and target polarization for $A_T$ and $A_L$.}
\label{tab:ax-az-error-budget}
\end{center}
\end{table}

% ^^^^^^^^^^^^^^^^^^^^^^^^^^^^^^^^^^^^^^^^^^^^^^^^^^^^^^^^^^^^^^^^^^^^^^^^^^^^^^^

					% Include the Chapter 5 text (chapter_5.tex)
	% vvvvvvvvvvvvvvvvvvvvvvvvvvvvvvvvvvvvvvvvvvvvvvvvvvvvvvvvvvvvvvvvvvvvvvvvvvvvvvvv
% Chapter 5 (chapter_5.tex)
%
% Chapter 5 of Elena Long's Ph.D. Dissertation
%
% To be completed: March, 2012
%
% ^^^^^^^^^^^^^^^^^^^^^^^^^^^^^^^^^^^^^^^^^^^^^^^^^^^^^^^^^^^^^^^^^^^^^^^^^^^^^^^^

\chapter{Results and Discussion}	% Chapter Title
\label{results}					% Chapter Label
\normalsize					% Return to Normal font size

\large
\section {Asymmetry Measurements} 
\label{ch4-asyms}
\normalsize
% Vertical 3He(e,e'n) Asymmetries
% vvvvvvvvvvvvvvvvvvvvvvvvvvvvvvvvvvvvvvvvvvvvvvvvvvvvvvvvvvvvvvvvvvvvvvvvvvvvvvvv
Three different asymmetries were measured for this dissertation. Of them, there are two types: target single-spin asymmetries and beam-target double-spin asymmetries. Although both use the same mathematical form for the asymmetries,
\begin{equation}
	A = \frac{1}{P} \frac{Y_{\uparrow} - Y_{\downarrow}}{Y_{\uparrow} + Y_{\downarrow}},
	\label{eq-asym}
\end{equation}
the variables are subtly different. In the case of the single-spin asymmetries ($A_y^0$), 
\begin{equation}
	Y_{\uparrow(\downarrow)} = \frac{N_{T\uparrow(\downarrow)}}{C_{T\uparrow(\downarrow)} LT_{T\uparrow(\downarrow)}},
\end{equation}
where $P$ is the target polarization, $N_{T\uparrow(\downarrow)}$ is the number of neutrons counted with the target spin oriented up (down), $C_{T\uparrow(\downarrow)}$ is the charge accumulated with the target spin up (down), and $LT_{T\uparrow(\downarrow)}$ is the live-time with the target spin up (down). 

In the case of the double-spin asymmetries ($A_T$ and $A_L$),
\begin{equation}
	Y_{\uparrow(\downarrow)} = \frac{N_{B\uparrow(\downarrow)}}{C_{B\uparrow(\downarrow)} LT_{B\uparrow(\downarrow)}},
\end{equation}
where $P$ is the product of the target polarization and the beam polarization, $N_{B\uparrow(\downarrow)}$ is the number of neutrons counted with the beam helicity oriented up (down), $C_{B\uparrow(\downarrow)}$ is the charge accumulated with the beam helicity up (down), and $LT_{B\uparrow(\downarrow)}$ is the live-time with the beam helicity up (down).

% ^^^^^^^^^^^^^^^^^^^^^^^^^^^^^^^^^^^^^^^^^^^^^^^^^^^^^^^^^^^^^^^^^^^^^^^^^^^^^^^^

\large
\section {Vertical $^3\mathrm{He}^{\uparrow}(e,e'n$) Asymmetries} 
\label{ch4-vert}
\begin{figure}
	\centering
	\includegraphics[width=6 in]{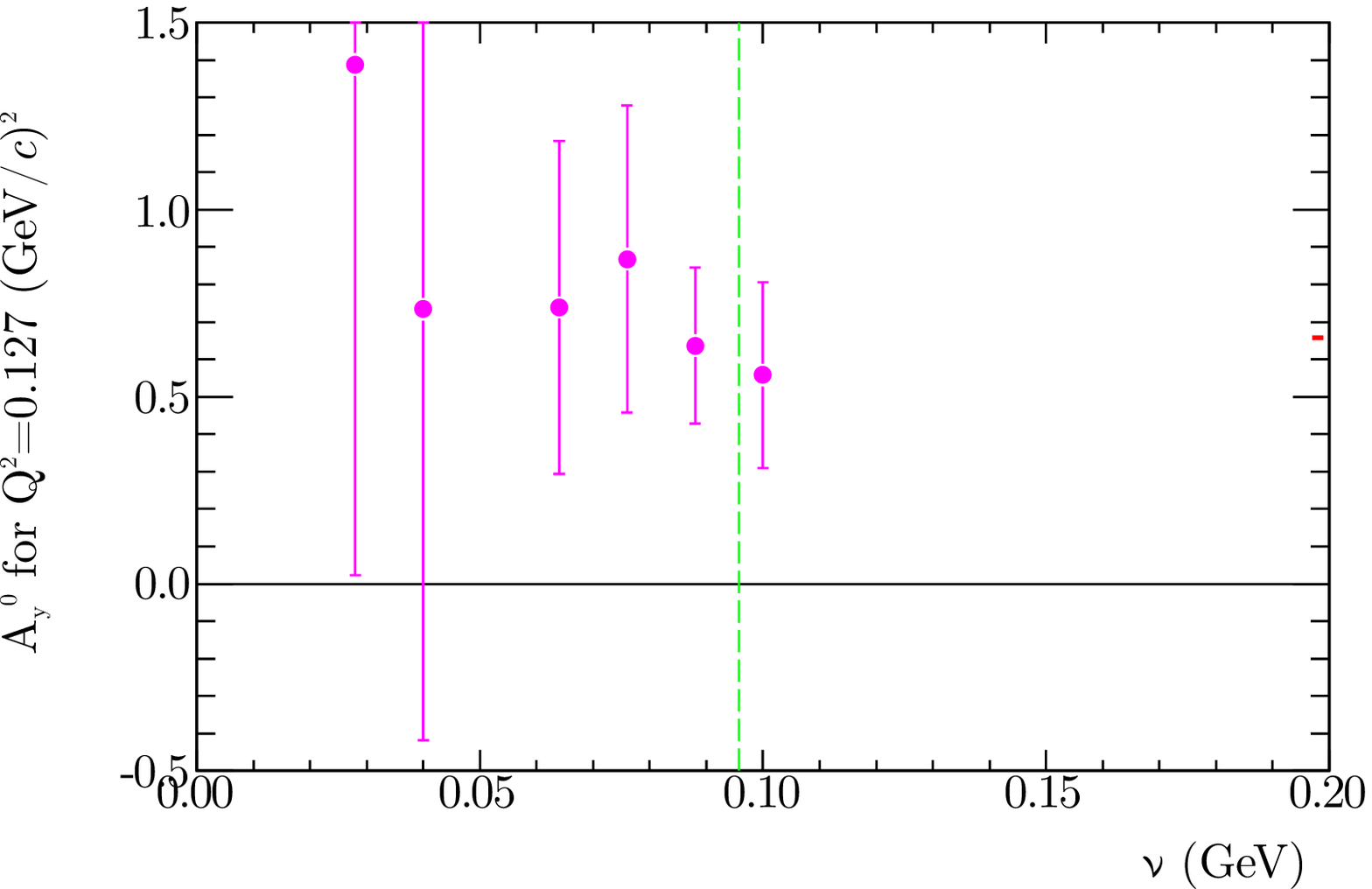}
	\caption [$A_y^0$ Measurements for Q$^2=0.127$ (GeV/$c$)$^2$] {This plot shows the current measurements for $A_y^0$ when Q$^2=0.127$ (GeV/$c$)$^2$. The green dashed line shows the central value of the quasi-elastic peak.}
	\label{ay-final-plot-q2-01}
\end{figure}

\begin{figure}
	\centering
	\includegraphics[width=6 in]{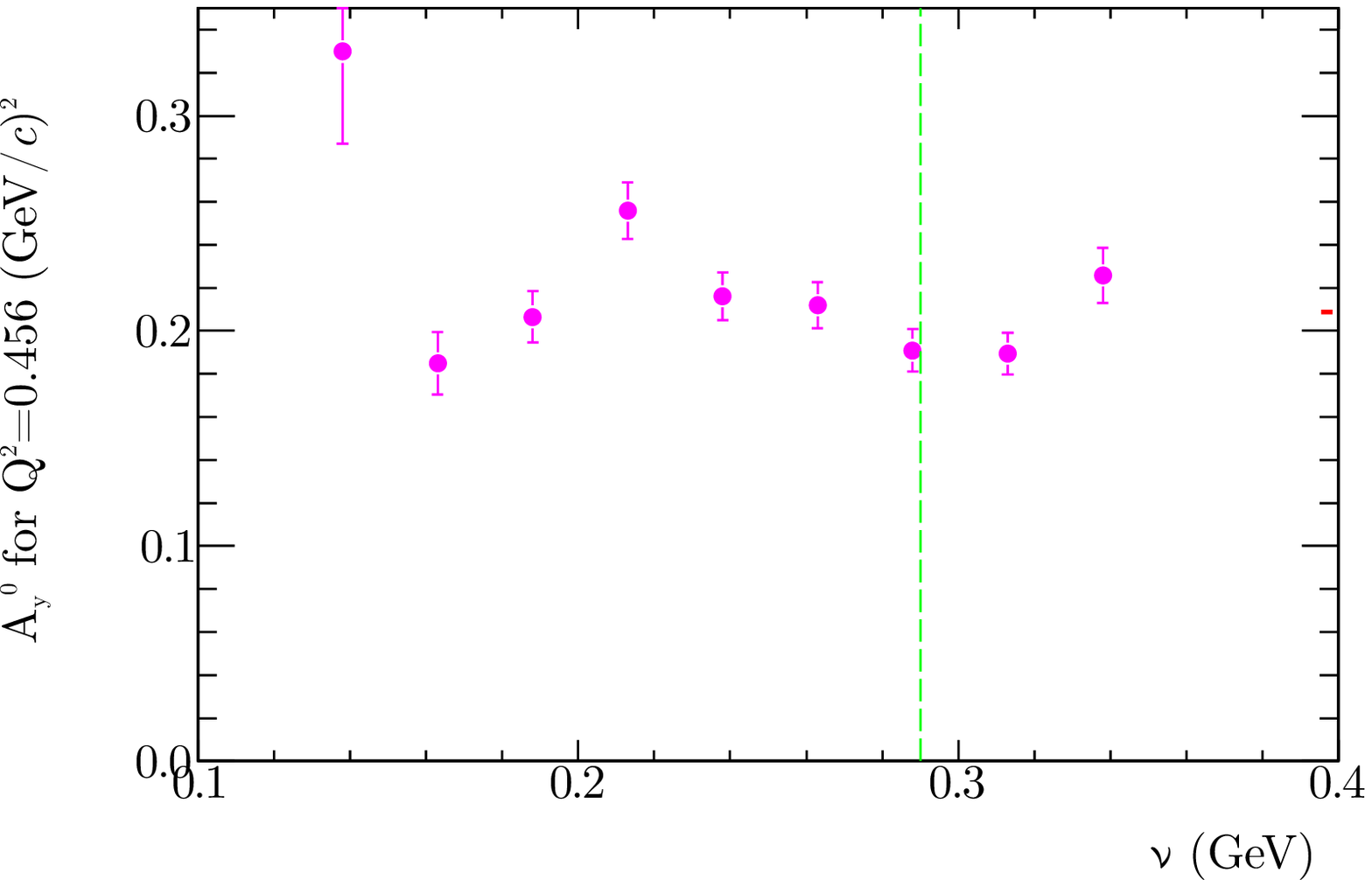}
	\caption [$A_y^0$ Measurements for Q$^2=0.456$ (GeV/$c$)$^2$] {This plot shows the current measurements for $A_y^0$ when Q$^2=0.456$ (GeV/$c$)$^2$. The green dashed line shows the central value of the quasi-elastic peak.}
	\label{ay-final-plot-q2-05}
\end{figure}

\begin{figure}
	\centering
	\includegraphics[width=6 in]{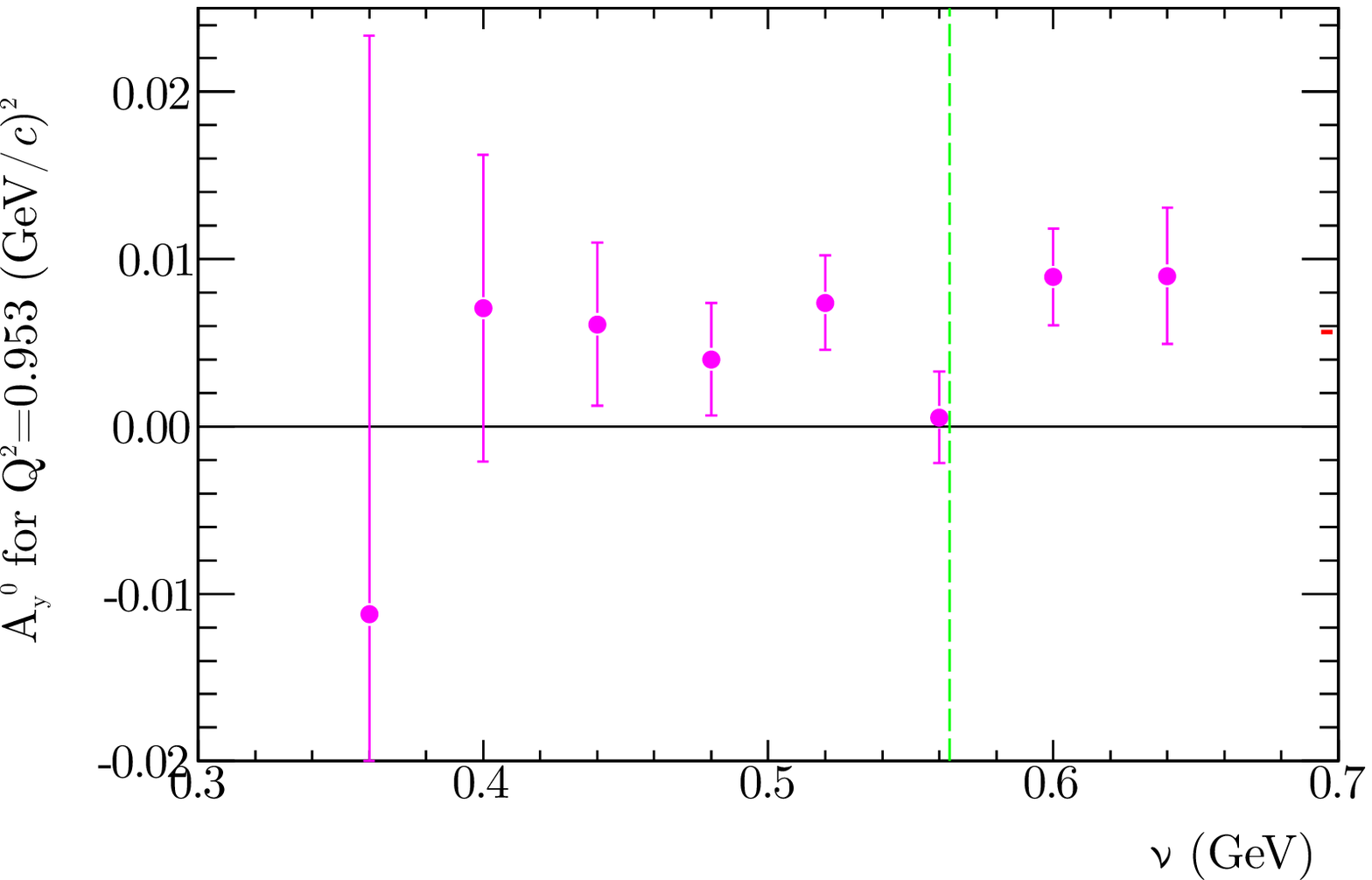}
	\caption [$A_y^0$ Measurements for Q$^2=0.953$ (GeV/$c$)$^2$] {This plot shows the current measurements for $A_y^0$ when Q$^2=0.953$ (GeV/$c$)$^2$. The green dashed line shows the central value of the quasi-elastic peak.}
	\label{ay-final-plot-q2-1}
\end{figure}

\begin{figure}
	\centering
	\includegraphics[width=6 in]{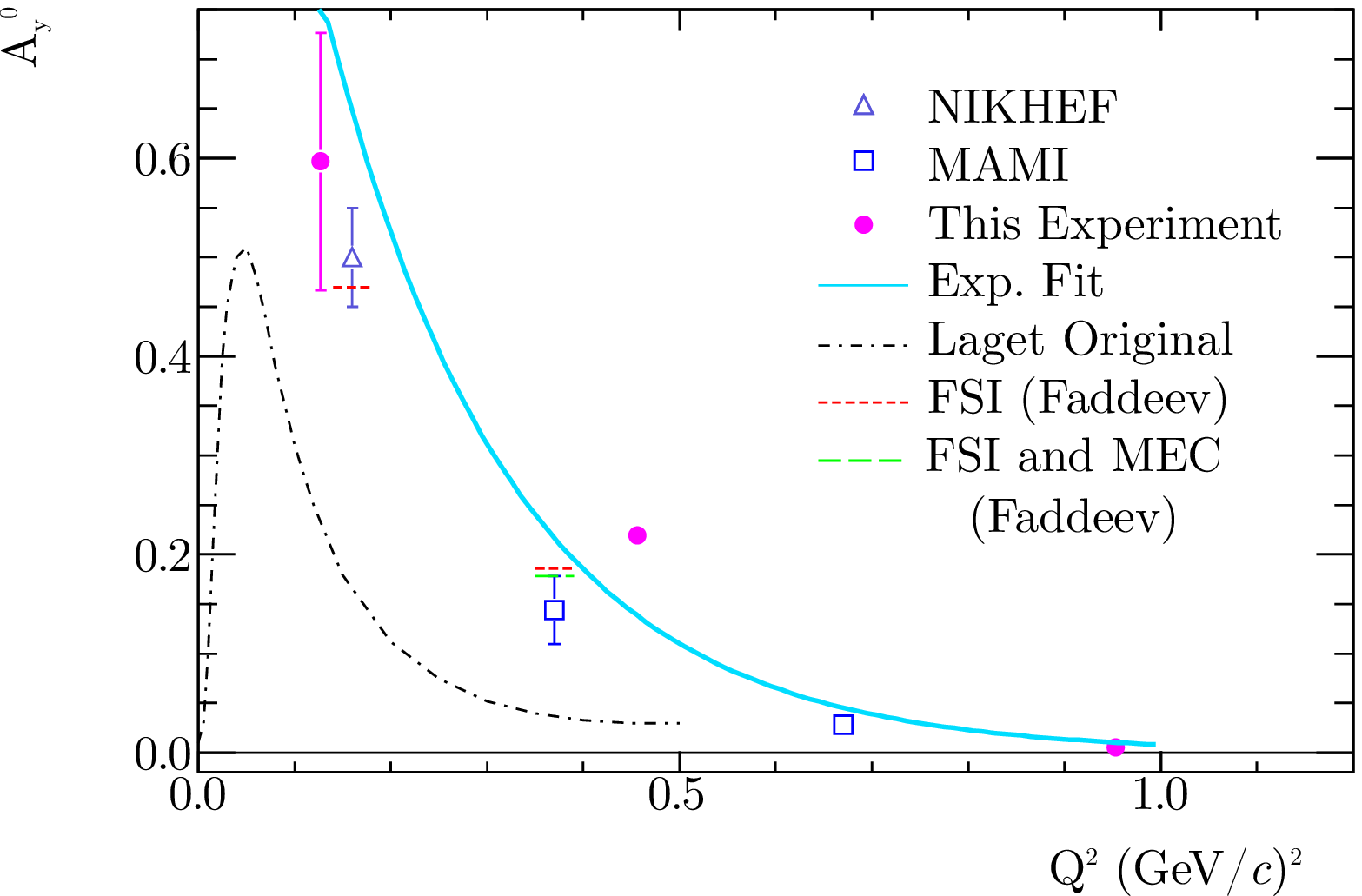}
	\caption [$A_y^0$ World Data] {This plot shows the current measurements along with the world data for $A_y^0$. The points at 0.456  (GeV/$c)^2$ and 0.953 (GeV/$c)^2$ have error bars smaller than the size of the data point. The range in uncertainties for these points can be found in Table \ref{tab:ay-final}. Also shown is an exponential fit of the world data.}
	\label{ay-final-plot}
\end{figure}

\normalsize
% Vertical 3He(e,e'n) Asymmetries
% vvvvvvvvvvvvvvvvvvvvvvvvvvvvvvvvvvvvvvvvvvvvvvvvvvvvvvvvvvvvvvvvvvvvvvvvvvvvvvvv
The vertical $^3\mathrm{He}^{\uparrow}(e,e'n$) target single-spin asymmetry, $A_y^0$, was measured using the equipment discussed in Chapter \ref{experimentsetup} and the method discussed in Section \ref{ch4-asyms}. Particle identification, as described in Chapter \ref{particleid}, was used to select neutrons that were quasi-elastically knocked-out from vertically polarized $^3$He nuclei by incident electrons within the acceptance of the high resolution spectrometer. Error analysis and dilution factors were taken into account as discussed in Chapter \ref{dilution-uncertainties}. The results are presented in Tables \ref{tab:ay-final-nu} and  \ref{tab:ay-final}. They are plotted against the energy transferred, $\nu$, in Figures \ref{ay-final-plot-q2-01} through \ref{ay-final-plot-q2-1} and against the squared four-momentum transferred, Q$^2$, along with the world data and current theory estimates in Figure \ref{ay-final-plot}.

\begin{table}
\centering
\begin{tabular}{c|c|c|c|c}
Q$^2$ (GeV/$c)^2$  & $\nu$ (GeV) & $A_y^0$ & Stat. Uncertainty & Sys. Uncertainty \\ \hline
0.127	&	0.028	&	1.3142	&	1.3625	&	0.1449	\\
0.127	&	0.040	&	0.6948	&	1.1505	&	0.0849	\\
0.127	&	0.064	&	0.6992	&	0.4428	&	0.0394	\\
0.127	&	0.076	&	0.8220	&	0.4079	&	0.0343	\\
0.127	&	0.088	&	0.6026	&	0.2063	&	0.0260	\\
0.127	&	0.100	&	0.5284	&	0.2463	&	0.0299	\\ \hline
0.456	&	0.138	&	0.3228	&	0.0398	&	0.0163	\\
0.456	&	0.163	&	0.1810	&	0.0111	&	0.0091	\\
0.456	&	0.188	&	0.2021	&	0.0061	&	0.0102	\\
0.456	&	0.213	&	0.2504	&	0.0038	&	0.0127	\\
0.456	&	0.238	&	0.2114	&	0.0027	&	0.0107	\\
0.456	&	0.263	&	0.2074	&	0.0024	&	0.0105	\\
0.456	&	0.288	&	0.1869	&	0.0024	&	0.0094	\\
0.456	&	0.313	&	0.1853	&	0.0027	&	0.0094	\\
0.456	&	0.338	&	0.2211	&	0.0061	&	0.0112	\\ \hline
0.953	&	0.360	&	-0.0109	&	0.0346	&	0.0006	\\
0.953	&	0.400	&	0.0069	&	0.0092	&	0.0004	\\
0.953	&	0.440	&	0.0059	&	0.0049	&	0.0003	\\
0.953	&	0.480	&	0.0039	&	0.0034	&	0.0002	\\
0.953	&	0.520	&	0.0072	&	0.0028	&	0.0004	\\
0.953	&	0.560	&	0.0005	&	0.0027	&	0.0000	\\
0.953	&	0.600	&	0.0087	&	0.0029	&	0.0005	\\
0.953	&	0.640	&	0.0087	&	0.0040	&	0.0005	\\ \hline
\end{tabular}
\caption[$A_y^0$ Measurements vs. $\nu$.]{These are the values for $A_y^0$ that were measured in this experiment against the squared four-momentum transferred (Q$^2$), and the energy transfer ($\nu$).}
\label{tab:ay-final-nu}
\end{table}


\begin{table}
\centering
\begin{tabular}{c|c|c|c}
Q$^2$ (GeV/$c)^2$ & $A_y^0$ & Stat. Uncertainty & Sys. Uncertainty \\ \hline
0.127 & 0.72686 & 0.11466 & 0.08731 \\ \hline
0.456 & 0.20234 & 0.00102 & 0.00189 \\ \hline
0.953 & 0.00518 & 0.000686 & 0.000076 \\ 
\end{tabular}
\caption[$A_y^0$ Measurements.]{These are the values for $A_y^0$ that were measured in this experiment.}
\label{tab:ay-final}
\end{table}

As discussed in Chapter \ref{introduction}, $A_y^0$ is useful for extracting information on the final-state interactions and meson-exchange currents from neutrons knocked-out of polarized $^3$He. The original Laget calculation, which was calculated using the PWIA with contributions from FSI and MEC, indicates that FSI and MEC were expected to contribute to $A_y^0$ largely at low Q$^2$ and drop off at higher Q$^2$. Although the magnitude of these contributions was vastly underestimated, the qualitative understanding agrees with the data presented. Full Faddeev calculations provided by the Bochum group came much closer to predicting $A_y^0$ values to both the historical and current data, although still appear to underestimate these contributions around Q$^2$ of 0.46 (GeV/$c)^2$. Faddeev calculations are not available above this range, since relativistic effects are not included in the calculations. This experiment is also unique in that it includes results at Q$^2=0.953$ (GeV/$c)^2$ where no prior measurements exist. $A_y^0$ is only around 0.5$\%$ in this region, which indicates that at any higher Q$^2$ than this, FSI can be considered negligible and the PWIA is valid.

% A higher value of $A_y^0$ corresponds to a greater effect due to FSI and MEC. From the results of this dissertation, it is indicated that FSI and MEC contributions are significant at lower Q$^2$, then fall off approximately exponentially until they are negligible above Q$^2$ of 1 (GeV/$c$)$^2$.

% ^^^^^^^^^^^^^^^^^^^^^^^^^^^^^^^^^^^^^^^^^^^^^^^^^^^^^^^^^^^^^^^^^^^^^^^^^^^^^^^

\large
\section {Transverse $^3\vec{\mathrm{He}}(\vec{e},e'n$) Asymmetries} 
\label{ch4-trans}
\begin{figure}
	\centering
	\includegraphics[width=6 in]{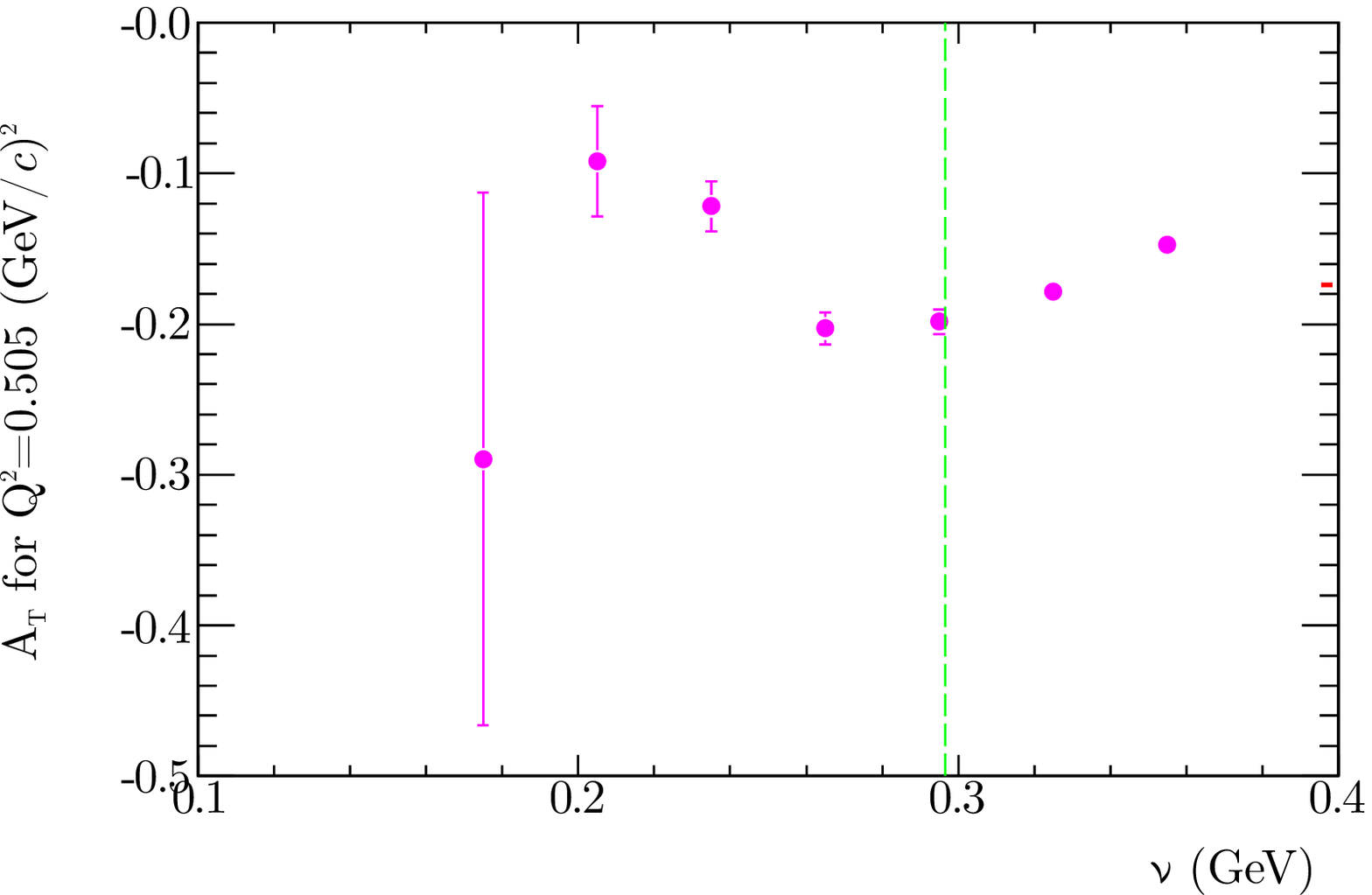}
	\caption [$A_T$ Measurements for Q$^2=0.505$ (GeV/$c$)$^2$] {This plot shows the current measurements for $A_T$ at Q$^2=0.505$ (GeV/$c$)$^2$. The green dashed line shows the central value of the quasi-elastic peak.}
	\label{at-final-plot-q2-05}
\end{figure}

\begin{figure}
	\centering
	\includegraphics[width=6 in]{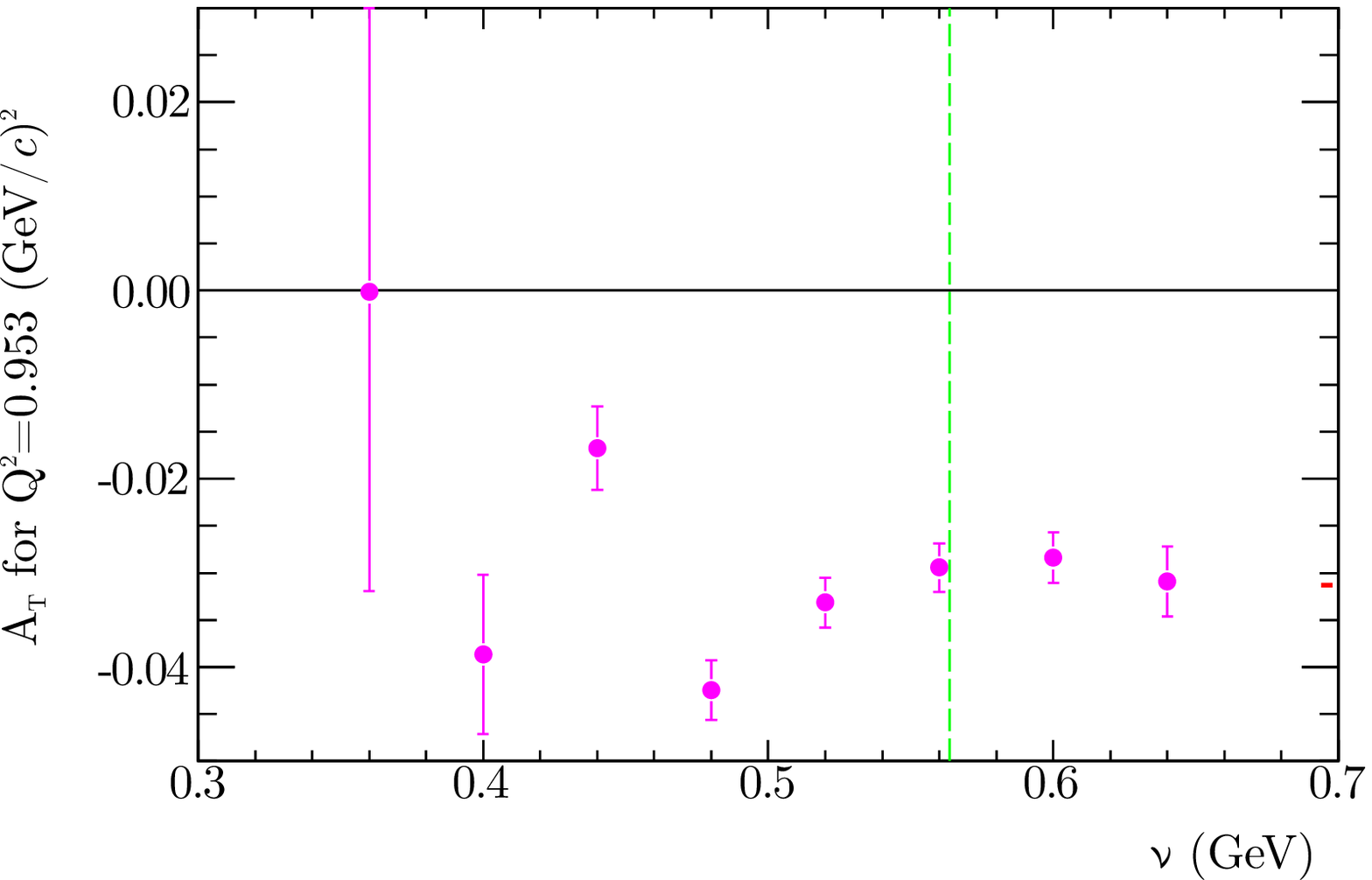}
	\caption [$A_T$ Measurements for Q$^2=0.953$ (GeV/$c$)$^2$] {This plot shows the current measurements for $A_T$ at Q$^2=0.953$ (GeV/$c$)$^2$. The green dashed line shows the central value of the quasi-elastic peak.}
	\label{at-final-plot-q2-1}
\end{figure}

\normalsize
% Transverse 3He(e,e'n) Asymmetries
% vvvvvvvvvvvvvvvvvvvvvvvvvvvvvvvvvvvvvvvvvvvvvvvvvvvvvvvvvvvvvvvvvvvvvvvvvvvvvvvv
The transverse $^3\vec{\mathrm{He}}(\vec{e},e'n$) beam-target double-spin asymmetry, $A_T$, was measured using the equipment discussed in Chapter \ref{experimentsetup} and the method discussed in Section \ref{ch4-asyms}. Particle identification, as described in Chapter \ref{particleid}, was used to select neutrons that were quasi-elastically knocked-out from transversely polarized $^3$He nuclei by incident electrons within the acceptance of the high resolution spectrometer. Error analysis and dilution factors were taken into account as discussed in Chapter \ref{dilution-uncertainties}. The results are presented in Table \ref{tab:at-final} and plotted against the energy transferred, $\nu$, in Figures \ref{at-final-plot-q2-05} and \ref{at-final-plot-q2-1} for Q$^2$ = 0.505 (GeV/$c)^2$ and 0.953 (GeV/$c)^2$, respectively.

\begin{table}
\centering
\begin{tabular}{c|c|c|c|c}
Q$^2$ (GeV/$c)^2$ & $\nu$ (GeV) & $A_T$ & Stat. Uncertainty & Sys. Uncertainty \\ \hline
0.505	&	0.175	&	$-0.2738$	&	0.1766	&	0.0072	\\
0.505	&	0.205	&	$-0.0868$	&	0.0365	&	0.0023	\\
0.505	&	0.235	&	$-0.1151$	&	0.0162	&	0.0030	\\
0.505	&	0.265	&	$-0.1918$	&	0.0094	&	0.0050	\\
0.505	&	0.295	&	$-0.1876$	&	0.0064	&	0.0049	\\
0.505	&	0.325	&	$-0.1686$	&	0.0057	&	0.0044	\\
0.505	&	0.355	&	$-0.1391$	&	0.0066	&	0.0036	\\ \hline
0.953	&	0.360	&	$-0.0002$	&	0.0318	&	0.0000	\\
0.953	&	0.400	&	$-0.0363$	&	0.0084	&	0.0010	\\
0.953	&	0.440	&	$-0.0157$	&	0.0044	&	0.0004	\\
0.953	&	0.480	&	$-0.0399$	&	0.0030	&	0.0010	\\
0.953	&	0.520	&	$-0.0311$	&	0.0025	&	0.0008	\\
0.953	&	0.560	&	$-0.0276$	&	0.0024	&	0.0007	\\
0.953	&	0.600	&	$-0.0267$	&	0.0026	&	0.0007	\\
0.953	&	0.640	&	$-0.0290$	&	0.0036	&	0.0008	\\ \hline
\end{tabular}
\caption[$A_T$ Measurements.]{These are the values for $A_T$ that were measured in this experiment.}
\label{tab:at-final}
\end{table}

% \begin{figure}
%	\centering
%	\includegraphics[width=15 cm]{./long_AT.eps}
%	\caption [$A_T$ Measurements] {$A_T$ Measurements. This plot shows the current measurement for $A_T$.}
%	\label{at-final-plot}
% \end{figure}

Although theoretical methods for calculating this asymmetry are available from Misak and the Bochum group to compare with experimental values for $A_T$, they have not yet been calculated at the kinematics presented in this dissertation. However, these measurements will provide an important test of the calculations when available. It is important to note that measurements for both Q$^2$ values are non-zero and negative. For Q$^2$ of 0.505 (GeV/$c)^2$, the asymmetry fluctuates around the mean of $-0.1738 \pm 0.0039$. For Q$^2$ of 0.953 (GeV/$c)^2$, the asymmetry is much smaller but remains non-zero and fluctuates around the mean of $-0.0313 \pm 0.0012$.

% \pagebreak

% ^^^^^^^^^^^^^^^^^^^^^^^^^^^^^^^^^^^^^^^^^^^^^^^^^^^^^^^^^^^^^^^^^^^^^^^^^^^^^^^

\large
\section {Longitudinal $^3\vec{\mathrm{He}}(\vec{e},e'n$) Asymmetries} 
\label{ch4-long}
\begin{figure}
	\centering
	\includegraphics[width=6 in]{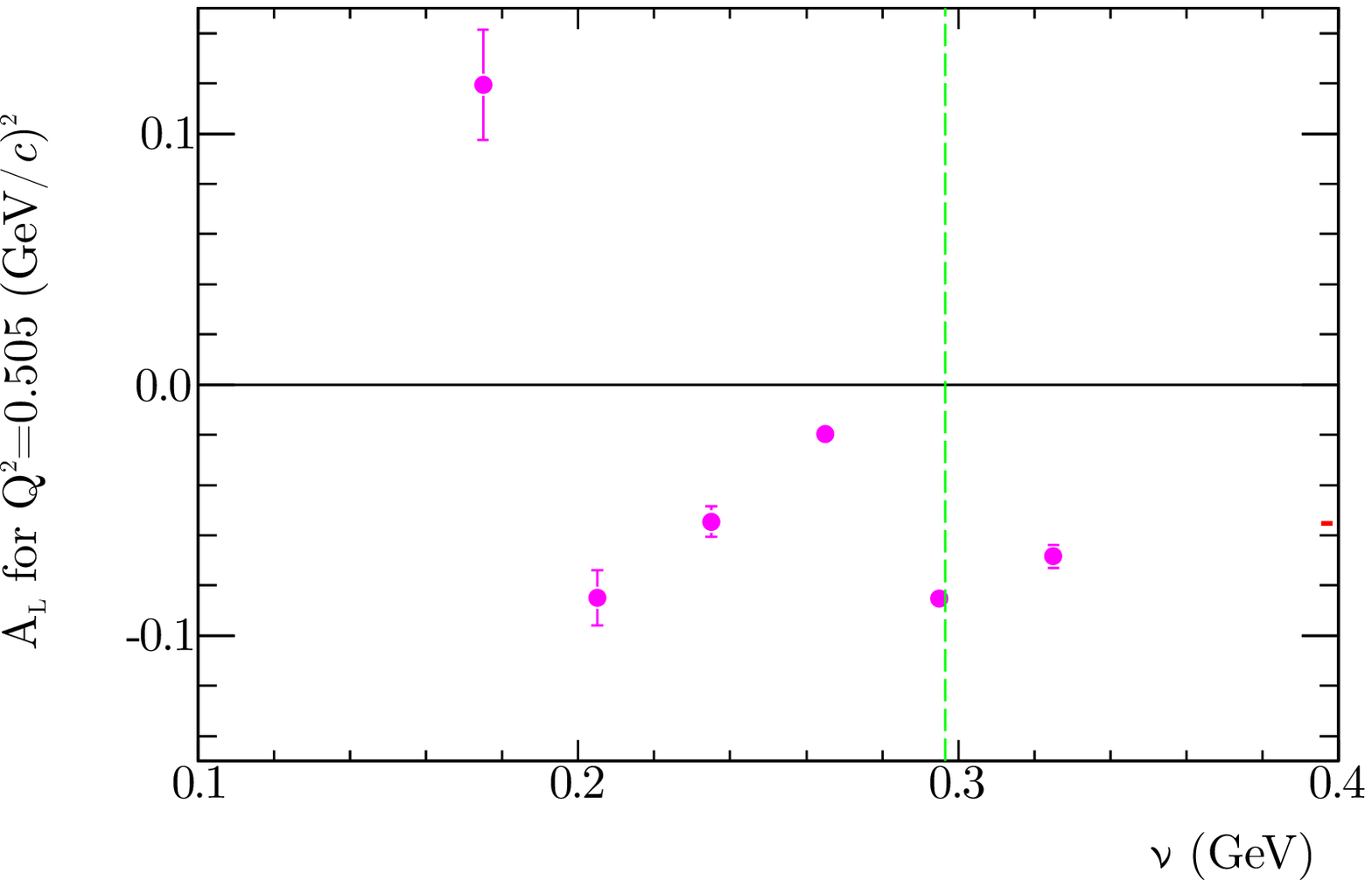}
	\caption [$A_L$ Measurements for Q$^2=0.505$ (GeV/$c$)$^2$] {This plot shows the current measurements for $A_L$ at Q$^2=0.505$ (GeV/$c$)$^2$. The green dashed line shows the central value of the quasi-elastic peak.}
	\label{al-final-plot-q2-05}
\end{figure}

\begin{figure}
	\centering
	\includegraphics[width=6 in]{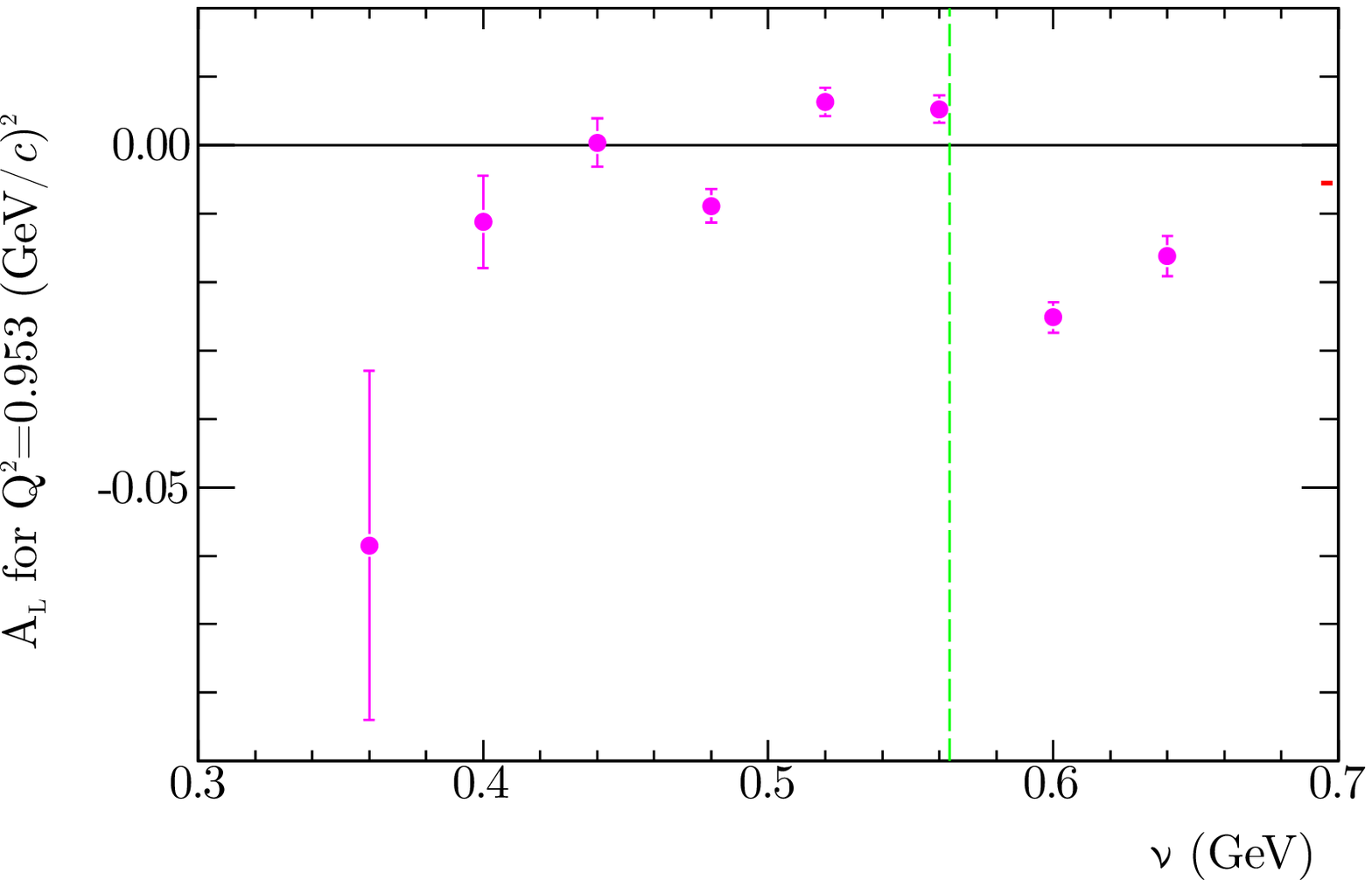}
	\caption [$A_L$ Measurements for Q$^2=0.953$ (GeV/$c$)$^2$] {This plot shows the current measurements for $A_L$ at Q$^2=0.953$ (GeV/$c$)$^2$. The green dashed line shows the central value of the quasi-elastic peak.}
	\label{al-final-plot-q2-1}
\end{figure}

\normalsize
% Longitudinal 3He(e,e'n) Asymmetries
% vvvvvvvvvvvvvvvvvvvvvvvvvvvvvvvvvvvvvvvvvvvvvvvvvvvvvvvvvvvvvvvvvvvvvvvvvvvvvvvv
The longitudinal $^3\vec{\mathrm{He}}(\vec{e},e'n$) beam-target double-spin asymmetry, $A_L$, was measured using the equipment discussed in Chapter \ref{experimentsetup} and the method discussed in Section \ref{ch4-asyms}. Particle identification, as described in Chapter \ref{particleid}, was used to select neutrons that were quasi-elastically knocked-out from longitudinally polarized $^3$He nuclei by incident electrons within the acceptance of the high resolution spectrometer. Error analysis and dilution factors were taken into account as discussed in Chapter \ref{dilution-uncertainties}. The results are presented in Table \ref{tab:al-final} and plotted against the energy transferred, $\nu$, in Figures \ref{al-final-plot-q2-05} and \ref{al-final-plot-q2-1} for Q$^2$ = 0.505 (GeV/$c)^2$ and 0.953 (GeV/$c)^2$, respectively.

\begin{table}
\centering
\begin{tabular}{c|c|c|c|c}
Q$^2$  (GeV/$c)^2$ & $\nu$ (GeV) & $A_L$ & Stat. Uncertainty & Sys. Uncertainty \\ \hline
0.505	&	0.145	&	-0.7804	&	0.1003	&	0.0224	\\
0.505	&	0.175	&	0.1160	&	0.0217	&	0.0033	\\
0.505	&	0.205	&	-0.0824	&	0.0108	&	0.0024	\\
0.505	&	0.235	&	-0.0530	&	0.0060	&	0.0015	\\
0.505	&	0.265	&	-0.0191	&	0.0041	&	0.0005	\\
0.505	&	0.295	&	-0.0828	&	0.0037	&	0.0024	\\
0.505	&	0.325	&	-0.0664	&	0.0042	&	0.0019	\\ \hline
0.953	&	0.400	&	-0.0105	&	0.0068	&	0.0003	\\
0.953	&	0.440	&	0.0003	&	0.0036	&	0.0000	\\
0.953	&	0.480	&	-0.0084	&	0.0024	&	0.0002	\\
0.953	&	0.520	&	0.0059	&	0.0020	&	0.0002	\\
0.953	&	0.560	&	0.0049	&	0.0020	&	0.0001	\\
0.953	&	0.600	&	-0.0236	&	0.0021	&	0.0007	\\
0.953	&	0.640	&	-0.0152	&	0.0029	&	0.0004	\\
0.953	&	0.360	&	-0.0549	&	0.0254	&	0.0016	\\ \hline
\end{tabular}
\caption[$A_L$ Measurements.]{These are the values for $A_L$ that were measured in this experiment.}
\label{tab:al-final}
\end{table}

% \begin{figure}
%	\centering
%	\includegraphics[width=15 cm]{./long_AL.eps}
%	\caption [$A_L$ Measurements] {$A_L$ Measurements. This plot shows the current measurement for $A_L$.}
%	\label{al-final-plot}
% \end{figure}

Although theoretical methods for calculating this asymmetry are available from Misak and the Bochum group to compare with experimental values for $A_T$, they have not been been calculated at the kinematics presented in this dissertation. However, these measurements will provide an important test of the calculations when available. It is important to note that measurements for both Q$^2$ values are non-zero and change sign with $\nu$. For Q$^2$ of 0.505 (GeV/$c)^2$, the asymmetry at low $\nu$ is both large and positive, which indicates misidentified contributions from the elastic $^3$He peak. As $\nu$ increases, the quasi-elastic region is reached where $A_L$ becomes negative and fluctuates around the mean of $-0.0551 \pm 0.0023$. For Q$^2$ of 0.953 (GeV/$c)^2$, the asymmetry is much smaller but also changes sign. The quasi-elastic region is positive and fluctuates around 0.005. At higher $\nu$ the asymmetry becomes negative and fluctuates around $-0.035$, which is most likely due to a resonance of the neutron causing the spin to flip.

% \pagebreak

% ^^^^^^^^^^^^^^^^^^^^^^^^^^^^^^^^^^^^^^^^^^^^^^^^^^^^^^^^^^^^^^^^^^^^^^^^^^^^^^^

\large
\section {Summary} 
\label{summary}
\normalsize
% Summary
% vvvvvvvvvvvvvvvvvvvvvvvvvvvvvvvvvvvvvvvvvvvvvvvvvvvvvvvvvvvvvvvvvvvvvvvvvvvvvvvv

For this dissertation, polarized $^3$He($e,e'n$) asymmetries were measured with the target polarized in three orthogonal directions. The target single-spin asymmetry was measured with the target polarized vertically, $A_y^0$, while beam-target double-spin asymmetries were measured with the target polarized transversely, $A_T$, and longitudinally, $A_L$. For $A_y^0$, this experiment provides the most precise measurements to date at Q$^2=0.456$ (GeV/$c)^2$ and provides the first measurement in the high Q$^2$ range at 0.953 (GeV/$c)^2$. This experiment also provides the first measurements of $A_T$ and $A_L$ performed for these kinematics and this is also the first time that all three measurements have been performed in the same experiment. The $A_y^0$ measurements are in general agreement with earlier measurements from NIKHEF and MAMI. Comparisons with early theory calculations show qualitative agreement, however all theoretical calculations currently underestimate the measurement as one goes to higher Q$^2$. The non-zero results indicate final state interactions and meson exchange current contributions that are under-predicted by available theoretical calculations.

Although theoretical calculations are possible for both $A_T$ and $A_L$, they have not been carried out at the kinematics presented. When available, these new measurements will provide important tests of those theory calculations.

In summary, these measurements put new constraints on theoretical predictions, which to will lead to a better modeling of the $^3$He($e,e'n$) reaction, especially in regards to the contributions of FSI and MEC. These effects are important for using the reaction to extract the neutron information and for better understanding the $^3$He wave function in general.

% ^^^^^^^^^^^^^^^^^^^^^^^^^^^^^^^^^^^^^^^^^^^^^^^^^^^^^^^^^^^^^^^^^^^^^^^^^^^^^^^

					% Include the Chapter 6 text (chapter_6.tex)

\newpage

%	\input{een-polarization-direction-definitions}

%	\input{een-layout-with-text.pdf_tex}
%	 \import{Figures/}{een-layout-with-text.pdf_tex}

	% *********************************************************************

	\end{spacing}								% End of Main Thesis Text
% ^^^^^^^^^^^^^^^^^^^^^^^^^^^^^^^^^^^^^^^^^^^^^^^^^^^^^^^^^^^^^^^^^^^^^^^^^^^^^^^^

\newpage

% Put Bibliographic information last
% vvvvvvvvvvvvvvvvvvvvvvvvvvvvvvvvvvvvvvvvvvvvvvvvvvvvvvvvvvvvvvvvvvvvvvvvvvvvvvvv
%	\fancyhead[RE]{\rightmark}
%	\fancyhead[LO]{\leftmark}	
	\bibliographystyle{prsty}						% Indicate citation in Phys. Rev. Lett. style
	\fancyhead[LO]{\rightmark}	
	\fancyhead[RE]{\leftmark}
	\fancyhead[LE,RO]{\thepage}
	\fancyfoot[LC,RC]{}
	\pagestyle{fancy}
	\newpage
%	\setlength{\topmargin}{-0.1875 in}
%	\setlength{\oddsidemargin}{0.4375 in} 
%	\setlength{\evensidemargin}{0.0625 in}
%	\setlength{\textheight}{8.5 in}

%	\ChTitleVar{\vspace{-0.75 in}\raggedleft \Large\rm}

	\bibliography{bibliography}			% Use the file ``bibliography.bib'' for the biblio
%	\input{bibliography}
	%\bibliographystyle{apalike}					% This uses (name, year) as the citation style
	%\bibliographystyle{apsrev}					% Indicate the sort by citation order style
% ^^^^^^^^^^^^^^^^^^^^^^^^^^^^^^^^^^^^^^^^^^^^^^^^^^^^^^^^^^^^^^^^^^^^^^^^^^^^^^^^

% Include the appendicies here
% vvvvvvvvvvvvvvvvvvvvvvvvvvvvvvvvvvvvvvvvvvvvvvvvvvvvvvvvvvvvvvvvvvvvvvvvvvvvvvvv
	%\addcontentsline{toc}{chapter}{}
	\appendix
	\addcontentsline{toc}{chapter}{Appendices}
	\fancyhead[LO]{APPENDIX \thechapter}
	\fancyhead[RE]	{\leftmark}
	\fancyhead[LE,RO]{\thepage}
	\fancyfoot[LC,RC]{}
\chapter[Quasi-Elastic Family Collaboration List]{Quasi-Elastic Family (E05-102, E05-015, E08-005) Collaboration List} % Chapter Title
\label{Collaboration_list}           % Chapter Label
\normalsize                          % Return to Normal font size
The Quasi-Elastic Family of Experiments (E05-102, E05-015, E08-005) collaborators (in alphabatical order) along with their respective home institutions are listed below (83 people from 31 different institutions).

\begin{center}
\noindent 
	K. Allada$^{1}$,
	B. Anderson$^{2}$,
	J. R. M. Annand$^{3}$,
	T. Averett$^{4}$,
	W. Boeglin$^{5}$,
	P. Bradshaw$^{4}$,
	A. Camsonne$^{1}$,
	M. Canan$^{6}$,
	G. Cates$^{9}$,
	C. Chen$^{7}$,
	J. P. Chen$^{1}$,
	E. Chudakov$^{1}$,
	R. De Leo$^{8}$,
	X. Deng$^{9}$,
	A. Deur$^{1}$,
	C. Dutta$^{10}$,
	L. El Fassi$^{11}$,
	D. Flay$^{12}$,
	S. Frullani$^{13}$,
	F. Garibaldi$^{13}$,
	H. Gao$^{14}$,
	S. Gilad$^{15}$,
	R. Gilman$^{11}$,
	O. Glamazdin$^{34}$,
	S. Golge$^{6}$,
	J. Gomez$^{1}$,
	O. Hansen$^{1}$,
	D. Higinbotham$^{1}$,
	T. Holmstrom$^{28}$,
	J. Huang$^{15}$,
	H. Ibrahim$^{32}$,
	C. W. de Jager$^{1}$,
	E. Jensen$^{16}$,
	X. Jiang$^{17}$,
	G. Jin$^{9}$,
	M. Jones$^{1}$,
	H. Kang$^{18}$,
	J. Katich$^{4}$,
	H. P. Khanal$^{5}$,
	P. King$^{19}$,
	W. Korsch$^{10}$,
	J. LeRose$^{1}$,
	R. Lindgren$^{9}$,
	E. Long$^{2}$,
	H.-J. Lu$^{20}$,
	W. Luo$^{21}$,
	P. Markowitz$^{5}$,
	M. Meziane$^{4}$,
	R. Michaels$^{1}$,
	M. Mihovilovic$^{22}$,
	B. Moffit$^{1}$,
	P. Monaghan$^{7}$,
	N. Muangma$^{15}$,
	S. Nanda$^{1}$,
	B. E. Norum$^{9}$,
	K. Pan$^{15}$,
	D. Parno$^{23}$,
	E. Piasetzky$^{24}$,
	M. Posik$^{12}$,
	V. Punjabi$^{30}$,
	A. J. R. Puckett$^{17}$,
	X. Qian$^{14}$,
	Y. Qiang$^{1}$,
	X. Qui$^{21}$,
	S. Riordan$^{9}$,
	A. Saha$^{1}$,
	B. Sawatzky$^{1}$,
	M. Shabestari$^{9}$,
	A. Shahinyan$^{26}$,
	B. Shoenrock$^{25}$,
	S. Sirca$^{27}$,
	J. St. John$^{28}$,
	R. Subedi$^{29}$,
	V. Sulkosky$^{15}$,
	W. A. Tobias$^{9}$,
	W. Tireman$^{25}$,
	G. M. Urciuoli$^{13}$,
	D. Wang$^{9}$,
	K. Wang$^{9}$,
	Y. Wang$^{33}$,
	J. Watson$^{1}$,
	B. Wojtsekhoski$^{1}$,
	Z. Ye$^{7}$,
	X. Zhan$^{15}$,
	Y.-W. Zhang$^{11}$,
	Y. Zhang$^{21}$,
	X. Zheng$^{9}$,
	B. Zhao$^{4}$,
	L. Zhu$^{7}$\\
  (The Jefferson Laboratory E05-102, E05-015, E08-005, and Hall A Collaborations)
  % \\[5mm]

\noindent
	\small{$^{1}$Thomas Jefferson National Accelerator Facility, Newport News, VA 23606, USA}\\                      
	\small{$^{2}$Kent State University, Kent, OH, 44242, USA}\\ 
	\small{$^{3}$Glasgow University, Glasgow, G12 8QQ, Scotland, United Kingdom}\\ 
	\small{$^{4}$The College of William and Mary, Williamsburg, VA, 23187, USA}\\
	\small{$^{5}$Florida International University, Miami, FL, 33181, USA}\\
	\small{$^{6}$Old Dominion University, Norfolk, VA, 23508, USA}\\
	\small{$^{7}$Hampton University , Hampton, VA, 23669, USA}\\
	\small{$^{8}$Universite di Bari, Bari, 70121 Italy}\\ 
	\small{$^{9}$University of Virginia, Charlottesville, VA, 22908, USA}\\
	\small{$^{10}$University of Kentucky, Lexington, KY, 40506, USA}\\
	\small{$^{11}$Rutgers University, New Brunswick, NJ, 08901, USA}\\
	\small{$^{12}$Temple University, Philadelphia, PA, 19122, USA}\\
	\small{$^{13}$Istituto Nazionale Di Fisica Nucleare, INFN/Sanita, Roma, Italy}\\ 
	\small{$^{14}$Duke University, Durham, NC, 27708, USA}\\
	\small{$^{15}$Massachusetts Institute of Technology, Cambridge, MA, 02139, USA}\\
	\small{$^{16}$Christopher Newport University, Newport News, VA, 23606, USA}\\
	\small{$^{17}$Los Alamos National Laboratory, Los Alamos, NM, 87545, USA}\\
	\small{$^{18}$Seoul National University, Seoul, Korea}\\
	\small{$^{19}$Ohio University, Athens, OH, 45701, USA}\\
	\small{$^{20}$Huangshan University, People's Republic of China}\\
	\small{$^{21}$Lanzhou University, Lanzhou, Gansu, 730000, People's Republic of China}\\
	\small{$^{22}$Jozef Stefan Institute, Ljubljana 1000, Slovenia}\\
	\small{$^{23}$Carnegie Mellon University, Pittsburgh, PA, 15213, USA}\\
	\small{$^{24}$Tel Aviv University, Tel Aviv 69978, Israel}\\
	\small{$^{25}$Northern Michigan University, Marquette, MI, 49855, USA}\\
	\small{$^{26}$Yerevan Physics Institute, Yerevan, Armenia}\\
	\small{$^{27}$University of Ljubljana, Ljubljana, 1000, Slovenia}\\
	\small{$^{28}$Longwood College, Farmville, VA, 23909, USA}\\
	\small{$^{29}$George Washington University, Washington, D.C., 20052, USA}\\
	\small{$^{30}$Norfolk State University, Norfolk, VA, 23504, USA}\\
	\small{$^{31}$Randolph Macon College, Ashland, VA, 23005, USA}\\
	\small{$^{32}$Cairo University, Cairo, Giza 12613, Egypt}\\
	\small{$^{33}$University of Illinois at Urbana-Champaign, Urbana, IL, 61801, USA}\\
	\small{$^{34}$Kharkov Institute of Physics and Technology, Kharkov 61108, Ukraine}\\

%	\small{$^{}$}\\
\end{center}
  
%\end {center}

				% Appendix with the collaboration info 
	%
\chapter{Nucleon Form Factors} % Chapter Title
\label{ap:form-factors}           % Chapter Label
\normalsize                          % Return to Normal font size

%\begin{wrapfigure}{l}{0.3\textwidth}
%	\centering
%	\includegraphics[width=1.1 in]{./hand-numbers.eps}
%	\caption [HAND Plane and Bar Labels] {HAND Plane and Bar Labels. Tables \ref{tab:hand-vetoes-p1} through \ref{tab:hand-vetoes-p4} label which bars were used as vetoes for a hit in any given bar. This figure labels each of the planes, along the top, and each bar, on the bar.}
%	\label{hand-numbers}
%\end{wrapfigure}
As mentioned in Section \ref{pwia}, the nucleon form factors are useful quantities for measuring the contributions of charge and magnetization within the nucleons. There are two related ways of describing the form factors known as the Pauli and Dirac form factors and the Sachs form factors. 

Matrix elements of the nucleon electromagnetic current operator, $J^{\mu}$, are of the form
\begin{equation}	
	\bra{N(p',s')}J^{\mu}\ket{N(p,s)} = \bar{u}(p',s')e\Gamma^{\mu}u(p,s),
\end{equation}
where $u$ is a Dirac spinor, $p$ $(p')$ is the initial (final) momentum, and $s$ and $s'$ are spin four-vectors. The vertex function, $\Gamma^{\mu}$, is described as
\begin{equation}	
	\Gamma^{\mu} = F_1(Q^2)\gamma^{\mu} + \kappa 
F_2(Q^2)\frac{i\sigma^{\mu\nu}q_{\nu}}{2m},
\end{equation}
where $e$ is the charge of an electron, $m$ is the nucleon mass, $\kappa$ is the anomalous part of the magnetic moment, $\gamma^{mu}$ and $\sigma^{\mu\nu}$ are the usual Dirac matrices, and $F_1$ and $F_2$ are the Dirac and Pauli form factors \cite{Kelly:2002if}. The Sachs form factors are linear combinations of the Dirac and Pauli form factors, such that
\begin{equation}
	G_{E} = F_1 - \tau \kappa F_2,
\end{equation}
\begin{equation}
	G_{M} = F_1 + \kappa F_2,
\end{equation}
where $\tau$ is $Q^2/4m^2$ \cite{Kelly:2002if}.  

In the Breit frame, the electron transfers momentum, $\vec{q_B}$, but not energy, $\nu=0$. This causes $Q^2=\vec{q_b}^2$. In this frame, the electromagnetic current separates into electric and magnetic contributions, which are described by the Sachs form factors, as \cite{Kelly:2002if}
\begin{equation}
	u(p',s')\Gamma^{\mu}u(p,s) = \chi^{\dagger}_{s'}\left( G_E + \frac{i\vec{\sigma}\times \vec{q_B}}{2m}G_M \right)\chi_s,
\end{equation}
where $\chi_s$ is a two-component Pauli spinor \cite{Kelly:2002if}. Additionally, the current $J^\mu$ is described in terms of the form factors by \cite{Perdrisat:2006hj}
\begin{equation}
	J^0 = e2M\chi'^{\dagger}\chi(F_1 - \tau F_2) = e2M\chi'^{\dagger}\chi G_E,
\end{equation}
\begin{equation}
	\vec{J} = ie\chi'^{\dagger}(\vec{\sigma}\times \vec{q_B})\chi(F_1 + F_2) = ie\chi'^{\dagger}(\vec{\sigma}\times \vec{q_B})\chi G_M.
\end{equation}

				% Appendix with the veto bar info 
	%
\chapter{Veto Bars used for HAND} % Chapter Title
\label{hand-vetoes}           % Chapter Label
\normalsize                          % Return to Normal font size

\begin{SCfigure}[1]
	\centering
	\includegraphics[width=1.5 in]{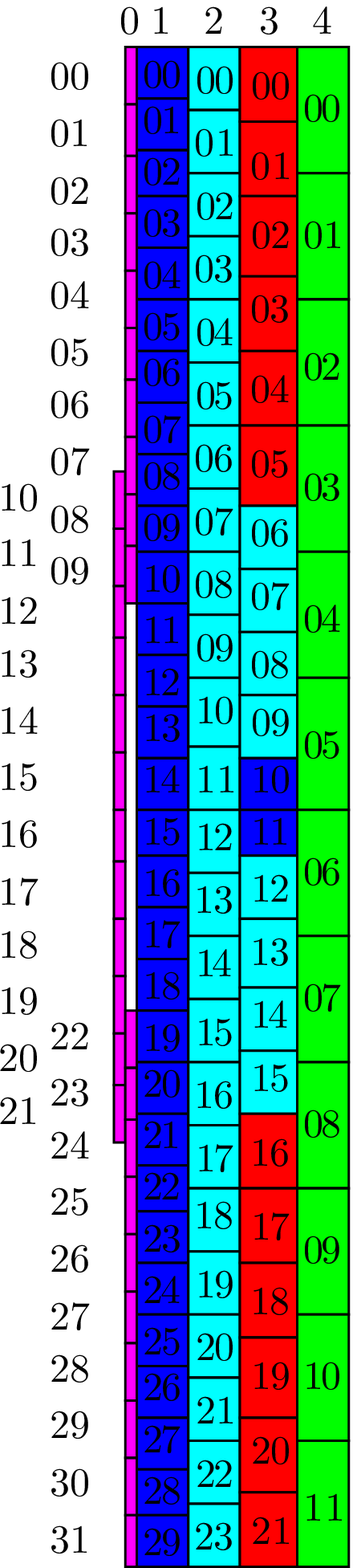}
	\caption [HAND Plane and Bar Labels] {HAND Plane and Bar Labels. Tables \ref{tab:hand-vetoes-p1} through \ref{tab:hand-vetoes-p4} label which bars were used as vetoes for a hit in any given bar. This figure labels each of the planes, along the top, and each bar, on the bar.}
	\label{hand-numbers}
\end{SCfigure}

%\begin{wrapfigure}{l}{0.3\textwidth}
%	\centering
%	\includegraphics[width=1.1 in]{./hand-numbers.eps}
%	\caption [HAND Plane and Bar Labels] {HAND Plane and Bar Labels. Tables \ref{tab:hand-vetoes-p1} through \ref{tab:hand-vetoes-p4} label which bars were used as vetoes for a hit in any given bar. This figure labels each of the planes, along the top, and each bar, on the bar.}
%	\label{hand-numbers}
%\end{wrapfigure}
As discussed in Section \ref{veto-bars}, particle identification of neutrons in the Hall A Neutron Detector required the use of ``veto" bars in order to separate neutrons from protons. This cannot be done through timing information alone, as the time-of-flight peaks overlap for the protons and neutrons. Tables \ref{tab:hand-vetoes-p1} through \ref{tab:hand-vetoes-p4} show, in detail, which bars were used as vetoes for a hit in any given scintillator bar. The tables list the bars and described in Figure \ref{hand-numbers}.

\begin{table}
\begin{center}
\begin{tabular}{c|c||c|c|c|c|c|c|c|c|c|c|c|c}
\multicolumn{2}{c||}{TDC} &\multicolumn{2}{c|}{Veto 1} & \multicolumn{2}{c|}{Veto 2} & \multicolumn{2}{c|}{Veto 3} & \multicolumn{2}{c|}{Veto 4} & \multicolumn{2}{c|}{Veto 5} & \multicolumn{2}{c}{Veto 6}\\ \hline
Pl & Bar & Pl & Bar & Pl & Bar & Pl & Bar & Pl & Bar & Pl & Bar & Pl & Bar \\ \hline
1 & 0 & 0 & 0 & 0 & 1 & 1 & 1 & & & & & & \\
1 & 1 & 0 & 0 & 0 & 1 & 1 & 0 & 1 & 2 & & & & \\
1 & 2 & 0 & 1 & 0 & 2 & 1 & 1 & 1 & 3 &  &  &  &  \\
1 & 3 & 0 & 2 & 0 & 3 & 1 & 2 & 1 & 4 &  &  &  &  \\
1 & 4 & 0 & 3 & 0 & 4 & 1 & 3 & 1 & 5 &  &  &  &  \\
1 & 5 & 0 & 4 & 0 & 5 & 1 & 4 & 1 & 6 &  &  &  &  \\
1 & 6 & 0 & 5 & 0 & 6 & 1 & 5 & 1 & 7 &  &  &  &  \\
1 & 7 & 0 & 6 & 0 & 7 & 1 & 6 & 1 & 8 &  &  &  &  \\
1 & 8 & 0 & 7 & 0 & 8 & 0 & 10 & 1 & 7 & 1 & 9 &  &  \\
1 & 9 & 0 & 8 & 0 & 9 & 0 & 10 & 0 & 11 & 1 & 8 & 1 & 10 \\
1 & 10 & 0 & 9 & 0 & 11 & 0 & 12 & 1 & 9 & 1 & 11 &  &  \\
1 & 11 & 0 & 9 & 0 & 12 & 0 & 13 & 1 & 10 & 1 & 12 &  &  \\
1 & 12 & 0 & 13 & 0 & 14 & 1 & 11 & 1 & 13 &  &  &  &  \\
1 & 13 & 0 & 13 & 0 & 14 & 0 & 15 & 1 & 12 & 1 & 14 &  &  \\
1 & 14 & 0 & 14 & 0 & 15 & 0 & 16 & 1 & 13 & 1 & 15 &  &  \\
1 & 15 & 0 & 15 & 0 & 16 & 0 & 17 & 1 & 14 & 1 & 16 &  &  \\
1 & 16 & 0 & 16 & 0 & 17 & 0 & 18 & 1 & 15 & 1 & 17 &  &  \\
1 & 17 & 0 & 17 & 0 & 18 & 0 & 19 & 1 & 16 & 1 & 18 &  &  \\
1 & 18 & 0 & 18 & 0 & 19 & 0 & 20 & 0 & 22 & 1 & 17 & 1 & 19 \\
1 & 19 & 0 & 19 & 0 & 20 & 0 & 22 & 1 & 18 & 1 & 20 &  &  \\
1 & 20 & 0 & 20 & 0 & 21 & 0 & 22 & 0 & 23 & 1 & 19 & 1 & 21 \\
1 & 21 & 0 & 21 & 0 & 23 & 0 & 24 & 1 & 20 & 1 & 22 &  &  \\
1 & 22 & 0 & 24 & 0 & 25 & 1 & 21 & 1 & 23 &  &  &  &  \\
1 & 23 & 0 & 25 & 0 & 26 & 1 & 22 & 1 & 24 &  &  &  &  \\
1 & 24 & 0 & 26 & 0 & 27 & 1 & 23 & 1 & 25 &  &  &  &  \\
1 & 25 & 0 & 27 & 0 & 28 & 1 & 24 & 1 & 26 &  &  &  &  \\
1 & 26 & 0 & 27 & 0 & 28 & 0 & 29 & 1 & 25 & 1 & 27 &  &  \\
1 & 27 & 0 & 28 & 0 & 29 & 0 & 30 & 1 & 26 & 1 & 28 &  &  \\
1 & 28 & 0 & 29 & 0 & 30 & 0 & 31 & 1 & 27 & 1 & 29 &  &  \\
1 & 29 & 0 & 30 & 0 & 31 & 1 & 28 &  &  &  &  &  &  \\
\end{tabular}
\caption[Veto Bars Used for HAND Plane 1.]{This table shows, for any given scintillator bar of HAND in the first plane, which surrounding bars were used in the veto cut. Each is labeled by Plane (Pl) and Bar number. The maximum number of vetoes for any given bar is six, however most of the bars use less. This is why there are blank spaces.}
\label{tab:hand-vetoes-p1}
\end{center}
\end{table}

\begin{table}
\begin{center}
\begin{tabular}{c|c||c|c|c|c|c|c|c|c|c|c|c|c}
\multicolumn{2}{c||}{TDC} &\multicolumn{2}{c|}{Veto 1} & \multicolumn{2}{c|}{Veto 2} & \multicolumn{2}{c|}{Veto 3} & \multicolumn{2}{c|}{Veto 4} & \multicolumn{2}{c|}{Veto 5} & \multicolumn{2}{c}{Veto 6}\\ \hline
Pl & Bar & Pl & Bar & Pl & Bar & Pl & Bar & Pl & Bar & Pl & Bar & Pl & Bar \\ \hline
2 & 0 & 1 & 0 & 1 & 1 & 2 & 1 &  &  &  &  &  &  \\
2 & 1 & 1 & 1 & 1 & 2 & 2 & 0 & 2 & 2 &  &  &  &  \\
2 & 2 & 1 & 2 & 1 & 3 & 2 & 1 & 2 & 3 &  &  &  &  \\
2 & 3 & 1 & 3 & 1 & 4 & 1 & 5 & 2 & 2 & 2 & 4 &  & \\ 
2 & 4 & 1 & 4 & 1 & 5 & 1 & 6 & 2 & 3 & 2 & 5 &  &  \\
2 & 5 & 1 & 6 & 1 & 7 & 2 & 4 & 2 & 6 &  &  &  &  \\
2 & 6 & 1 & 7 & 1 & 8 & 2 & 5 & 2 & 7 &  &  &  &  \\
2 & 7 & 1 & 8 & 1 & 9 & 1 & 10 & 2 & 6 & 2 & 8 &  & \\ 
2 & 8 & 1 & 9 & 1 & 10 & 1 & 11 & 2 & 7 & 2 & 9 &  &  \\
2 & 9 & 1 & 11 & 1 & 12 & 2 & 8 & 2 & 10 &  &  &  &  \\
2 & 10 & 1 & 12 & 1 & 13 & 2 & 9 & 2 & 11 & 1 & 11 & 1 & 14 \\
2 & 11 & 1 & 13 & 1 & 14 & 1 & 15 & 2 & 10 & 2 & 12 &  &  \\
2 & 12 & 1 & 14 & 1 & 15 & 1 & 16 & 2 & 11 & 2 & 13 &  &  \\
2 & 13 & 1 & 16 & 1 & 17 & 2 & 12 & 2 & 14 & 2 & 14 &  &  \\
2 & 14 & 1 & 17 & 1 & 18 & 2 & 13 & 2 & 15 & 2 & 15 &  &  \\
2 & 15 & 1 & 18 & 1 & 19 & 1 & 20 & 2 & 14 & 2 & 16 &  &  \\
2 & 16 & 1 & 19 & 1 & 20 & 1 & 21 & 2 & 15 & 2 & 17 &  &  \\
2 & 17 & 1 & 21 & 1 & 22 & 2 & 16 & 2 & 18 & 2 & 18 &  &  \\
2 & 18 & 1 & 22 & 1 & 23 & 2 & 17 & 2 & 19 &  &  &  &   \\
2 & 19 & 1 & 23 & 1 & 24 & 1 & 25 & 2 & 18 & 2 & 20 &  & \\ 
2 & 20 & 1 & 24 & 1 & 25 & 1 & 26 & 2 & 19 & 2 & 21 &  &  \\
2 & 21 & 1 & 26 & 1 & 27 & 2 & 20 & 2 & 22 &  &  &  &  \\
2 & 22 & 1 & 27 & 1 & 28 & 2 & 21 & 2 & 23 &  &  &  &  \\
2 & 23 & 1 & 28 & 1 & 29 & 2 & 22 &  &  &  &  &  &  \\
\end{tabular}
\caption[Veto Bars Used for HAND Plane 2.]{This table shows, for any given scintillator bar of HAND in the second plane, which surrounding bars were used in the veto cut. Each is labeled by Plane (Pl) and Bar number. The maximum number of vetoes for any given bar is six, however most of the bars use less. This is why there are blank spaces.}
\label{tab:hand-vetoes-p2}
\end{center}
\end{table}

\begin{table}
\begin{center}
\begin{tabular}{c|c||c|c|c|c|c|c|c|c|c|c}
\multicolumn{2}{c||}{TDC} &\multicolumn{2}{c|}{Veto 1} & \multicolumn{2}{c|}{Veto 2} & \multicolumn{2}{c|}{Veto 3} & \multicolumn{2}{c|}{Veto 4} & \multicolumn{2}{c}{Veto 5} \\ \hline
Pl & Bar & Pl & Bar & Pl & Bar & Pl & Bar & Pl & Bar & Pl & Bar \\ \hline
3 & 0 & 2 & 0 & 2 & 1 & 3 & 1 &  &   &  & \\
3 & 1 & 2 & 1 & 2 & 2 & 3 & 0 & 3 & 2 &  &  \\
3 & 2 & 2 & 2 & 2 & 3 & 3 & 1 & 3 & 3 & &  \\
3 & 3 & 2 & 3 & 2 & 4 & 3 & 2 & 3 & 4 &  &  \\
3 & 4 & 2 & 4 & 2 & 5 & 2 & 6 & 3 & 3 & 3 & 5  \\
3 & 5 & 2 & 5 & 2 & 6 & 2 & 7 & 3 & 4 & 3 & 6  \\
3 & 6 & 2 & 7 & 2 & 8 & 3 & 5 & 3 & 7 &  &   \\
3 & 7 & 2 & 8 & 2 & 9 & 3 & 6 & 3 & 8 &  &   \\
3 & 8 & 2 & 9 & 2 & 10 & 3 & 7 & 3 & 9  &   & \\
3 & 9 & 2 & 10 & 2 & 11 & 3 & 8 & 3 & 10   &  & \\
3 & 10 & 2 & 11 & 2 & 12 & 3 & 9 & 3 & 11  &  & \\
3 & 11 & 2 & 11 & 2 & 12 & 3 & 10 & 3 & 12  &  &  \\
3 & 12 & 2 & 12 & 2 & 13 & 2 & 14 & 3 & 11 & 3 & 13 \\
3 & 13 & 2 & 13 & 2 & 14 & 3 & 12 & 3 & 14 &  &   \\
3 & 14 & 2 & 14 & 2 & 15 & 3 & 13 & 3 & 15 &  &   \\
3 & 15 & 2 & 15 & 2 & 16 & 3 & 14 & 3 & 16 &  &   \\
3 & 16 & 2 & 16 & 2 & 17 & 2 & 18 & 3 & 15 & 3 & 17  \\
3 & 17 & 2 & 17 & 2 & 18 & 2 & 19 & 3 & 16 & 3 & 18  \\
3 & 18 & 2 & 19 & 2 & 20 & 3 & 17 & 3 & 19 & &  \\
3 & 19 & 2 & 20 & 2 & 21 & 3 & 18 & 3 & 20 &  &  \\
3 & 20 & 2 & 21 & 2 & 22 & 3 & 19 & 3 & 21 &  &  \\
3 & 21 & 2 & 22 & 2 & 23 & 3 & 20 &  &  &  &  \\
\end{tabular}
\caption[Veto Bars Used for HAND Plane 3.]{This table shows, for any given scintillator bar of HAND in the third plane, which surrounding bars were used in the veto cut. Each is labeled by Plane (Pl) and Bar number. The maximum number of vetoes for any given bar is six, however most of the bars use less. This is why there are blank spaces.}
\label{tab:hand-vetoes-p3}
\end{center}
\end{table}

\begin{table}
\begin{center}
\begin{tabular}{c|c||c|c|c|c|c|c|c|c|c|c|c|c}
\multicolumn{2}{c||}{TDC} &\multicolumn{2}{c|}{Veto 1} & \multicolumn{2}{c|}{Veto 2} & \multicolumn{2}{c|}{Veto 3} & \multicolumn{2}{c|}{Veto 4} & \multicolumn{2}{c|}{Veto 5} & \multicolumn{2}{c}{Veto 6}\\ \hline
Pl & Bar & Pl & Bar & Pl & Bar & Pl & Bar & Pl & Bar & Pl & Bar & Pl & Bar \\ \hline
4 & 0 & 3 & 0 & 3 & 1 & 3 & 1 & 4 & 1 &  &  &  &  \\
4 & 1 & 3 & 1 & 3 & 2 & 3 & 3 & 4 & 0 & 4 & 2  &  & \\
4 & 2 & 3 & 3 & 3 & 4 & 3 & 5 & 4 & 1 & 4 & 3  &  & \\
4 & 3 & 3 & 4 & 3 & 5 & 3 & 6 & 4 & 2 & 4 & 4  &  & \\
4 & 4 & 3 & 6 & 3 & 7 & 3 & 8 & 4 & 3 & 4 & 5  &  & \\
4 & 5 & 3 & 8 & 3 & 9 & 3 & 10 & 3 & 11 & 4 & 4 & 4 & 6 \\
4 & 6 & 3 & 10 & 3 & 11 & 3 & 12 & 3 & 13 & 4 & 5 & 4 & 7 \\
4 & 7 & 3 & 13 & 3 & 14 & 3 & 15 & 4 & 6 & 4 & 8  &  &\\
4 & 8 & 3 & 15 & 3 & 16 & 3 & 17  & 4 & 7 & 4 & 9  &  &\\
4 & 9 & 3 & 16 & 3 & 17 & 3 & 18 & 4 & 8 & 4 & 10  &  &\\
4 & 10 & 3 & 18 & 3 & 19 & 3 & 20 & 4 & 9 & 4 & 11  &  &\\
4 & 11 & 3 & 20 & 3 & 21 & 3 & 21 & 4 & 10  &  &\\
\end{tabular}
\caption[Veto Bars Used for HAND Plane 4.]{This table shows, for any given scintillator bar of HAND in the fourth plane, which surrounding bars were used in the veto cut. Each is labeled by Plane (Pl) and Bar number. The maximum number of vetoes for any given bar is six, however most of the bars use less. This is why there are blank spaces.}
\label{tab:hand-vetoes-p4}
\end{center}
\end{table}

				% Appendix with the veto bar info 

% ^^^^^^^^^^^^^^^^^^^^^^^^^^^^^^^^^^^^^^^^^^^^^^^^^^^^^^^^^^^^^^^^^^^^^^^^^^^^^^^^

% Include the author's vita here
% vvvvvvvvvvvvvvvvvvvvvvvvvvvvvvvvvvvvvvvvvvvvvvvvvvvvvvvvvvvvvvvvvvvvvvvvvvvvvvvv
	%\vitaauthor{Elena Amanda Long}
	%\begin{thesisauthorvita}
	%\input{vita.tex}							% Include the vita
	%\end{thesisauthorvita}
% ^^^^^^^^^^^^^^^^^^^^^^^^^^^^^^^^^^^^^^^^^^^^^^^^^^^^^^^^^^^^^^^^^^^^^^^^^^^^^^^^

% Include the University Microfiche Abstract here
% vvvvvvvvvvvvvvvvvvvvvvvvvvvvvvvvvvvvvvvvvvvvvvvvvvvvvvvvvvvvvvvvvvvvvvvvvvvvvvvv
	\fontsize{10}{12}\selectfont
	%\begin{umiabstract}
	%\input{umiabstract.tex}						% Include UMI abstract
	%\end{umiabstract}
	\normalsize
% ^^^^^^^^^^^^^^^^^^^^^^^^^^^^^^^^^^^^^^^^^^^^^^^^^^^^^^^^^^^^^^^^^^^^^^^^^^^^^^^^

% If you are indexing include the index format
% vvvvvvvvvvvvvvvvvvvvvvvvvvvvvvvvvvvvvvvvvvvvvvvvvvvvvvvvvvvvvvvvvvvvvvvvvvvvvvvv
	\begin{spacing}{1.0}
	%\printindex								% Print the Index
	\end{spacing}
% ^^^^^^^^^^^^^^^^^^^^^^^^^^^^^^^^^^^^^^^^^^^^^^^^^^^^^^^^^^^^^^^^^^^^^^^^^^^^^^^^

\end{document}